\documentclass[a4paper,11pt]{article}
\pdfoutput=1 

\usepackage{jheppub} 

\usepackage[T1]{fontenc}
\usepackage{multirow,bbold,slashed,wasysym,tikzfeynman,subcaption,booktabs}
\usepackage{xstring}
\usepackage{letltxmacro}
\usepackage{siunitx}
\usepackage{placeins}
\usepackage{tikz}
\usepackage{adjustbox}
\usetikzlibrary{shapes.geometric, arrows, fit}
\usetikzlibrary{automata,positioning}
\usetikzlibrary{shapes.multipart} 

\usepackage{xparse}
\usepackage{todonotes}

\makeatletter
\AddToHook{cmd/appendix/before}{\def\cref@section@alias{appendix}\def\cref@subsection@alias{appendix}}
\makeatother

\allowdisplaybreaks
 
\definecolor{nicered}{rgb}{0.7,0.1,0.1} 
\definecolor{nicegreen}{rgb}{0.1,0.5,0.1}
\definecolor{niceblue}{rgb}{0.0,0.1,0.7}
\hypersetup{colorlinks,citecolor=niceblue,linkcolor=niceblue,urlcolor=niceblue}
\usepackage[capitalize]{cleveref}
\usepackage{lmodern}

\usepackage[normalem]{ulem}

\def \bm#1{\mbox{\boldmath$#1$\unboldmath}}
\def \beq{\begin{equation}}
\def \eeq{\end{equation}}
\def \bea{\begin{eqnarray}}
\def \eea{\end{eqnarray}}

\newcommand*{\powheg}{{\tt POWHEG}}
\newcommand*{\powhegbox}{{\tt POWHEG-BOX}}

\newcommand*{\wbsm}{\ensuremath{\omega_{{\rm BSM}}}}
\newcommand*{\wbsmopt}{\ensuremath{\omega_{{\rm BSM}}^{{\rm opt}}}}
\newcommand*{\wint}{\ensuremath{\omega_{{\rm Int}}}}
\newcommand*{\wintopt}{\ensuremath{\omega_{{\rm Int}}^{{\rm opt}}}}

\newcommand*{\sm}{\ensuremath{\mathrm{SM}}}
\newcommand*{\bsm}{\ensuremath{{\rm BSM}}}
\newcommand*{\inter}{\ensuremath{{\rm Int}}}
\newcommand*{\LOinter}{\ensuremath{\rm LOInt}}

\newcommand*{\btilde}{\ensuremath{\tilde{B}}}

\newcommand*{\Bbar}{\bar{B} (\Phi_{B})}
\newcommand*{\Btilde}{\tilde{B} (\Phi)}
\newcommand*{\PSB}{\Phi_{B}}
\newcommand*{\PSrad}{\Phi_{\rm rad}}
\newcommand*{\PS}{\Phi}

\newcommand*{\myoplus}{+}
\newcommand*{\myominus}{-}
\newcommand*{\myopm}{\pm}

\newcommand*{\noPS}{\text{no}\,\text{-PS}}

\title{Matrix element method at NLO: \\ A fine proof of concept in \texttt{\textbf{POWHEG}}}

\author[a]{Ulrich Haisch,}
\author[a,b]{Jakob Linder,}
\author[a,b]{Luc Schnell,}
\author[a]{Marius Wiesemann}
\author[a,b]{\\ and Giulia Zanderighi}

\affiliation[a]{Max Planck Institute for Physics, \\ Boltzmannstr.~8, 85748 Garching, Germany}
\affiliation[b]{Physik Department T31, Technische Universit{\"a}t München, \\ James-Franck-Straße 1, D-85748
Garching, Germany}

\emailAdd{haisch@mpp.mpg.de}
\emailAdd{linder@mpp.mpg.de}
\emailAdd{schnell@mpp.mpg.de}
\emailAdd{wieseman@mpp.mpg.de}
\emailAdd{zanderighi@mpp.mpg.de}

\preprint{MPP-2026-101} 

\abstract{The matrix element method~(MEM) provides a fully probabilistic approach to confront experimental events with theory, retaining all correlations in the scattering matrix element. While leading-order MEM is widely used and automated, extending it to next-to-leading order~(NLO) in QCD is challenging due to infrared divergences, negative weights, extra final-state partons, and multi-dimensional phase-space integration. We~demonstrate that the~{\tt POWHEG}~method offers a practical path to MEM at NLO~accuracy. By projecting real-emission events onto Born kinematics via the mappings inherited from the~$\tilde{B} (\Phi)$ function, our method consistently includes the hardest QCD radiation while preserving the NLO-accurate normalization. As a proof of concept, we apply the method to fully leptonic $W^+W^-$ production in the Standard Model effective field theory~(SMEFT), considering a CP-even dimension-six triple-gauge-boson operator. At the parton level, the resulting NLO MEM exploits spin and polarization correlations in the final-state lepton kinematics, yielding enhanced discrimination between the SM and BSM hypotheses compared with conventional cut-and-count observables. Our results demonstrate the feasibility of incorporating NLO QCD corrections into the MEM and establish a framework for future applications to a broad range of collider processes.}

\begin{document} 

\maketitle
\flushbottom

\section{Introduction}
\label{sec:introduction}

The matrix element method (MEM)~\cite{Kondo:1988yd,Kondo:1991dw,Dalitz:1991wa,Dalitz:1992np,Kondo:1993in} is a powerful and elegant approach for extracting the maximum amount of information from experimental events. Instead of relying on indirect observables or approximations, MEM directly confronts each event with first-principles theoretical predictions, providing a probabilistic way to test hypotheses. For~detailed overviews of MEM and its modern machine-learning~(ML) adaptations, see, e.g.~,~Refs.~\cite{Fiedler:2010sg,volobouev2011matrixelementmethodhep,Gainer:2013iya,Albertsson:2018maf}. Historically, the method was pioneered by the D{\O} collaboration for the precision measurement of the top-quark mass~\cite{D0:2004lvh,D0:2004rvt}, and it quickly became a standard tool at the Tevatron, applied to both top-quark and Higgs physics~\cite{CDF:2006nne,D0:2006fah,CDF:2009uss,CDF:2009qjw,CDF:2010zcm,D0:2011rkb,D0:2011fla,CDF:2011yhv,CDF:2011wdm,D0:2016ull}. Today,~MEM is widely used at the~LHC, in both its original formulation~\cite{CMS:2013fjq,CMS:2014quz,ATLAS:2014euz,ATLAS:2014kct,CMS:2014nkk,CMS:2015enw,ATLAS:2015utn,ATLAS:2015jmq,CMS:2015cal,CMS:2017dib,CMS:2018fdh,CMS:2019ekd,CMS:2020cga} and in modern~full~ML-based~implementations~\cite{ATLAS:2024jry,ATLAS:2025clx}. Numerous phenomenological studies also exist, including~Refs.~\cite{Gao:2010qx,Bolognesi:2012mm,Andersen:2012kn,Freitas:2012uk,Artoisenet:2013vfa,Anderson:2013afp,Campbell:2013una,Schouten:2014yza,Gritsan:2016hjl,Elahi:2017ppe,Betancur:2017kqe,Gritsan:2020pib,Bahl:2021dnc,Haisch:2021hvy,Haisch:2022rkm,Butter:2022vkj,Heimel:2023mvw,Balzani:2023jas,Haisch:2023aiz,Mastandrea:2024irf,Ghosh:2025fma,Silva:2025hzo,Bahl:2025mib}, in which MEM-inspired methods are applied to collider physics.

What makes MEM so powerful is its ability to retain the full physics information encoded in the matrix element. Each event is assigned a weight under a given theoretical hypothesis, producing a likelihood that measures how well the data aligns with theory. At leading order~(LO), calculating these per-event weights is straightforward, allowing full automation in tools such as {\tt MadWeight}~\cite{Artoisenet:2008zz,Artoisenet:2010cn} and {\tt MoMEMta}~\cite{Brochet:2018pqf}. However, with the high-precision datasets of the LHC, LO predictions are often not enough for stringent tests of the Standard Model~(SM) or searches for beyond-the-SM~(BSM) physics. Extending the MEM to next-to-leading order~(NLO) in QCD~\cite{Alwall:2010cq,Campbell:2012cz,Campbell:2012ct,Campbell:2013hz,Martini:2015fsa,Baumeister:2016maz,Kraus:2019qoq,Kraus:2019myc,Martini:2023ylv,Tartarin:2025gbt,Tartarin:2026uoh} is a natural step toward exploiting existing higher-order calculations with observables that capture the dynamics of the underlying process beyond LO. However, this extension introduces substantial technical challenges. NLO weights require a consistent combination of virtual and real contributions, both of which contain infrared~(IR) divergences. The appearance of negative weights in NLO-accurate event samples further complicates their probabilistic interpretation. In~addition, real emissions introduce extra final-state partons, making the mapping between parton-level configurations and measurable observables more involved. The evaluation of NLO weights also entails multi-dimensional phase-space integrations and numerically stable treatments of large cancellations, while preserving a meaningful event-by-event probabilistic interpretation. As~a~result, a~fully general NLO~MEM has remained~elusive.

In this article, we demonstrate that the \powheg~method~\cite{Nason:2004rx,Frixione:2007vw}, as implemented in the \powhegbox~framework~\cite{Alioli:2010xd,Jezo:2015aia} offers a practical and
theoretically well-founded path to MEM at NLO accuracy. The key ingredient is the $\Btilde$ function, which encodes the NLO-accurate event weight evaluated at fixed kinematics. It is this quantity according to which \powheg~events are generated. Since each event carries an underlying Born configuration sampled from the $\Btilde$-weighted distribution, the MEM likelihood can be constructed directly as a ratio of~$\Btilde$ evaluated under different signal and background hypotheses, without any additional reweighting procedure. Real-emission events are projected onto their underlying Born kinematics via the \powheg~sector maps, ensuring that the effects of the hardest QCD emission are consistently included and that no double counting is introduced. The NLO-accurate normalization of the cross section is preserved throughout. A detailed discussion of the MEM implementation within the \powheg~framework is given in~Section~\ref{sec:MEMNLOPS}.

 To illustrate the application of this approach, we consider fully leptonic
$W^+W^-$ production within the framework of the SM effective field theory~(SMEFT)~\cite{Buchmuller:1985jz,Grzadkowski:2010es,Brivio:2017vri,Isidori:2023pyp},
focusing on a CP-even dimension-six triple-gauge-coupling operator that
modifies the polarization structure of the $W$~bosons. This process provides
a well-defined and phenomenologically relevant proof-of-concept application:
the angular structure of the $W$-boson decays contains rich spin information
that is sensitive to BSM modifications of the gauge sector. As demonstrated
in Section~\ref{sec:proofofconcept}, our NLO~MEM implementation exploits the
full kinematic information available at the parton level, including the
spin- and polarization-dependent correlations encoded in the final-state
lepton kinematics, and provides enhanced discrimination between the SM and
BSM hypotheses compared with conventional cut-and-count observables. In
particular, the characteristic angular correlations among the final-state
leptons, including their azimuthal correlations, provide sensitive probes of
BSM effects in diboson production. These results demonstrate the potential
of the NLO~MEM framework and establish a basis for extending the approach to
a broader range of collider processes and, ultimately, to realistic
experimental applications incorporating the treatment of experimentally
unobserved degrees of freedom. We conclude in Section~\ref{sec:conclusions}
with a summary of our results and an outlook for future applications of the
method. Additional technical details and supplementary material are provided
in a series of appendices.

\section{MEM at NLO}
\label{sec:MEMNLOPS}

In this section, we provide a detailed description of how the \powheg~method can be employed to perform MEM analyses at NLO~accuracy. We begin with a brief review of the conceptual role of the $\Bbar$ and $\Btilde$ functions within the \powheg~framework, which underlies the generation of Born configurations and the consistent inclusion of both virtual and real-emission NLO~corrections. For more detailed information on the \powheg~method, we refer the reader to~Refs.~\cite{Frixione:2007vw,Alioli:2010xd}. We~then illustrate how $\Btilde$ can be used to construct an extended likelihood ratio that fully captures~a~BSM~signal, the SM background, and their interference. Our~discussion further covers the construction of suitable classifiers and their implementation within the {\tt POWHEG-BOX}, including relevant technical aspects, subtleties, and~limitations.

\subsection{\powheg~method in a nutshell}
\label{sec:POWHEGnutshell}

Within the \powhegbox~framework, the $\Bbar$ function is a central ingredient of the NLO matching formalism and provides the basis for event generation. It represents the differential cross section evaluated at fixed underlying Born kinematics $\PSB$, inclusive over unresolved real radiation phase space $\PSrad$ and incorporating the full NLO QCD corrections.

In practice, the \powheg~algorithm generates phase-space configurations according to the function $\Btilde$, which can be regarded as the fully differential counterpart of~$\Bbar$. Its precise definition is given in~Eq.~(4.13) of~Ref.~\cite{Frixione:2007vw}. Schematically it reads
\beq \label{eq:Btilde}
\begin{split}
\Btilde & = B(\PSB) + V_{\rm fin}(\PSB) \\[2.5mm] 
& \phantom{xx} + \sum_{\alpha} 
\left|\frac{\partial \PSrad}{\partial X_{\rm rad}}\right|
\, \Big[ R(\PSB, \PSrad) - C(\PSB, \PSrad) \Big]_{\bar\Phi^\alpha = \PSB} \\[1mm]
& \phantom{xx} + \sum_{\alpha_\oplus} \frac{1}{z}\left|\frac{\partial z}{\partial X_{\rm rad}}\right| G_{\rm \oplus}^{\alpha_\oplus}(\Phi_{B,\oplus})
+ \sum_{\alpha_\ominus} \frac{1}{z}\left|\frac{\partial z}{\partial X_{\rm rad}}\right| G_{\rm \ominus}^{\alpha_\ominus}(\Phi_{B,\ominus}) \,,
\end{split}
\eeq
where the individual terms correspond to the Born squared matrix element $B(\PSB)$, the IR-finite virtual contribution $V_{\rm fin}(\PSB)$, the subtracted real-emission term $R(\PSB,\PSrad) - C(\PSB,\PSrad)$, and the collinear remnant functions $G_{\rm \oplus}^{\alpha_\oplus}(\Phi_{B,\oplus})$ and $G_{\rm \ominus}^{\alpha_\ominus}(\Phi_{B,\ominus})$. Detailed definitions of these ingredients can be found in~Refs.~\cite{Frixione:2007vw,Alioli:2010xd}. In the \powhegbox~implementation, the subtraction counterterms are constructed using the Frixione-Kunszt-Signer~(FKS) subtraction scheme~\cite{Frixione:1995ms}. The index $\alpha$ labels the partition of phase space into singular regions used to isolate IR divergences, while $\alpha_{\oplus}$ and $\alpha_{\ominus}$ denote the collinear sectors. The~variables
\beq \label{eq:Xrad}
X_{\rm rad} = \left \{ X_{\rm rad}^{(1)}, X_{\rm rad}^{(2)}, X_{\rm rad}^{(3)} \right \} \,,
\eeq
denote the radiation variables associated with each singular region, defined on a unit hypercube and are used to parametrize the real-emission phase space. The longitudinal momentum fraction $z$ entering the collinear remnants is expressed in terms of the same variables. The square brackets around the subtracted real-emission term in~Eq.~(\ref{eq:Btilde}) indicate that, for each singular region $\alpha$, only configurations of the full phase space $\PS = \{\PSB, \PSrad\}$ whose underlying Born kinematics $\bar\Phi^\alpha$ coincide with $\PSB$ are included.

Integrating $\Btilde$ over the radiation variables $X_{\rm rad}$ reproduces $\Bbar$:
\beq \label{eq:BtildetoBbar}
\bar{B}(\PSB) = \int \! d X_{\rm rad} \, \Btilde \,.
\eeq
As a result, upon integration over the phase space $\PS$, the total NLO cross section is recovered.

Given a fixed underlying Born configuration $\PSB$, the hardest emission is generated by sampling the radiation phase space according to the kernel
\beq \label{eq:RoverB}
\frac{R(\PSB, \PSrad)}{B(\PSB)} \,,
\eeq
supplemented by a Sudakov form factor that resums the no-emission probability above the scale set by $\PSrad$. This construction ensures that the hardest-emission distribution is NLO accurate, that inclusive observables reproduce the NLO cross section, and that IR limits are correctly preserved. 

The initial-state flavor configuration $f_b$, implicitly encoded in $\Btilde$, is then selected with probability
\beq \label{eq:flavorprob}
\frac{\tilde{B}_{f_b}(\Phi)}{\sum_{f_b^\prime} \tilde{B}_{f_b^\prime}(\Phi)} \,,
\eeq 
where the sum runs over all allowed initial-state flavor combinations. In this way, the \powheg~algorithm samples the joint distribution in phase and flavor space according to their relative contributions to $\Btilde$.

Since the Sudakov form factor is positive definite and the ratio in~Eq.~(\ref{eq:RoverB}) is non-negative by construction, the \powheg~procedure yields predominantly positive-weight events. The only source of negative weights is the function $\Btilde$ itself, which, as apparent from Eq.~(\ref{eq:Btilde}), can become negative in regions of phase space where large negative virtual corrections $V_{\rm fin}(\PSB)$ outweigh the positive Born contribution $B(\PSB)$ and the real-emission term~$R(\PSB,\PSrad)$. Since $\Btilde$ enters as the underlying differential weight in the generation, any such negativity propagates directly to the event weights, with cancellations occurring only after integration over $\PSrad$. The implications of negative-weight events for our NLO~MEM implementation are discussed in general terms in the following subsection and in practice in~Section~\ref{sec:analysis}.

\subsection{NLO~MEM classifiers}
\label{sec:classifiers}

The core objective of the MEM is to optimally distinguish events originating from different underlying probability distributions. In practice, this approach is typically used to determine or constrain theory parameters by disentangling their distinct contributions to the differential cross section. In the present analysis, it is employed to separate events consistent with SM dynamics from those exhibiting BSM effects. Statistically, this amounts to constructing or approximating the likelihood ratio between competing hypotheses, which provides the most powerful test in the Neyman-Pearson sense~\cite{Neyman:1933wgr}. This naturally connects the MEM to classical event-weighting techniques~\cite{Barlow:1986ek} and to modern ML~methods for classification and likelihood-free inference~\cite{Pivk:2004ty,Baldi:2014kfa,Cranmer:2015bka,Brehmer:2018kdj,Brehmer:2018eca,Brehmer:2018hga,Brehmer:2019xox,Cranmer_2020}, where event-level weights, matrix-element information, or suitably trained classifiers serve as approximations to the optimal likelihood~ratio.

In our MEM approach, the $\Btilde$ function defined in~Eq.~(\ref{eq:Btilde}) is used to define normalized event densities for each contribution via
\beq \label{eq:prob_distribution}
\mathcal{P}_{\bullet}(\PS) = \frac{\tilde{B}_{\bullet}(\PS)}{\sigma_{\bullet}} \,, \qquad \sigma_{\bullet} = \int \! d\PS^\prime \, \tilde{B}_{\bullet}(\PS^\prime)\,,
\eeq
where the label $\bullet \in \{\sm, \bsm, \inter\}$ distinguishes the SM, pure BSM, and SM-BSM interference contributions and $\sigma_{\bullet}$ are the corresponding NLO-accurate fiducial cross sections. In the~SMEFT setup discussed in~Section~\ref{sec:proofofconcept}, the SM term corresponds to the part of the amplitude independent of the Wilson coefficients, the BSM~contribution is quadratic in the Wilson coefficients, and the interference term is linear in them. This decomposition allows each contribution to be associated with a well-defined differential event density, enabling a fully differential analysis of how the different terms shape the observed distributions. Notice, in particular, that the normalized event densities in~Eq.~(\ref{eq:prob_distribution}) are, by construction, insensitive to the values of the Wilson coefficients.

To separate the three types of events, it is generally sufficient to define two independent classifiers, denoted by $\wbsm(\PS)$ and $\wint(\PS)$, as functions over the full phase space $\PS$. Ideally, each observable is large in regions of phase space favored by its corresponding event type and small where other
contributions dominate. Following~Ref.~\cite{Barlow:1986ek}, we define
\beq \label{eq:mem_weight}
\begin{split}
\wbsm(\PS) & = \frac{\mathcal{P}_{\bsm}(\PS)}{a \hspace{0.5mm} \mathcal{P}_{\sm}(\PS) + b \hspace{0.75mm} \mathcal{P}_{\inter}(\PS) + \mathcal{P}_{\bsm}(\PS)} \,, \\[1mm]
\wint(\PS) & = \frac{b \hspace{0.75mm} \mathcal{P}_{\inter}(\PS)}{a \hspace{0.5mm} \mathcal{P}_{\sm}(\PS) + b \hspace{0.75mm} \mathcal{P}_{\inter}(\PS) + \mathcal{P}_{\bsm}(\PS)} \,.
\end{split}
\eeq
These observables become optimal when the real normalization constants $a$ and $b$ are chosen equal to the corresponding cross-section ratios
\beq \label{eq:optimal}
a = \frac{\sigma_{\sm}}{\sigma_{\bsm}} \,, \qquad b = \frac{\sigma_{\inter}}{\sigma_{\bsm}} \,,
\eeq
with $\sigma_{\bullet}$ defined in~Eq.~(\ref{eq:prob_distribution}). In this case, $\wbsm(\PS)$ and $\wint(\PS)$ provide a measure of the relative compatibility of an event with the $\bsm$ and $\inter$ hypotheses, directly linked to the underlying matrix element.

Two points deserve emphasis. First, as noted in~Ref.~\cite{Barlow:1986ek}, the classifiers in~Eq.~(\ref{eq:mem_weight}) are relatively insensitive to the precise choice of $a$ and $b$: suboptimal values mainly reduce the discriminating power without invalidating the approach. Second, with the optimal constants of~Eq.~(\ref{eq:optimal}), the observables reduce to simple ratios of the $\tilde{B}_{\bullet}(\PS)$ functions
\beq \label{eq:optimal_mem_weight}
\begin{split}
\wbsmopt(\PS) & = \frac{\tilde{B}_{\bsm}(\PS)}{\tilde{B}_{\sm}(\PS) + \tilde{B}_{\inter}(\PS) + \tilde{B}_{\bsm}(\PS)} \,, \\[1mm]
\wintopt(\PS) & = \frac{\tilde{B}_{\inter}(\PS)}{\tilde{B}_{\sm}(\PS) + \tilde{B}_{\inter}(\PS) + \tilde{B}_{\bsm}(\PS)} \,.
\end{split}
\eeq
Although these ``optimal observables'' reduce to probabilistic classifiers in the positive-definite limit, finite-order NLO corrections can induce localized regions of phase space where the numerator or denominator becomes negative. Notice that this construction allows one to express the inclusive likelihood ratio between the SM and the full SM plus BSM hypothesis directly in terms of $\wbsm(\PS)$ and $\wint(\PS)$, thereby providing a MEM-based analog of the mixture models used in ML-based approaches~\cite{Cranmer:2015bka,ATLAS:2025clx,Ghosh:2025fma}.

The distributions $\mathcal{P}_{\bullet}(\PS)$ in~Eq.~(\ref{eq:prob_distribution}) are properly normalized by construction. In the absence of negative-weight events, the observables in~Eq.~(\ref{eq:mem_weight}) admit a direct probabilistic interpretation as event-level classifiers for the $\bsm$ and $\inter$ contributions. However, the function $\Btilde$ can become locally negative, such that the corresponding $\mathcal{P}_{\bullet}(\PS)$ should more precisely be interpreted as normalized signed event densities rather than strictly positive probability distributions. Consequently, $\wbsmopt(\PS)$ and $\wintopt(\PS)$ should be viewed as generalized likelihood-ratio observables that quantify the relative compatibility of a phase-space point with the $\sm$, $\bsm$, and $\inter$ hypotheses. Importantly, since $\wbsmopt(\PS)$ and $\wintopt(\PS)$ are constructed from the $\Btilde$ function in~Eq.~(\ref{eq:Btilde}), which satisfies~Eq.~(\ref{eq:BtildetoBbar}), these classifiers preserve full NLO accuracy.

Negative-weight events are a well-known feature of NLO~MC calculations, arising, as discussed in the previous subsection, in regions of phase space where large negative virtual corrections $V_{\rm fin}(\PSB)$ overwhelm the positive Born contribution $B(\PSB)$ and the real-emission term $R(\PSB,\PSrad)$, driving $\Btilde$ negative. In practice, the fraction of negative-weight events in the \powhegbox~framework is small --- for the process considered in~Section~\ref{sec:proofofconcept}, less than~$1\%$ --- but non-zero. A general method called {\tt ESME}~\cite{vanBeekveld:2025lpz} has recently been proposed that eliminates negative weights entirely, and various other approaches have been discussed in the literature~\cite{Nason:2007vt,Jadach:2015mza,Jadach:2016qti,Frederix:2020trv,Nachman:2020fff,Andersen:2021mvw,Danziger:2021xvr,Nason:2021xke,Andersen:2023cku,Frederix:2023hom,Sarmah:2024hdk,Andersen:2024mqh,Shyamsundar:2025nzn,Shyamsundar:2025mfw,Nachman:2025lid,Palmer:2025jmb,Sarmah:2025vnb,Gambhir:2025lka}. Once {\tt ESME} or any alternative positive-definite event generation method is interfaced with the \powhegbox~framework, this issue would be effectively resolved. As~noted in~Ref.~\cite{Barlow:1986ek}, proper normalization of the signed event densities is sufficient to preserve the discriminating power of the observables within the perturbative framework employed here, provided the fraction of negative-weight events remains small. In this regime, negative-weight events neither compromise the optimality of $\wbsm(\PS)$ and $\wint(\PS)$ nor significantly affect their interpretation as measures of relative hypothesis compatibility. In practice, as we demonstrate in~Section~\ref{sec:analysis} and Appendix~\ref{app:interferenceCut}, partitioning the phase space can reduce the impact of negative weights associated with SM-BSM interference, thereby improving the stability and performance of the classifiers. Since phase-space partitioning is a process-dependent consideration rather than an intrinsic feature of our NLO~MEM method, we defer its detailed discussion to the relevant process-specific sections.

Before proceeding, we note that~Refs.~\cite{Tartarin:2025gbt,Tartarin:2026uoh} recently proposed an NLO~MEM method also built on the \powhegbox~framework. Unlike our approach, which assigns event-level probabilities via the $\Btilde$ function as in~Eq.~(\ref{eq:prob_distribution}), their method evaluates the MEM integrand directly using the NLO cross section point by point, including the Born, virtual, and real-emission contributions, but omitting the subtraction counterterm and the collinear remnants, cf.~Eq.~(\ref{eq:Btilde}). To maintain an IR-safe likelihood across the full phase space, a transverse-momentum cutoff and a collinear treatment for the radiated parton are imposed to regulate soft and collinear regions. In contrast, our approach avoids these regulators altogether, naturally preserving the NLO-accurate normalization of the cross section through the \powheg~construction.~Refs.~\cite{Tartarin:2025gbt,Tartarin:2026uoh} apply their method to double-Higgs production via gluon-gluon fusion, a process in which the initial-state parton flavors entering the matrix elements do not need to be reconstructed --- a common challenge in~NLO~MEM implementations that we address explicitly below.

\subsection{Practical implementation}
\label{sec:implementation}

As discussed in Section~\ref{sec:POWHEGnutshell}, events in the \powhegbox~are generated by sampling phase-space points and flavor configurations according to the probability density~$\tilde{B}(\Phi)$. Constructing the classifiers in Eq.~(\ref{eq:mem_weight}), which depend on~$\tilde{B}_{\bullet}(\Phi)$, from a given event therefore effectively amounts to reversing the \powheg~event-generation procedure. When applied to experimentally measured events, however, only the momenta of the observed final-state particles are directly accessible. The underlying partonic event configuration must therefore be reconstructed, including both the initial- and final-state kinematics as well as the flavors of the incoming partons. In addition, the reconstructed event must be mapped onto the corresponding underlying Born and real-emission configurations in order to consistently evaluate the NLO ingredients entering the classifier.

\paragraph{Mapping to underlying Born variables:} 

A crucial step in reconstructing the $\tilde{B}_{\bullet}(\Phi)$ functions is the mapping of a full phase-space configuration (for instance of a measured event), which includes additional radiation, onto the underlying Born phase space $\Phi_B$. For this purpose, we adapt the procedure introduced in~Ref.~\cite{Frixione:2007vw}, which we briefly review here. 

For fully leptonic $W^+ W^-$ production in proton-proton collisions, as considered below in~Section~\ref{sec:WWproduction}, our first goal is to reconstruct the corresponding underlying Born configuration for the process 
\beq \label{eq:process}
p(x_{\myoplus} \hspace{0.5mm} K_{\myoplus}) \, p(x_{\myominus} \hspace{0.5mm} K_{\myominus}) \rightarrow \nu_{e} (k_{3}) \, e^{+} (k_{4}) \, \mu^{-} (k_{5}) \, \bar{\nu}_\mu (k_{6}) \, X(k_{\rm rad}) \, ,
\eeq
where $X(k_{\rm rad})$ represents QCD radiation originating from the initial state, with total momentum $k_{\rm rad}$. Depending on the level of modeling, $X(k_{\rm rad})$ may correspond to a single emitted parton or, more generally, to a collection of particles. In an experimental event, these particles would be part of the recorded final state. Here, we assume that~$X(k_{\rm rad})$ is generated by the parton-shower~(PS) evolution of an initial-state parton.

The goal of the mapping is to project the total momentum of the final-state system, excluding radiation
\beq \label{eq:ktot}
k_{\rm tot} = \sum_{i=3}^{6} k_i \, ,
\eeq
onto a purely longitudinal configuration
\beq \label{eq:mapping_requirements}
k_{\rm tot} \; \longrightarrow \; \bar{k}_{\rm tot} = \sum_{i=3}^{6} \bar{k}_i = \begin{pmatrix} \bar{E}_{\rm tot} \\ 0 \\ 0 \\ \bar{k}_{\rm tot}^z \end{pmatrix} \, ,
\eeq
thereby removing its transverse momentum. Here, $\bar{E}_{\rm tot}$ denotes the total energy of the reconstructed system in the laboratory frame, while $\bar{k}_{\rm tot}^z$ is its longitudinal momentum along the beam ($z$) axis. We can now introduce the momentum fractions~$\bar{x}_{\myoplus}$ and~$\bar{x}_{\myominus}$ of the incoming protons, which are determined by momentum conservation at the level of the underlying Born kinematics
\beq \label{eq:xbars}
\bar{x}_{\myoplus} \hspace{0.5mm} K_{\myoplus} + \bar{x}_{\myominus} \hspace{0.5mm} K_{\myominus} = \bar{k}_{\rm tot} \,, \qquad K_{\myopm} = \frac{\sqrt{S}}{2}\, \begin{pmatrix} 1 \\ 0 \\ 0 \\ \pm 1 \end{pmatrix} \,,
\eeq
where $K_{\myopm}$ denote the momenta of the incoming protons in a center-of-mass (COM) frame with energy $\sqrt{S}$. Notice, that $\bar{x}_{\myoplus}$ and $\bar{x}_{\myominus}$ are given explicitly below.

The mapping is performed by removing the transverse component of $k_{\rm tot}$ through a sequence of boosts. First, a longitudinal boost ${\cal B}_{\rm L}$ is applied to eliminate the $z$-component
\beq \label{eq:boostL}
k_{\rm tot} \;\longrightarrow\; \mathcal{B}_{\rm L} \hspace{0.25mm} k_{\rm tot} \,, \quad (\mathcal{B}_{\rm L} \hspace{0.25mm} k_{\rm tot})^z = 0 \,, \quad \vec{\beta}_{\rm L} = \begin{pmatrix} 0 \\[1mm] 
0 \\[1mm] 
\displaystyle \frac{k_{\rm tot}^z}{k_{\rm tot}^0} \end{pmatrix} \,,
\eeq
where $\vec{\beta}_{\rm L}$ denotes the velocity of the longitudinal boost along the beam axis, and $k_{\rm tot}^0$ is the energy of $k_{\rm tot}$. Next, a transverse boost ${\cal B}_{\rm T}$ with velocity $\vec{\beta}_{\rm T}$ is applied to remove the $x$- and $y$-components
\beq \label{eq:boostT}
\mathcal{B}_{\rm L} \hspace{0.25mm} k_{\rm tot} \;\longrightarrow\; \mathcal{B}_{\rm T} \hspace{0.25mm} \mathcal{B}_{\rm L} \hspace{0.25mm} k_{\rm tot} \,, \quad (\mathcal{B}_{\rm T} \hspace{0.25mm} \mathcal{B}_{\rm L} \hspace{0.25mm} k_{\rm tot})^{x,y} = 0 \,, \quad \vec{\beta}_{\rm T} = \frac{1}{(\mathcal{B}_{\rm L} \hspace{0.25mm} k_{\rm tot})^0} \, \begin{pmatrix} \left (\mathcal{B}_{\rm L} \hspace{0.25mm} k_{\rm tot} \right )^x \\[1mm]
\left (\mathcal{B}_{\rm L} \hspace{0.25mm} k_{\rm tot} \right )^y \\[1mm]
0 \end{pmatrix} \,.
\eeq
Finally, the longitudinal component is restored using the inverse boost $\mathcal{B}_{\rm L}^{-1}$ with velocity $-\vec{\beta}_{\rm L}$ as defined in \cref{eq:boostL}
\beq \label{eq:boostrestore}
\mathcal{B}_{\rm L}^{-1} \hspace{0.25mm} \mathcal{B}_{\rm T} \hspace{0.25mm} \mathcal{B}_{\rm L} \hspace{0.25mm} k_{\rm tot} = \bar{k}_{\rm tot} \,,
\eeq
thereby completing the mapping onto the underlying Born configuration and defining a unique $\bar{k}_{\rm tot}$.

Notice that this particular choice of mapping ensures that both the invariant mass
\beq \label{eq:mtot}
m^2_{\rm tot} = k_{\rm tot}^2 = \bar k_{\rm tot}^2 \,,
\eeq
and the rapidity of the system defining the underlying Born configuration
\beq \label{eq:ytot}
y_{\rm tot} = \frac{1}{2} \ln \left ( \frac{k_{\rm tot}^0 + k_{\rm tot}^z}{k_{\rm tot}^0 - k_{\rm tot}^z} \right ) = \frac{1}{2} \ln \left ( \frac{\bar k_{\rm tot}^0 + \bar k_{\rm tot}^z}{\bar k_{\rm tot}^0 - \bar k_{\rm tot}^z} \right ) \,,
\eeq
are preserved under the mapping~(\ref{eq:boostrestore}).

Given the final-state momenta $\bar{k}_{i}$ with $i = 3, 4, 5, 6$ of the underlying Born configuration, the associated momentum fractions $\bar{x}_{\myopm}$ can be determined using Eq.~(\ref{eq:xbars}), yielding
\beq \label{eq:InitialStateMomentaRec}
\bar{x}_{\myopm} \hspace{0.5mm} = \frac{m_{\rm tot}}{\sqrt{S}} \, e^{\pm y_{\rm tot}} \,,
\eeq
where $m_{\rm tot}$ and $y_{\rm tot}$ are defined in~Eqs.~(\ref{eq:mtot}) and (\ref{eq:ytot}), respectively.

\begin{figure}[t!]
\centering

\begin{tikzpicture}[
 node distance=2.5cm,
 process/.style = {rectangle, rounded corners, minimum width=10cm, minimum height=1.5cm, text centered, text width=10cm, draw=black, fill=orange!30},
 arrow/.style = {ultra thick,->,>=stealth}
 ]

\node (pro0) [process] {Map final-state momenta $k_i$ to underlying Born momenta $\bar{k}_i$ using~Eq.~(\ref{eq:boostrestore})};

\node (pro1) [process, below of=pro0] {Assign underlying Born initial-state momentum fractions $\bar{x}_{\myopm}$ using~Eq.~(\ref{eq:InitialStateMomentaRec})};

\node (pro2) [process, below of=pro1] {Parametrize radiation momentum $k_{\rm rad}$ using~\cref{eq:kradagain} and randomly sample polar angle cosine $y \!\!\!\! \in \!\!\!\! [-1,1]$};

\node (pro3) [process, below of=pro2] {Compute initial-state momentum fractions $x_{\myopm}$ for real-emission configuration using Eqs.~(\ref{eq:PowhegRadParametrisation}),~(\ref{eq:xpxm_reconstructed}) and~(\ref{eq:xiDeterminedfrom4l})};

\draw [arrow] (pro0) -- (pro1);
\draw [arrow] (pro1) -- (pro2);
\draw [arrow] (pro2) -- (pro3);

\end{tikzpicture} 
\vspace{4mm}
\caption{Schematic overview of the procedure to reconstruct the full phase-space kinematics required to evaluate $\tilde{B}(\Phi)$ from the final-state momenta $k_i$ with $i = 3,4,5,6$. For~fully leptonic $W^+ W^-$ production, as discussed in~Section~\ref{sec:WWproduction}, these momenta correspond to the final-state charged leptons and neutrinos.} \label{fig:mapping_flow_chart} 
\end{figure}
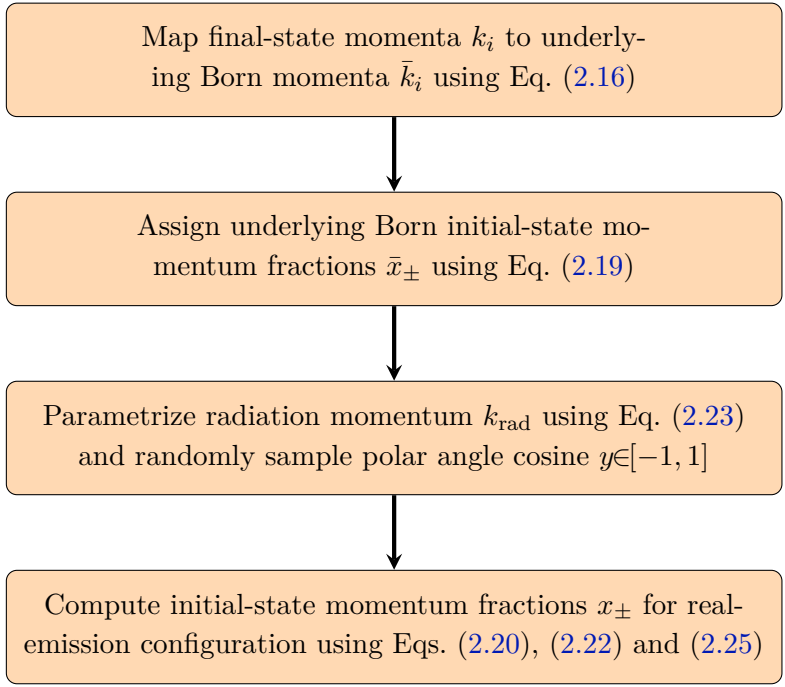

\paragraph{Real radiation:} In order to evaluate the $\tilde{B}(\Phi)$ function the full real phase-space configuration needs to be determined, i.e.~including the radiation variables and the initial-state momentum fractions of the real event.
To this end, we will use that the radiation momentum $k_{\rm rad}$ recoils against the system formed by the leptons in the real event.

In the COM frame of the incoming partons, $x_{\myoplus} \hspace{0.5mm} K_{\myoplus} + x_{\myominus} \hspace{0.5mm} K_{\myominus}$, the radiation momentum can be parametrized as
\beq \label{eq:PowhegRadParametrisation}
k_{\rm rad} = \frac{\sqrt{s}}{2} \, \xi \begin{pmatrix} 1 \\[1mm] 
\sqrt{1 - y^{2}} \, \sin\phi \\[1mm] 
\sqrt{1 - y^{2}} \, \cos\phi \\[1mm] 
y \end{pmatrix} \,, \qquad s = x_{\myoplus} \hspace{0.5mm} x_{\myominus} \hspace{0.5mm} S \,, \qquad y = \cos \theta \,,
\eeq
where $\xi$ is the energy fraction of the radiation, $\theta$ is the polar angle between $k_{\rm rad}$ and the positive $z$-axis, and $\phi$ the azimuthal angle. Applying the sequence of boosts in~Eq.~(\ref{eq:boostrestore}) to the right-hand side of
\beq \label{eq:ktotagain}
k_{\rm tot} = x_{\myoplus} \hspace{0.5mm} K_{\myoplus} + x_{\myominus} \hspace{0.5mm} K_{\myominus} - k_{\rm rad} \,,
\eeq
one finds that the energy fractions $x_{\myopm}$ of the initial-state partons in the event including radiation are related to the Born energy fractions $\bar{x}_{\myopm}$ through
\beq \label{eq:xpxm_reconstructed}
x_{\myopm} = \frac{\bar{x}_{\myopm}}{\sqrt{1 - \xi}} \; \sqrt{\frac{2 - \xi \left ( 1 \mp y \right )}{2 - \xi \left ( 1 \pm y \right )}} \,.
\eeq
Hence, knowledge of the radiation momentum $k_{\rm rad}$ is sufficient to uniquely determine the initial-state momenta of the real phase-space configuration. 

Using that the transverse momentum of the radiation balances that of the four-lepton system, $k_{\rm rad}$ can be written as
\beq
\label{eq:kradagain}
k_{\rm rad} = \begin{pmatrix} \displaystyle \sqrt{\frac{\left ( k_{\rm tot}^x \right )^{2} + \left ( k_{\rm tot}^y \right )^{2}}{1 - y^{2}}} \\[1mm] 
-k_{\rm tot}^x \\[1mm] 
-k_{\rm tot}^y \\[1mm]
\displaystyle y \, \sqrt{\frac{\left ( k_{\rm tot}^x \right )^{2} + \left ( k_{\rm tot}^y \right )^{2}}{1 - y^{2}}} \end{pmatrix} \,,
\eeq
where the energy and longitudinal momentum components have been expressed in terms of the transverse recoil and the variable $y$.

By comparing the energy component $k_{\rm rad}^{0}$ in the two parametrizations, Eqs.~(\ref{eq:PowhegRadParametrisation}) and~(\ref{eq:kradagain}), and using
\beq \label{eq:sagain}
s = x_{\myoplus} \hspace{0.5mm} x_{\myominus} \hspace{0.5mm} S = \frac{\bar{x}_{\myoplus} \hspace{0.5mm} \bar{x}_{\myominus}}{1 - \xi} S \,,
\eeq
as implied by Eq.~(\ref{eq:xpxm_reconstructed}), one can express $\xi$ directly in terms of $k_{\rm tot}$ and $y$ as
\beq\label{eq:xiDeterminedfrom4l}
\xi = \frac{\chi}{2} \left( \sqrt{1 + \frac{4}{\chi}} - 1 \right) \,, \qquad
\chi = \frac{4}{\bar{x}_{\myoplus} \hspace{0.5mm} \bar{x}_{\myominus} \hspace{0.5mm} S} \, \frac{(k_{\rm tot}^x)^2 + (k_{\rm tot}^y)^2}{1 - y^2} \,.
\eeq
The real-emission kinematics are now fully specified up to the parameter~$y$, which is not constrained by the reconstructed four-lepton system and is sampled uniformly in the interval~$y \in [-1,1]$. For the practitioner, Figure~\ref{fig:mapping_flow_chart} provides a concise summary of the main steps of the phase-space mapping.

\paragraph{Initial-state flavor reconstruction:} The final step required to evaluate the~classifiers~$\wbsm(\Phi)$ and~$\wint(\Phi)$ defined in~Eq.~(\ref{eq:mem_weight}) is the determination of the initial-state flavor configuration used in the computation of the $\tilde{B}_{\bullet}(\Phi)$ functions. Instead of summing over all possible flavor combinations, a single configuration is chosen randomly according to the probability distribution in~Eq.~(\ref{eq:flavorprob}), following the same approach employed in the \powheg~method for the evaluation of the inclusive NLO~cross section. In practice, this selection is performed using~$\tilde{B}_{\sm}(\Phi)$. In~Appendix~\ref{app:flavorselection}, we demonstrate that, generally, performing a MC flavor selection in this way yields more stable results than explicitly summing over all initial flavors. We also discuss the choice of $\tilde{B}_{\rm SM}(\Phi)$ compared to $\tilde{B}_{\rm BSM}(\Phi)$ and $\tilde{B}_{\mathrm{SMEFT}_{8}}(\Phi)$, where $\mathrm{SMEFT}_{8}$ denotes the combined $\sm$, $\bsm$, and $\inter$ contributions, providing a posteriori validation of our selection.

\paragraph{Subtleties and limitations:} The algorithm described above provides a robust framework for computing the $\tilde{B}_{\bullet}(\PS)$ contributions. In contrast to existing NLO~MEM implementations, e.g.~Refs.~\cite{Alwall:2010cq,Campbell:2012cz,Campbell:2012ct, Campbell:2013hz,Martini:2015fsa,Baumeister:2016maz,Kraus:2019qoq}, we do not integrate over the unknown initial-state momenta but instead fix them, and determine the three unit-cube variables $X_{\rm rad}$ of \cref{eq:Xrad} through a mapping that effectively inverts the \powheg~procedure for initial-state radiation~(ISR). This~currently restricts the implementation to ISR-dominated processes. We emphasize, however, that this is not a fundamental limitation of the method. Since the \powhegbox~treats initial- and final-state radiation on an equal footing within the FKS subtraction framework, extending the approach to final-state radiation~(FSR), or to processes involving both simultaneously, is relatively straightforward in principle, as demonstrated for example in~Refs.~\cite{Mazzitelli:2020jio,Mazzitelli:2021mmm}. Moreover, multiple valid options exist for reconstructing the radiation momentum $k_{\rm rad}$, not just the one we selected. Appendix~\ref{app:kradReconstructionSelections} illustrates alternatives consistent with energy-momentum conservation and shows minimal impact on the $\wbsm(\PS)$ and $\wint(\PS)$ distributions, indicating that our ISR reconstruction procedure is practical and robust for NLO~MEM analyses.

In order to determine the NLO-accurate fiducial cross sections $\sigma_{\bullet}$ taken as the normalization of the probability distributions in~Eq.~(\ref{eq:prob_distribution}), the weights of all generated phase-space points need to be summed. These weights include the Jacobian factors associated with each phase-space point, cf.~Eq.~(\ref{eq:Btilde}). They are determined during the integration grid optimization of the underlying {\tt MINT}~\cite{Nason:2007vt} algorithm. Starting~from a set of events alone, this factor cannot be reconstructed for an individual phase-space point. The factor cancels, however, in the normalized observables defined in~Eqs.~(\ref{eq:mem_weight}) and (\ref{eq:optimal_mem_weight}), provided that the normalization in~Eq.~(\ref{eq:prob_distribution}) is determined in advance from a full~\powhegbox~run.

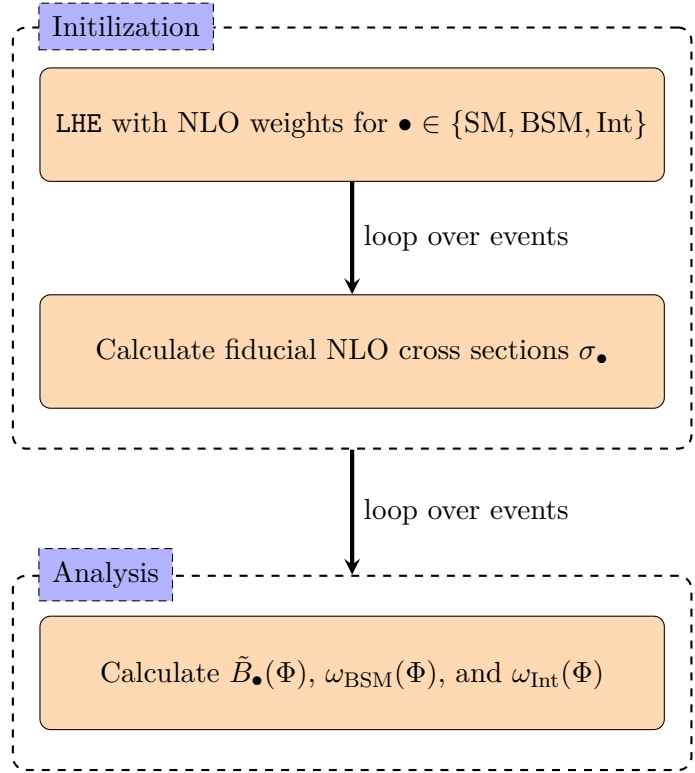
\begin{figure}[t!]
\centering
\begin{tikzpicture}[
 node distance=2cm,
 process/.style = {rectangle, rounded corners, minimum width=8cm, minimum height=1.5cm, text centered, text width=8cm, draw=black, fill=orange!30},
 arrow/.style = {ultra thick,->,>=stealth},
 myfit/.style={draw,dashed,black,thick, inner xsep=10pt, inner ysep=15pt, rounded corners=5pt},
 mytitle/.style={draw,densely dashed,black, fill=blue!30, inner sep=5pt, right, xshift=10pt}]

\node (pro0) [process] {{\tt LHE}~with NLO~weights for~$\bullet \in \{ {\rm SM}, {\rm BSM}, {\rm Int} \}$};

\node (pro1) [process, below of=pro0, yshift=-1cm] {Calculate~fiducial NLO~cross sections~$\sigma_\bullet$};

\node (pro2) [process, below of=pro1, yshift=-2.25cm] {Calculate $\tilde{B}_{\bullet} (\Phi)$, $\wbsm (\Phi)$, and $\wint (\Phi)$};

\draw [arrow] (pro0) -- node[anchor=west] {loop over events} (pro1);

\node[fit=(pro1)(pro0),myfit] (myfit1) {};
\node[mytitle] at (myfit1.north west) {Initilization};

\node[fit=(pro2),myfit] (myfit2) {};
\node[mytitle] at (myfit2.north west) {Analysis};

\draw[arrow] (myfit1) -- node[anchor=west] {loop over events} (myfit2);

\end{tikzpicture} 
\vspace{4mm}
\caption{Schematic workflow of the NLO~MEM~``afterburner''. Starting from an {\tt LHE}~file with NLO~event weights for the $\bullet \in \{ {\rm SM}, {\rm BSM}, {\rm Int} \}$ contributions, fiducial NLO~cross sections~$\sigma_\bullet$ are computed in an initialization phase, followed by an event-by-event analysis phase in which the differential NLO~weights $\tilde{B}_{\bullet}(\Phi)$ and the classifiers defined in~Eq.~(\ref{eq:mem_weight}) are evaluated. Further explanations can be found in the main text.}
\label{fig:mem_flow_chart}
\end{figure}

\paragraph{MEM post-event generation:} So far, we have focused primarily on the technical details underlying our NLO~MEM approach. We now provide a broader overview of how these components can be assembled into a general NLO~MEM framework. This framework can be applied as an ``afterburner'' to pre-generated events, allowing the computation of event-by-event weights $\wbsm(\Phi)$ and $\wint(\Phi)$ without the need to regenerate the full~NLO~event~sample.

A schematic workflow of our ``afterburner'' procedure is illustrated in~Figure~\ref{fig:mem_flow_chart}. The starting point is a Les~Houches~Event ({\tt LHE}) file~\cite{Alwall:2006yp}, generated using a \powheg~code capable of providing a list of NLO weights, and in particular weights associated to the $\sm$, $\bsm$, and their interference $\inter$. The procedure proceeds in two main stages. In the first, initialization stage, the total NLO cross sections are computed for each contribution from the individual NLO~event weights provided in the {\tt LHE} file. These cross sections serve as normalization factors, ensuring that both the overall event sample size and the relative contributions of the different components are properly accounted for. In the subsequent event-by-event analysis stage, the differential NLO~weights $\tilde{B}_{\bullet}(\Phi)$ are evaluated for each event using the mapped underlying Born kinematics, reconstructed initial-state momenta, and assigned initial-state flavors, as described previously. Simultaneously, the classifiers~$\wbsm(\Phi)$ and $\wint(\Phi)$ defined in~Eq.~(\ref{eq:mem_weight}) are constructed for each event.

Thanks to its modular design, the ``afterburner'' realization of the NLO~MEM is highly flexible and largely process-independent. Aside from the subtleties and limitations discussed above, it requires only a dedicated \powhegbox~implementation that computes NLO~weights for the \sm, \bsm, and $\inter$ contributions. Once~such an implementation is available, the ``afterburner'' can process pre-generated event samples and efficiently evaluate the classifiers~$\wbsm(\Phi)$ and~$\wint(\Phi)$ across the full phase space with NLO~accuracy. Furthermore, the method is not limited to \powhegbox~events but, since the phase-space mapping naturally accommodates additional radiation beyond the Born level --- whether described by a single parton or a multi-particle system --- the ``afterburner'' can also be applied to external samples, including fully showered events, both at parton or hadron level, and to experimental data, provided that a corresponding \powhegbox~implementation or an alternative way to compute $\tilde{B}_\bullet(\Phi)$, the relevant phase-space mappings, and the NLO cross sections are available.

\section{SMEFT effects in diboson production}
\label{sec:proofofconcept}

The combination of precise measurements and accurate SM predictions makes diboson processes a sensitive probe of SMEFT effects. Numerous studies~\cite{Melia:2011tj,Degrande:2012wf,Falkowski:2015jaa,Falkowski:2016cxu,Helset:2017mlf,Baglio:2017bfe,Azatov:2017kzw,Panico:2017frx,Franceschini:2017xkh,Chiesa:2018lcs,Liu:2018pkg,Grojean:2018dqj,Baglio:2018bkm,Azatov:2019xxn,Baglio:2019uty,Baglio:2020oqu,Ellis:2020unq,Degrande:2021zpv,Degrande:2023iob,Aoude:2023hxv,Degrande:2024bmd,ElFaham:2024uop,Banerjee:2024eyo,Thomas:2024dwd,Haisch:2025jqr,ElFaham:2025fow,Pelliccioli:2026ltl} highlight their sensitivity to both modified triple-gauge-boson couplings and interactions between fermions and electroweak~(EW) gauge bosons, the latter complementing EW precision measurements. Helicity selection rules suppress the interference between SM and dimension-six contributions~\cite{Azatov:2016sqh}, motivating ``interference-resurrecting'' strategies: chirality-sensitive azimuthal angles~\cite{Azatov:2017kzw,Panico:2017frx,Franceschini:2017xkh,Azatov:2019xxn}, associated jet production~\cite{Azatov:2017kzw}, NLO~QCD corrections~\cite{Panico:2017frx,Franceschini:2017xkh,Azatov:2019xxn}, fiducial lepton cuts~\cite{Azatov:2019xxn}, and off-shell modeling beyond the narrow width approximation~\cite{Helset:2017mlf}. Polarization-sensitive observables~\cite{ElFaham:2024uop,Haisch:2025jqr} have also been explored to further enhance diboson sensitivity. As we will argue below, these features make diboson processes an ideal test case for applying~MEM.

To fix our notation and conventions, we define the~SMEFT Lagrangian of dimension-six operators as
\beq \label{eq:LSMEFT}
{\cal L}_{\rm SMEFT} = \sum_i \frac{C_i (\mu_R)}{\Lambda^2} \, Q_i \,,
\eeq
where $C_i (\mu_R)$ are dimensionless Wilson coefficients at the renormalization scale $\mu_R$, multiplying the corresponding effective operators $Q_i$. In this work, we assume all Wilson coefficients are real. The symbol $\Lambda$ denotes the common BSM~scale suppressing the~SMEFT operators.

While NLO~QCD predictions for the full set of relevant dimension-six operators exist~\cite{Baglio:2017bfe,Chiesa:2018lcs,Baglio:2018bkm,Azatov:2019xxn,Baglio:2019uty,Baglio:2020oqu,Degrande:2024bmd,ElFaham:2024uop,Thomas:2024dwd,Haisch:2025jqr}, we focus here on a CP-even dimension-six triple-gauge-coupling operator. In the Warsaw basis~\cite{Grzadkowski:2010es}, this operator is
\beq \label{eq:W3}
Q_W = \epsilon_{ijk} \hspace{0.25mm} W^{i, \nu}_{\mu} \hspace{0.5mm} W^{j, \lambda}_{\nu} \hspace{0.5mm} W^{k, \mu}_{\lambda} \,,
\eeq
with the $SU(2)_L$ field-strength tensor
\beq \label{eq:fieldstrength}
W_{\mu \nu}^i = \partial_\mu W_\nu^i - \partial_\nu W_\mu^i - g_2 \hspace{0.25mm} \epsilon_{ijk} \hspace{0.25mm} W_\mu^j \hspace{0.5mm} W_\nu^k \,,
\eeq
where $W_\mu^i$ denotes the $SU(2)_L$ gauge field, $g_2$ the corresponding coupling, and $\epsilon_{ijk}$ represents the totally antisymmetric Levi-Civita symbol with $\epsilon_{123} = +1$. Notice that the operator defined in~Eq.~\eqref{eq:W3} is particularly interesting phenomenologically because interference between SM and SMEFT amplitudes is suppressed by the aforementioned helicity selection rules, as different helicity configurations dominate the LO amplitudes in each~case. Compared to other SMEFT operators constructed from Higgs fields and EW gauge field-strength tensors, the study of $Q_W$ is particularly interesting because its Wilson coefficient is less strongly constrained by indirect measurements, such as the $W$-boson mass or the Higgs decay to diphotons~\cite{Haisch:2025jqr}, making EW diboson production at the LHC the most sensitive probe.

In the following, we adopt the benchmark value for the Wilson coefficient of the CP-even dimension-six triple-gauge-coupling operator
\beq \label{eq:CWbenchmark}
\frac{C_W}{\Lambda^2} = \frac{1}{{\rm TeV}^2} \,,
\eeq
where the chosen value agrees with that employed in the recent studies~\cite{ElFaham:2024uop,Haisch:2025jqr}, thereby enabling a direct comparison with the results presented in this work.

The dimension-six operator introduced in~Eq.~(\ref{eq:W3}) alters the structure of the triple-gauge-boson interaction vertices. We describe these interactions using the conventions of~\cite{Hagiwara:1986vm}, restricting ourselves to the Lorentz structures impacted by this operator:
\beq \label{eq:LWWV}
{\cal L}_{WWV} \hspace{0.25mm} \supset \hspace{0.25mm} i \hspace{0.25mm} g_{WWV} \, \frac{\lambda_V}{m_W^2} \, W_{\mu}^{+ \hspace{0.25mm} \nu} \hspace{0.25mm} W_{\nu}^{- \hspace{0.25mm} \rho} \hspace{0.25mm} {V_\rho}^\mu \,, \qquad V = Z, \gamma \,. 
\eeq
The corresponding overall couplings and field-strength tensors are defined as
\beq \label{eq:gWWV}
\begin{gathered}
g_{WWZ}= -\frac{c_w}{s_w} \hspace{0.5mm} e \,, \qquad g_{WW\gamma}= -e \,, \\[1mm]
W_{\mu \nu}^{\pm}= \partial_\mu W_\nu^\pm - \partial_\nu W_\mu^\pm \,, \qquad V_{\mu \nu}= \partial_\mu V_\nu - \partial_\nu V_\mu \,,
\end{gathered}
\eeq
where $e$ is the electromagnetic coupling, $s_w$ and $c_w$ denote the sine and cosine of the weak mixing angle, and $W_\mu^\pm$ and $V_\mu$ represent the physical gauge-boson fields. The parameters~$\lambda_V$ appearing in~Eq.~(\ref{eq:LWWV}) are directly related to the Wilson coefficient of the operator in~Eq.~(\ref{eq:W3}), yielding
\beq \label{eq:lambdaV}
\lambda_Z = \lambda_\gamma = -\frac{3 \hspace{0.125mm} e}{2 \hspace{0.125mm} s_w} \hspace{0.125mm} \frac{v^2}{\Lambda^2} \hspace{0.5mm} C_{W} \,.
\eeq
Here, $v$ denotes the Higgs vacuum expectation value~(VEV). The overall minus sign originates from the convention chosen for the non-abelian term in the $SU(2)_L$ field-strength tensor defined in Eq.~(\ref{eq:fieldstrength}). Using our benchmark choice for the Wilson coefficient $C_W$ in~Eq.~(\ref{eq:CWbenchmark}), together with the numerical input parameters specified in~Section~\ref{sec:WWproduction}, this gives the numerical result
\beq \label{eq:NlambdaV}
\lambda_Z = \lambda_\gamma = -5.93 \cdot 10^{-2} \,.
\eeq

\subsection[$W^+ W^-$ production]{$\bm{W^+ W^-}$ production}
\label{sec:WWproduction}

\begin{figure}[t!]
\centering
\begin{subfigure}[t]{0.4\textwidth}
\centering
\vspace{5pt}
\begin{tikzpicture}[line width=1.5 pt, scale=1.0]
	\draw[fermion] (1,0) -- (-0.5,0);
	\draw[fermion] (-0.5,2) -- (1,2);
	\draw[fermion] (1,2) -- (1,0);
	\draw[vector] (1,0) -- (2.5,0);
	\draw[vector] (1,2) -- (2.5,2);
 \node at (-0.75,0) {$q$};
	\node at (3.0,0) {$W^-$};	
 \node at (-0.75,2) {$q$};
	\node at (3.0,2) {$W^+$};
 \node at (0.6,1) {$q^\prime$};
\end{tikzpicture}
\end{subfigure}
\hspace{0.5cm}
\begin{subfigure}[t]{0.4\textwidth}
\centering
\vspace{0pt}
\begin{tikzpicture}[line width=1.5 pt, scale=1.3]
	\draw[fermionbar] (-140:1)--(0,0);
	\draw[fermion] (140:1)--(0,0);
	\draw[vector] (0:1)--(0,0);
	\node at (-140:1.2) {$q$};
	\node at (140:1.2) {$q$};
	\node at (0.45,0.3) {$\gamma, Z$};	
 \begin{scope}[shift={(1,0)}]
	\draw[vector] (-40:1)--(0,0);
	\draw[vector] (40:1)--(0,0);
	\node at (-40:1.4) {$W^-$};
	\node at (40:1.4) {$W^+$};
 \draw[fill=black] (-0.1,-0.1) rectangle (0.1,0.1);
 \end{scope}
\end{tikzpicture}
\end{subfigure}
\vspace{4mm} 
\caption{Representative Feynman diagrams for the LO contribution to $W^+ W^-$ production in the SM (left) and SMEFT (right). The black box denotes an insertion of the dimension-six operator introduced in~Eq.~\eqref{eq:W3}. SM diagrams involving triple-gauge boson interactions are not shown, nor are the decays of the $W$ bosons.}
\label{fig:diagrams}
\end{figure}
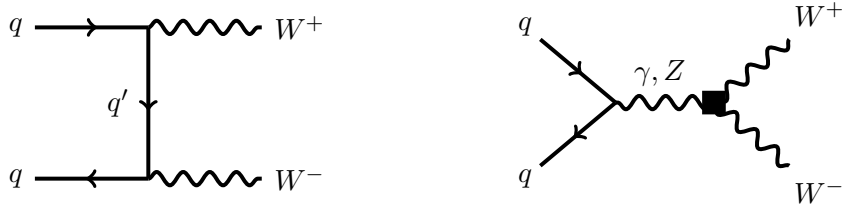

The subsequent MEM study considers fully leptonic $W^+W^-$ production at the LHC. Representative Feynman diagrams for the LO contributions in the SM and SMEFT are shown in Figure~\ref{fig:diagrams}. Compared with other EW diboson processes, $W^+W^-$ production benefits from a relatively large cross section, while the fully leptonic channel poses additional experimental challenges. In particular, top-quark backgrounds with the same final-state signature complicate the isolation of the $W^+W^-$ signal, while the presence of two neutrinos prevents a complete reconstruction of the event kinematics and restricts the set of directly accessible~observables.

The present study is deliberately formulated at the parton level. We assume that the momenta of the charged leptons, neutrinos, and additional partons generated through QCD radiation are fully specified. This idealized setup allows us to isolate the impact of incorporating NLO QCD corrections into the MEM without introducing the additional complications associated with event reconstruction and detector effects. Our aim is therefore not to provide a complete experimental analysis of the $WW\to2\ell2\nu$ final state, but to establish and study the NLO MEM in a controlled setting.

In a realistic experimental analysis, the two neutrinos would remain unobserved and their momenta would be marginalized over in the likelihood rather than uniquely reconstructed. This issue is already present at LO and is a generic feature of the MEM for processes involving invisible particles, as discussed, for example, in~Ref.~\cite{Brochet:2018pqf}. All configurations compatible with the measured charged-lepton momenta and missing transverse momentum would contribute, with the latter providing an experimental constraint on the unobserved degrees of freedom. In the narrow-width approximation, the on-shell conditions of the two intermediate $W$ bosons provide additional constraints and reduce, but do not eliminate, the remaining neutrino phase-space integration. A realistic treatment would additionally require a resolution function for the missing transverse momentum and, where appropriate, transfer functions describing detector effects. As emphasized in~Ref.~\cite{Brochet:2018pqf}, the missing-transverse-momentum resolution is in general correlated with the reconstruction uncertainties of the visible objects and therefore cannot simply be factorized from their transfer functions. ML techniques could provide a flexible way of modeling such correlated detector responses and constructing multi-dimensional transfer functions from simulated or calibration data. The~resulting marginalization reduces the kinematic information available to the MEM, but does not preclude useful discrimination between different hypotheses, as demonstrated, for example, in~studies of off-shell Higgs production in the $pp\to ZZ\to2\ell2\nu$~channel~\cite{Ghosh:2025fma}.

The parton-level treatment adopted here therefore provides a controlled proof of concept for the NLO MEM in a process with non-trivial invisible degrees of freedom. Our NLO~MEM ``afterburner'', illustrated in Figure~\ref{fig:mem_flow_chart}, does not currently support the additional marginalization over unobserved degrees of freedom or the correlated detector-level ingredients required for a realistic experimental analysis. We leave their implementation to future work and focus here on the NLO structure and phenomenological impact of the~MEM~itself.

\paragraph{Numerical input:} All SM input parameters are taken from the latest PDG review~\cite{ParticleDataGroup:2024cfk}. For the on-shell~(OS) masses and total widths of the EW~gauge bosons we use
\beq \label{eq:input1}
m_W^{\rm OS} = 80.369 \, {\rm GeV}\,, \hspace{2.5mm} m_Z^{\rm OS} = 91.188 \, {\rm GeV}\,, \hspace{2.5mm} \Gamma_W^{\rm OS} = 2.085 \, {\rm GeV}\,, \hspace{2.5mm} \Gamma_Z^{\rm OS} = 2.4955 \, {\rm GeV}\,. 
\eeq
These are converted to pole masses via~\cite{Bardin:1988xt}
\beq \label{eq:OStopole}
m_V = \frac{m_V^{\rm OS}}{\sqrt{1 + (\Gamma_V^{\rm OS}/m_V^{\rm OS})^2}}\,, \qquad \Gamma_V = \frac{\Gamma_V^{\rm OS}}{\sqrt{1 + (\Gamma_V^{\rm OS}/m_V^{\rm OS})^2}}\,, \qquad V=W,Z\,,
\eeq
yielding
\beq \label{eq:polemasses}
m_W = 80.342 \, {\rm GeV}\,, \hspace{2.5mm} m_Z = 91.154 \, {\rm GeV}\,, \hspace{2.5mm} \Gamma_W = 2.084 \, {\rm GeV}\,, \hspace{2.5mm} \Gamma_Z = 2.4946 \, {\rm GeV}\,. 
\eeq
In the $G_F$ scheme~\cite{Denner:2000bj}, the EW~input is defined by the weak-boson pole masses together with the Fermi constant:
\beq \label{eq:Fermiconstant}
G_F = 1.1663788 \cdot 10^{-5} \, {\rm GeV}^{-2}\,.
\eeq
The electromagnetic coupling, the cosine of the weak mixing angle, and the Higgs VEV are given by
\bea \label{eq:alphasineVEV}
\alpha = \frac{\sqrt{2} \hspace{0.125mm} s_w^2 \hspace{0.25mm} G_F \hspace{0.25mm} m_W^2}{\pi} = \frac{1}{132.22} \,, \hspace{2.5mm} c_w = \frac{m_W}{m_Z} = 0.8814 \,, \hspace{2.5mm} v = \frac{1}{\sqrt[4]{2} \hspace{0.25mm} \sqrt{G_F}} = 246.22 \, {\rm GeV} \,, \hspace{7mm}
\eea
where the numerical values are obtained using the pole masses and total widths of the EW~gauge bosons specified in~Eq.~(\ref{eq:polemasses}).

Our \powhegbox~simulations utilize the NLO~event generator presented in~Ref.~\cite{Melia:2011tj}. All~leptons are treated as massless, and calculations are performed in the five-flavor scheme, neglecting quark mixing. Parton distribution functions~(PDFs) and the strong coupling~$\alpha_s$ are accessed through the {\tt LHAPDF} interface~\cite{Buckley:2014ana}, employing the {\tt NNPDF31\_nlo\_as\_0118} set~\cite{NNPDF:2017mvq} with $\alpha_s(m_Z)=0.118$. The factorization and renormalization scales are chosen~as
\beq \label{eq:muchoice}
\mu_F = \mu_R = m_W = 80.342 \, {\rm GeV}\,.
\eeq

\paragraph{Selection cuts:} We consider two experimental setups for fully leptonic $W^+ W^-$ production. The first is an inclusive setup with no lepton cuts or jet vetoes. The second is a fiducial setup with selection requirements similar, though slightly simplified, to those used in the ATLAS analysis~\cite{ATLAS:2019rob}. These cuts were also employed in~Ref.~\cite{Lombardi:2021rvg}. The fiducial selections require exactly one oppositely charged electron ($e$) and muon ($\mu$) pair and no jets, with detailed transverse momentum and pseudorapidity ($\eta$) cuts given in~Table~\ref{tab:fiducial}. Jets~are reconstructed using {\tt FastJet}~\cite{Cacciari:2011ma} with the anti-$k_t$ algorithm~\cite{Cacciari:2008gp} and radius parameter~$R = 0.4$. Additional requirements on the missing transverse energy ($E_T^{\rm miss}$) and the transverse momentum of the $e\mu$ system ($p_T^{e\mu}$) suppress Drell-Yan contributions, while a cut on the invariant mass ($m_{e\mu}$) ensures that contamination from $pp \to h \to W^+ W^-$ remains below $1\%$ of the expected signal. At truth level, $E_T^{\rm miss}$ is defined as the magnitude of the vector sum of the transverse momenta of the two neutrinos. We consider both setups because fiducial cuts are known to significantly distort spin- and polarization-dependent correlations among final-state leptons (see, e.g.,~Refs.~\cite{ElFaham:2024uop,Haisch:2025jqr}). Since these correlations are encoded in the matrix element, studying the dependence of MEM performance on the selection criteria allows us to assess the extent to which its sensitivity is preserved under semi-realistic experimental conditions.

{
\def\arraystretch{1.25}
\begin{table}[t!]
\centering
\begin{tabular}{|l|c|}
\hline
variable & selection \\
\hline
$N_e$ & $= 1$, $p_{T,e} > 27 \, {\rm GeV}$, $|\eta_e| < 2.5$ \\
$N_\mu$ & $= 1$, $p_{T,\mu} > 27 \, {\rm GeV}$, $|\eta_\mu| < 2.5$ \\
$N_j$ & $= 0$, $p_{T,j} > 35 \, {\rm GeV}$, $|\eta_j| < 4.5$ \\
\hline
$E_T^{\rm miss}$ & $> 20 \, {\rm GeV}$ \\
$p_T^{e\mu}$ & $> 30 \, {\rm GeV}$ \\
$m_{e\mu}$ & $> 55 \, {\rm GeV}$ \\
\hline
\end{tabular}
\vspace{4mm}
\caption{Fiducial selection cuts for fully leptonic $W^+ W^-$ events with exactly one oppositely charged $e\mu$ pair. $N_e$, $N_\mu$, and $N_j$ denote the number of electrons, muons, and jets, respectively. For further details, consult the text.}
\label{tab:fiducial}
\end{table}
}

\paragraph{Cross sections:} To better understand the impact of higher-order QCD corrections in our MEM analysis, it will turn out to be useful to present cross sections and other key observables at different orders in perturbation theory for both the inclusive and fiducial selections in $W^+ W^-$ production in the channel with exactly one oppositely charged $e\mu$ pair. In~Table~\ref{tab:ww_xsec_summary}, we provide cross-section results for proton-proton collisions at a COM energy of $\sqrt{S} = 13 \, {\rm TeV}$, decomposed into the $\sm$, $\bsm$, and $\inter$ contributions, together with their sum. For the PS, we employ {\tt Pythia~8.244}~\cite{Sjostrand:2014zea}. Note that the $\bsm$ and $\inter$ contributions scale as~$1/\Lambda^4$ and $1/\Lambda^2$, respectively. The~results shown in the table are obtained by generating events for the full~sum of the SM and SMEFT contributions, including terms up to order $1/\Lambda^4$, using the benchmark value of the Wilson coefficient in~Eq.~(\ref{eq:CWbenchmark}), or equivalently Eq.~(\ref{eq:NlambdaV}), and subsequently reweighting them to extract the corresponding $\sm$, $\bsm$, and $\inter$ event weights. The same value of the Wilson coefficient and the same reweighting procedure are used throughout this study.

The overall picture emerging from~Table~\ref{tab:ww_xsec_summary} is that the fiducial selection cuts defined in~Table~\ref{tab:fiducial} lead to a substantial reduction of the cross section. For the NLO QCD predictions, for example, only about $10\%$ and $30\%$ of events survive in the $\sm$ (and hence also in the sum) and $\bsm$ contributions, respectively. In the case of the $\inter$ contribution, the inclusive cross section is negative, while the fiducial one is positive but has a magnitude of only about~$5\%$. To first approximation, these fractions carry over to the NLO$+$PS predictions, as the PS predominantly produces soft QCD radiation and therefore has only a limited impact on the acceptance of the fiducial signal region, which is slightly reduced for all contributions. The observed reduction is mainly driven by the imposed jet veto, which also explains why the ratios of fiducial to inclusive cross sections at LO differ from those at NLO and NLO$+$PS. This underscores that an accurate description of jet-veto effects requires the inclusion of higher-order QCD corrections in the matrix~elements. It is also interesting to note that for the inclusive selection, QCD corrections tend to increase the predictions, while for the fiducial selections, the opposite is the case. Finally, we mention that the SM~results agree within scale uncertainties with the predictions of~Ref.~\cite{Lombardi:2021rvg}, providing a useful cross-check of our MC setup and reweighting~procedure.

{
\def\arraystretch{1.25}
\begin{table}[t!]
\centering
\begin{tabular}{|l|cc|cc|cc|}
\hline
\multirow{2}{*}{$\sigma \, [{\rm fb}]$} & \multicolumn{2}{c|}{LO} & \multicolumn{2}{c|}{NLO} & \multicolumn{2}{c|}{NLO$+$PS} \\
\cline{2-7}
& inclusive & fiducial & inclusive & fiducial & inclusive & fiducial \\
\hline
SM & $869.42(9)$ & $161.0(2)$ & $1269.4(1)$ & $145.1(2)$ & $1269.4(1)$ & $139.19(6)$ \\
BSM & $39.32(8)$ & $28.26(7)$ & $41.61(9)$ & $12.78(5)$ & $41.61(9)$ & $12.18(2)$ \\
Int & $2.95(8)$ & $0.33(6)$ & $-7.79(8)$ & $0.49(3)$ & $-7.80(8)$ & $0.419(9)$ \\
sum & $911.7(1)$ & $189.6(2)$ & $1303.2(2)$ & $158.4(2)$ & $1303.2(2)$ & $151.79(6)$ \\
\hline
\end{tabular}
\vspace{4mm}
\caption{Cross sections in units of ${\rm fb}$ for $\sm$, $\bsm$, and $\inter$, together with their summed contributions at different orders in perturbation theory, are shown for both the inclusive and fiducial selections in $W^+ W^-$ production in the $e\mu$ channel. The quoted $\bsm$ and $\inter$ values correspond to the Wilson coefficient in~Eq.~(\ref{eq:CWbenchmark}), or equivalently~Eq.~(\ref{eq:NlambdaV}). The~numbers in brackets indicate the statistical uncertainties from the~MC~integration. Further details can be found in the main text.}
\label{tab:ww_xsec_summary}
\end{table}
}

\begin{figure}[t!]
\centering
\includegraphics[height=.35\textheight,page=2]{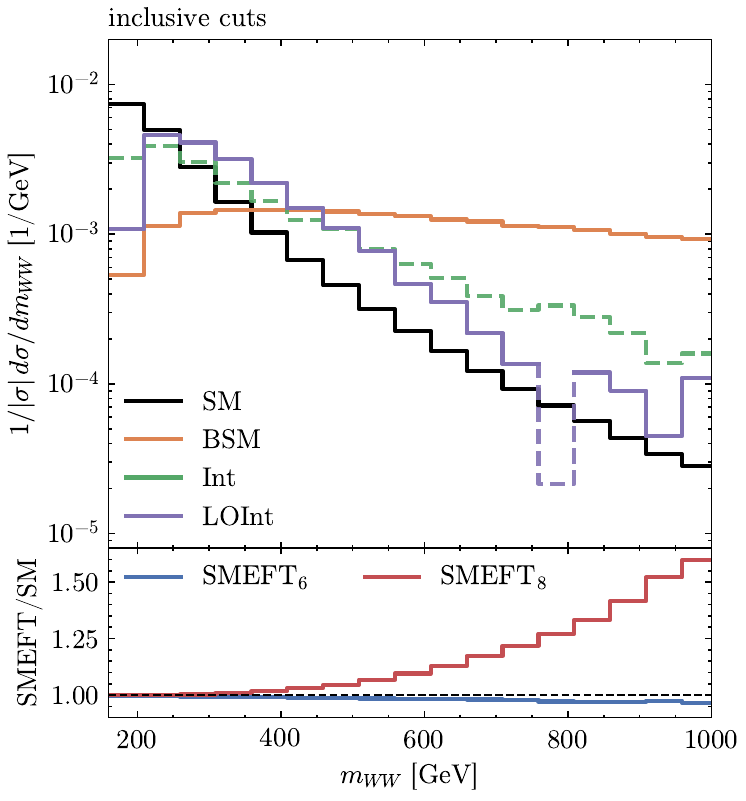} \hfill
\includegraphics[height=.35\textheight,page=2]{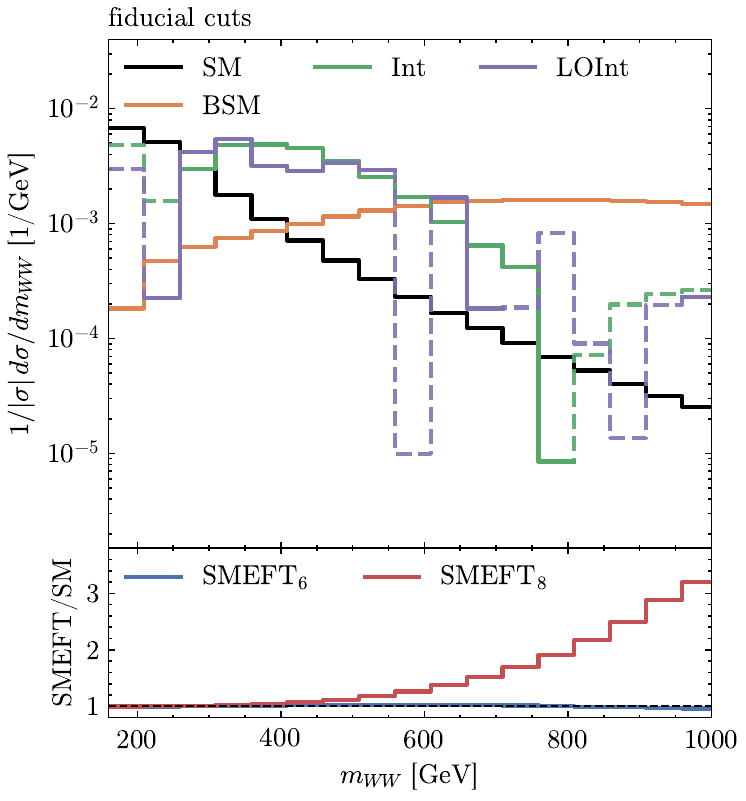}
\vspace{2mm}
\caption{Differential NLO$+$PS distributions for the invariant mass $m_{WW}$. The left and right panels correspond to the inclusive and fiducial setups, respectively. The upper section of each panel shows the $\sm$, $\bsm$, and $\inter$ predictions, normalized to the absolute value of the corresponding cross-section contributions. The $\LOinter$ curve shows the interference contribution at LO. Dashed lines denote negative cross-section values reflected onto the positive axis. The lower sections of each panel show the ratios of the ${\rm SMEFT}_6$ and ${\rm SMEFT}_8$ predictions to the~SM spectrum. See the main text for further explanations.}
\label{fig:xsec_mww}
\end{figure}

\paragraph{Diboson invariant mass:} Another observable worth examining is the invariant mass of the two intermediate $W$ bosons ($m_{WW}$). Our predictions for this spectrum are shown in~Figure~\ref{fig:xsec_mww}, with the left (right) panel corresponding to the inclusive (fiducial) selection. 
The upper sections of the panels show the $m_{WW}$ spectra for the $\sm$, $\bsm$, and $\inter$ contributions, normalized to the absolute values of the corresponding cross sections. The~$\LOinter$ curve shows the LO interference contribution, included to illustrate its sensitivity to QCD corrections and to the details of the experimental setup, most notably the fiducial cuts, as also evident from~Table~\ref{tab:ww_xsec_summary}. The lower sections display the ratios of the ${\rm SMEFT}_6$ and ${\rm SMEFT}_8$ predictions to the~$\sm$ spectrum.

From the results shown in~Figure~\ref{fig:xsec_mww}, it is evident that the fiducial cuts induce a non-trivial distortion of the $m_{WW}$ distribution for all three contributions. In particular, while the high-mass tail is strongly suppressed in the $\sm$ case, the $\bsm$ contribution exhibits a harder spectrum after applying the fiducial selections. For the $\inter$ contribution, the normalized $m_{WW}$ spectrum remains strictly negative in the inclusive setup, whereas it changes sign twice once the fiducial cuts are imposed. The $\LOinter$ contribution, by contrast, exhibits markedly different behavior: it is essentially always positive in the inclusive case, while under fiducial cuts it changes sign several times. This pronounced difference underscores the important role of the jet veto, which preferentially suppresses configurations with additional QCD radiation and thereby significantly reshapes the event kinematics. Since the pattern of QCD radiation is itself strongly affected by higher-order corrections, this also explains the substantial differences between the $\inter$ and $\LOinter$ predictions. As a result, the relative impact of SMEFT effects is enhanced with respect to the $\sm$ prediction in the high-$m_{WW}$ region, making the fiducial selection particularly advantageous for BSM searches in the~SMEFT framework. Quantitatively, at $m_{WW}=1\,{\rm TeV}$, we find ratios of approximately~$1.0$~($3.3$) for the ${\rm SMEFT}_6$~(${\rm SMEFT}_8$) contribution with the fiducial selection, compared with about~$1.0$~($1.6$) for the inclusive setup.

\begin{figure}[t!]
\centering
\includegraphics[height=.35\textheight,page=1]{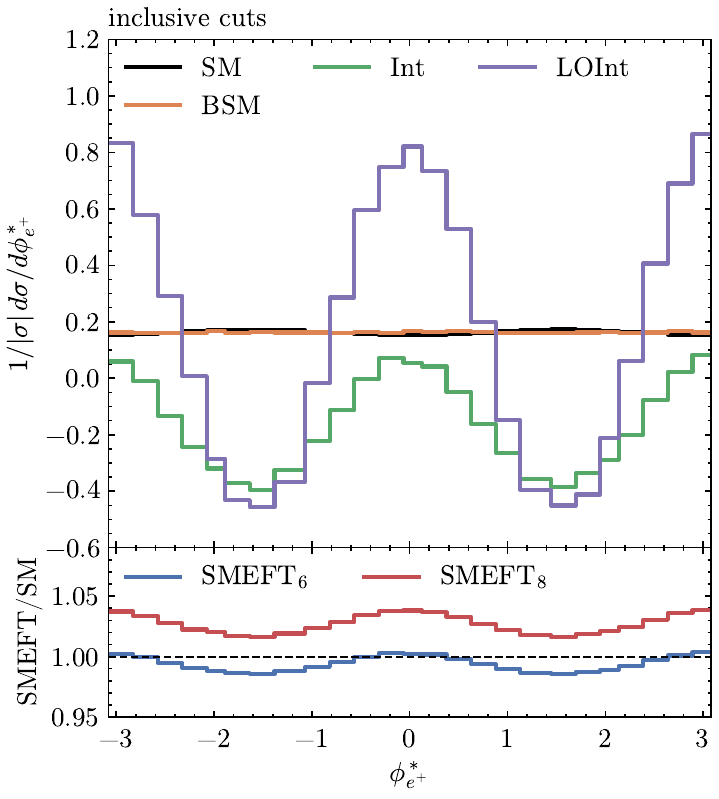} \hfill
\includegraphics[height=.35\textheight,page=1]{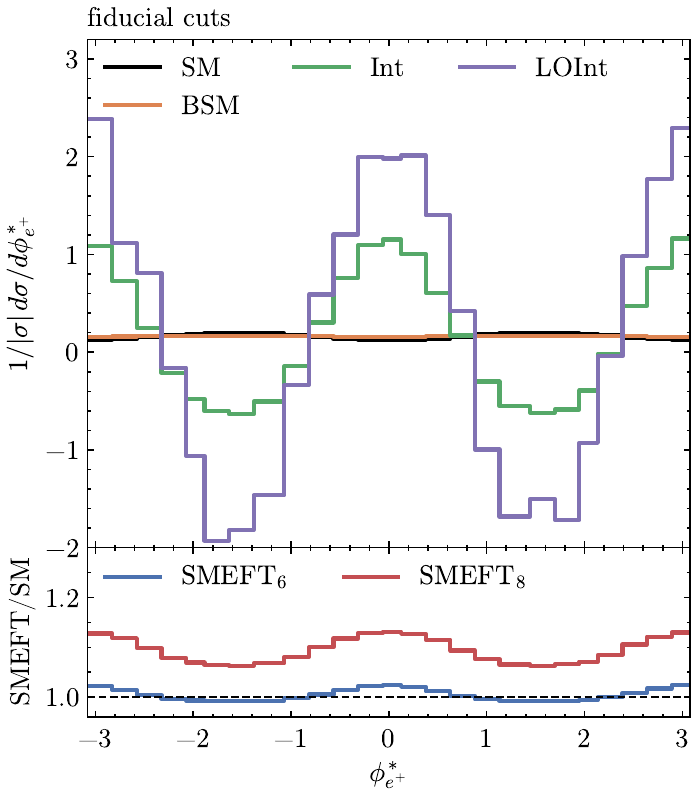}
\vspace{2mm}
\caption{Same as~Figure~\ref{fig:xsec_mww}, but for the observable $\phi_{e^+}^\ast$. See the main text for additional details.}
\label{fig:xsec_phi}
\end{figure}

\paragraph{Lepton azimuthal correlations:} A key feature of $W^+ W^-$ production is that the $W$-boson decays induce characteristic angular distributions of the final-state leptons, providing a sensitive probe of BSM effects. To illustrate how these angular distributions are affected by experimental cuts, we show in~Figure~\ref{fig:xsec_phi} the distribution of the positron azimuthal decay angle in the $W$-boson rest frame ($\phi_{e^+}^\ast$). The left and right panels correspond to the inclusive and fiducial selections, respectively. In the upper section of each panel, we display the $\phi_{e^+}^\ast$ spectra for the~$\sm$, $\bsm$, $\inter$, and $\LOinter$ contributions, normalized to the corresponding absolute cross sections, while the lower sections show the ratios of the ${\rm SMEFT}_6$ and ${\rm SMEFT}_8$ predictions relative to the~$\sm$.

A first observation from~Figure~\ref{fig:xsec_phi} is that the normalized $\phi_{e^+}^\ast$ distributions differ markedly between the $\sm$, $\bsm$, and $\inter$ contributions. In the SM, the dependence on $\phi_{e^+}^\ast$ arises from interference between $W$-boson helicity amplitudes in production and decay, leading to a mild modulation dominated by transverse-transverse interference, with subleading longitudinal-transverse terms. The $\inter$ contribution, corresponding to the interference between the SM amplitude and the $Q_W$ operator~given~in~Eq.~(\ref{eq:W3}), modifies these helicity amplitudes and distorts the SM interference pattern, resulting in significantly stronger shape changes in~$\phi_{e^+}^\ast$. In~contrast, the $\bsm$ contribution is the squared SMEFT amplitude and a positive-definite sum over helicity configurations. Although $Q_W$ mainly affects the transverse polarizations of the $W$ bosons, the absence of SM interference and helicity summation leads to a strong dilution of azimuthal correlations and a nearly flat $\phi_{e^+}^\ast$ dependence. It is clear from a comparison~of the left and right panels that the fiducial cuts have a non-trivial impact on the shape of the~$\phi_{e^+}^\ast$ spectrum for both the $\sm$ and $\inter$ contributions, while the $\bsm$ contribution is only mildly affected. Likewise, whether the interference contribution is evaluated at NLO or LO has a pronounced impact on the normalized $\phi_{e^+}^\ast$ distributions. As a result, and since the fiducial selections enhance the relative SMEFT yield, the ratio of both ${\rm SMEFT}_6$ and ${\rm SMEFT}_8$ to the $\sm$ shows a stronger modulation in the fiducial case than in the inclusive selection. Additional~$W^+ W^-$~distributions can be found in~Appendix~\ref{app:moredistributions}. Having presented predictions for key observables in $W^+ W^-$ production within the SM and SMEFT at different orders in QCD, and having examined the impact of semi-realistic experimental selections on the resulting distributions, we now turn to the MEM analysis.

\subsection{Interpretation of MEM classifiers}
\label{sec:analysis}

A key feature of the MEM is that event classifiers such as those defined in~Eqs.~(\ref{eq:mem_weight}),~(\ref{eq:optimal}), and~(\ref{eq:optimal_mem_weight}) are constructed from the full event kinematics. This allows physical observables to be studied differentially as functions of the classifier, thereby identifying the regions of the phase space that drive the discrimination and providing a direct mapping between classifier response and underlying kinematic configurations. In this way, the classifier serves both as a discriminant and as a diagnostic tool for the kinematic origin of the sensitivity to BSM~effects.

\paragraph{Negative weights and interference:} Negative-weight events are a feature of NLO calculations and, as discussed in~Section~\ref{sec:classifiers}, arise in localized regions of phase space where virtual corrections dominate over the Born and real-emission contributions. In~the~MEM construction for SMEFT contributions to $W^+ W^-$ production, these effects enter the event classifiers $\wbsm(\PS)$ and $\wint(\PS)$ through the underlying $\tilde{B}_{\bullet}(\PS)$ densities. The~classifier~$\wbsm(\PS)$ is primarily sensitive to the absolute rate deformation induced by the BSM-squared contribution, while $\wint(\PS)$ isolates the interference structure between SM and BSM amplitudes. In this sense, $\wint(\PS)$ provides access to sign-dependent physics information, whereas~$\wbsm(\PS)$ captures the positive-definite BSM modification of the total event density.

The interference contribution $\tilde{B}_{\inter}(\PS)$ plays a central role in the definition of $\wint(\PS)$. Unlike $\tilde{B}_{\sm}(\PS)$ and $\tilde{B}_{\bsm}(\PS)$, which are positive definite up to effects associated with negative-event weights, $\tilde{B}_{\inter}(\PS)$ is not sign definite and can change sign across phase space. While in $W^+ W^-$ production the integrated interference contribution $\sigma_{\inter}$ is typically small for the studied SMEFT Wilson coefficient, it remains observable dependent and can exhibit cancellations between positive and negative regions. This can affect the stability of the normalized density entering $\wint(\PS)$ and lead to increased statistical fluctuations in regions where such cancellations are significant.

To mitigate these effects, the event sample is partitioned according to the sign of the LO interference contribution in $W^+ W^-$ production, separating regions with positive and negative interference. This reduces cancellations between opposite-sign contributions within each partition and stabilizes the evaluation of $\wint(\PS)$. The classifier $\wbsm(\PS)$ is only indirectly affected, through the common normalization structure of Eq.~(\ref{eq:mem_weight}), and therefore remains more stable. The sign of the interference term is evaluated at LO, avoiding an additional ambiguity that would arise from combining negative NLO event weights with a negative interference contribution. The real normalization constants $a$ and $b$ appearing in~Eq.~(\ref{eq:mem_weight}) are defined separately in the two interference regions:
\beq \label{eq:ab}
\begin{gathered}
a = 3.09 \,, \qquad b = 0.44 \,, \qquad (\sigma_{\inter}^{\rm LO} > 0 ) \,, \\[1mm]
a = 5.33 \,, \qquad b = -0.47 \,, \qquad (\sigma_{\inter}^{\rm LO} < 0) \,.
\end{gathered}
\eeq
These values are chosen such that, when $\lambda_Z = \lambda_\gamma = -0.1$ is used in the normalized event densities of \cref{eq:prob_distribution}, the classifiers defined in~Eq.~(\ref{eq:mem_weight}) coincide, within each partition, with the ``optimal observables'' introduced in~Eq.~(\ref{eq:optimal_mem_weight}). Additional details on this point can be found in Appendix~\ref{app:cW_dependence}. We~find that removing the interference-sign partitioning leaves the method functional but leads to a noticeable loss in performance, in particular for the interference-sensitive observable~$\wint$, which exhibits increased migration outside its physical range. The impact on $\wbsm$ is significantly smaller. Further details are provided in~Appendix~\ref{app:interferenceCut}. Overall, the partitioning stabilizes interference cancellations and improves numerical robustness.

\begin{figure}[t!]
\centering
\includegraphics[height=.45\textheight,page=1]{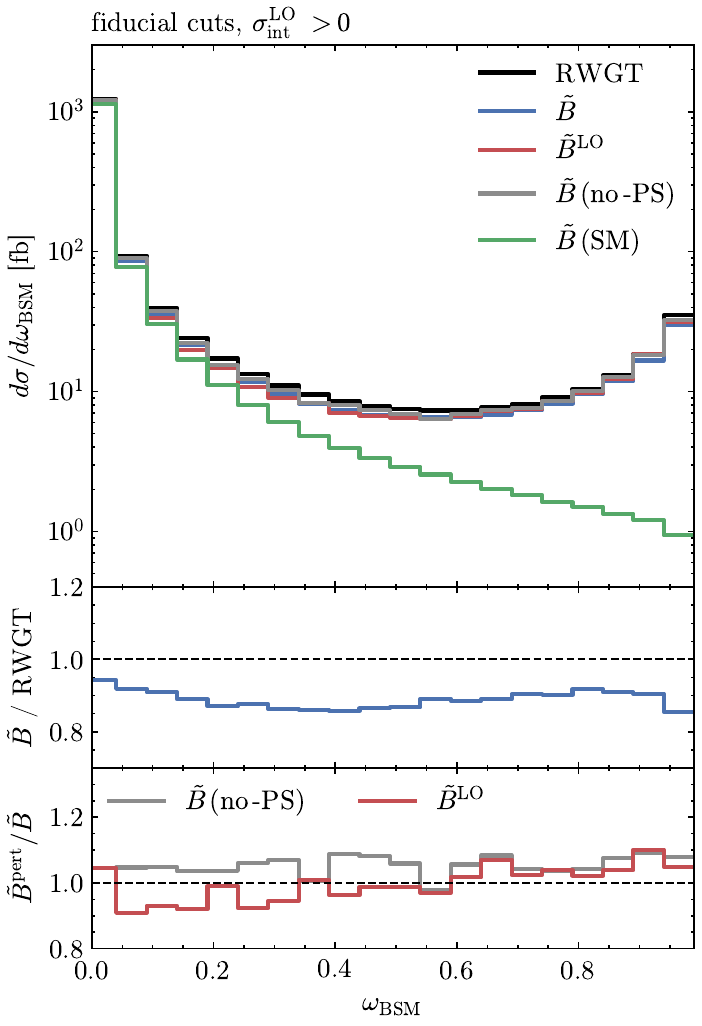} \hfill
\includegraphics[height=.45\textheight,page=5]{plots/hadronic_nlops_dl0p06_fiducial.pdf}
\vspace{2mm}
\caption{Fiducial $W^+ W^-$ cross sections as functions of the classifiers defined in~Eq.~(\ref{eq:mem_weight}). The phase space with positive LO interference is considered and, unless stated otherwise, showered events are used to compute the cross sections. The upper sections compare SMEFT predictions obtained via reweighting, labeled by RWGT, with those derived from different versions of the $\tilde B$ function. Here, $\tilde B$ denotes the full result based on NLO$+$PS events, $\tilde B^{\rm LO}$ the corresponding LO prediction, and $\tilde B \, (\noPS)$ the NLO result obtained without PS effects. The~curves labeled $\tilde B \, (\sm)$ correspond to the SM cross section evaluated with the same~$\tilde B$ function. The middle sections show the ratio of the full~$\tilde B$ prediction to the reweighted result, while the lower sections display the ratios of the~$\tilde B \, (\noPS)$ and $\tilde B^{\rm LO}$ predictions to the full $\tilde B$ result. See the main text for further details.}
\label{fig:MEMbasic}
\end{figure}

\paragraph{MEM classifiers at work:} To illustrate the behavior of the classifiers~(\ref{eq:mem_weight}) in practice, Figure~\ref{fig:MEMbasic} shows fiducial $W^+W^-$ cross sections as functions of $\wbsm$~(left panel) and~$\wint$~(right panel), considering only the phase space with positive LO interference. Additional distributions for negative LO interference are provided in Appendix~\ref{app:moreMEM}. The classifiers are constructed from NLO$+$PS SMEFT events via reweighting in the~\powhegbox, and correspond to Eq.~(\ref{eq:mem_weight}) with $a$ and $b$ chosen as in the first line of Eq.~(\ref{eq:ab}). In~addition to the reweighted prediction, the upper panels show several SMEFT predictions obtained using the NLO~MEM ``afterburner'' illustrated in~Figure~\ref{fig:mem_flow_chart}. The corresponding cross sections from NLO$+$PS events are denoted~$\tilde B$, while~$\tilde B^{\rm LO}$ represents the~LO prediction. The label $\tilde B \, (\noPS)$ refers to NLO samples before parton showering, and $\tilde B \, (\sm)$ denotes the corresponding SM prediction within the same framework. While the resulting differential cross sections depend on the underlying prediction method, the classifiers used as horizontal axes are identical in all cases, which is essential for a consistent probabilistic interpretation and a meaningful comparison between different theoretical descriptions. The~only exception is~$\btilde^{\mathrm{LO}}$, where the LO observable is evaluated on NLO$+$PS events. Notice that this situation mirrors LO~MEM applications to experimental data, where the classifier is constructed at LO~accuracy while the underlying physical events include quantum corrections.

The first notable observation from~Figure~\ref{fig:MEMbasic} is that the $\tilde B$ spectra lie systematically above the corresponding $\tilde B \, (\sm)$ distributions at large values of the classifiers. This indicates that the high-classifier region is dominated by phase-space configurations where SMEFT effects enhance the cross section relative to the SM, while the low-classifier region remains largely SM-like, where both predictions coincide within uncertainties. This confirms that the classifiers in~Eq.~(\ref{eq:mem_weight}) behave as intended. The effect is more pronounced for $\wbsm$ than for $\wint$, since for the benchmark in~Eq.~(\ref{eq:CWbenchmark}) the BSM-squared contribution dominates over the interference term, as reflected in~Table~\ref{tab:ww_xsec_summary}. The~different kinematic ranges of the two observables follow from their matrix-element structure. In~$\wbsm$, the BSM-squared term proportional to $|\mathcal{M}_{\bsm}|^2$ can dominate, leading to $\wbsm \simeq 1$. In contrast, $\wint$ is driven by the interference term $2 \hspace{0.25mm} {\rm Re}\left (\mathcal{M}_{\sm}\mathcal{M}_{\bsm}^\ast \right )$, which is bounded relative to the SM and BSM-squared contributions and can only become comparable in a restricted region where all three terms are similar in size. This yields a maximal value $\wint \simeq 1/2$, for positive LO interference, explaining the ranges $\wbsm \in [0,1]$ and $\wint \in [0,0.5]$, and why~$\wint$ is more sensitive to cancellations.

Another important observation from~Figure~\ref{fig:MEMbasic} is the comparison between the reweighted results and the different evaluations of the $\tilde B$ function on the BSM samples. Overall, the~SMEFT reweighted predictions show very good agreement with the full $\tilde B$ result across the entire range of classifiers, confirming the robustness of the NLO~MEM ``afterburner'' procedure described in~Section~\ref{sec:implementation}. A similarly good agreement is observed for $\tilde B^{\rm LO}$ and $\tilde B \,(\noPS)$ relative to $\tilde B$, with only mild deviations over most of the phase space. The~PS~therefore has only a minor impact on the evaluation of $\wbsm(\PS)$ and $\wint(\PS)$, as expected from the partial cancellation of shower effects in the ratios entering the MEM. Noticeable differences emerge only for larger values of $\wint\gtrsim0.45$, where the $\tilde B^{\rm LO}$ prediction begins to deviate more visibly from the full $\tilde B$ result. This reflects the increasing importance of higher-order QCD corrections in the kinematic configurations contributing to this region of $\wint$. Based on the same argument, we also expect only a limited dependence on the {\tt POWHEG} hardest-emission prescription, in particular on the treatment of hard radiation in the Sudakov exponent, as well as on the choice of shower parameters. While~these effects are expected to be small, their precise impact will depend on the specific experimental setup, most notably on the fiducial cuts. A comprehensive assessment should therefore be performed as part of a dedicated experimental analysis.

\subsection{Performance of MEM classifiers}
\label{sec:MEMperformance}

The performance of the MEM classifiers defined in~Eqs.~(\ref{eq:mem_weight}),~(\ref{eq:optimal}), and~(\ref{eq:optimal_mem_weight}) follows directly from their construction in terms of the full event kinematics. This provides a unified framework in which LO and NLO predictions can be compared, making it possible to quantify how higher-order QCD corrections affect the detailed shape of the distributions while leaving the overall SM-BSM discriminating power largely intact. Relative to standard kinematic observables, the classifiers offer a more direct and robust probe of the underlying dynamics. Moreover, applying cuts on the classifiers naturally enhances regions of phase space with increased BSM sensitivity, effectively acting as optimized selections that improve signal-to-background separation while preserving a transparent physical interpretation. In~the following, we illustrate all these aspects in the case of $W^+ W^-$ production.

\begin{figure}[t!]
\centering
\includegraphics[height=.25\textheight,page=2]{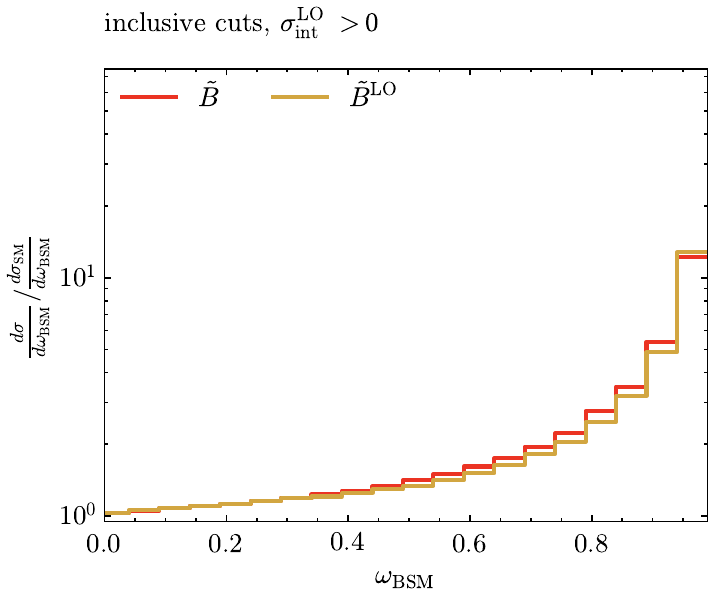} \hfill
\includegraphics[height=.25\textheight,page=2]{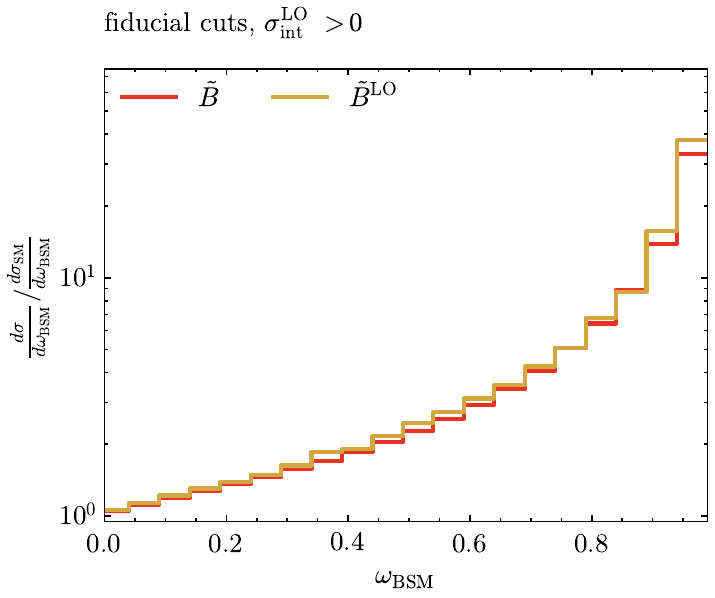}

\vspace{2mm}

\includegraphics[height=.25\textheight,page=6]{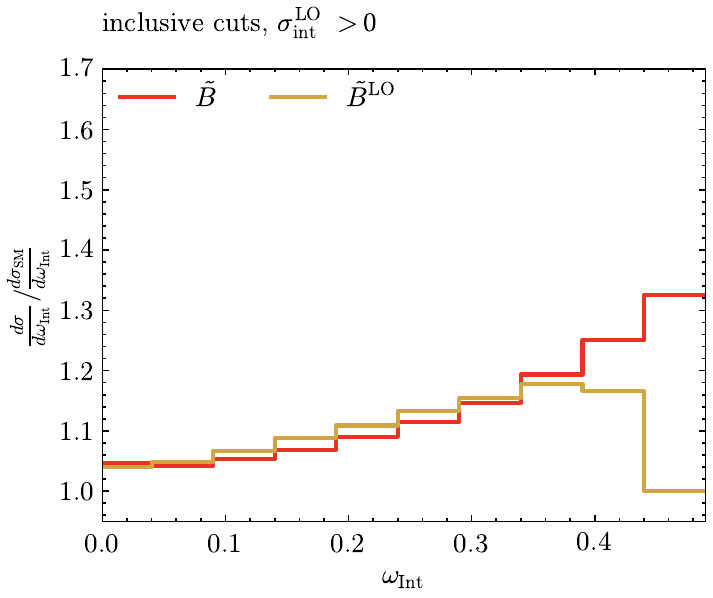} \hfill
\includegraphics[height=.25\textheight,page=6]{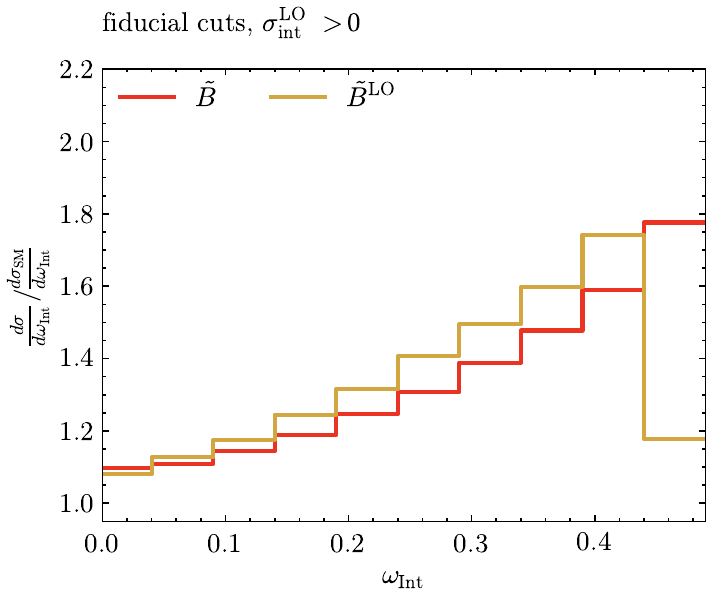}
\vspace{2mm}
\caption{Ratios of the~SMEFT to SM cross sections for $W^+ W^-$ production as functions of the classifiers defined in~Eq.~(\ref{eq:mem_weight}). Only events with a positive LO interference contribution are taken into account, and the results are obtained from showered events for both inclusive~(left) and fiducial~(right) selections. Here, $\tilde B$ denotes results obtained using the NLO $\tilde B$ function, while $\tilde B^{\rm LO}$ corresponds to the associated LO version. Additional details can be found in the main text.}
\label{fig:NLOvsLO}
\end{figure}

\paragraph{Impact of NLO effects:} To assess the impact of higher-order corrections on the sensitivity of the MEM approach to BSM effects, Figure~\ref{fig:NLOvsLO} shows ratios of SMEFT to SM cross sections for $W^+ W^-$ production as functions of the classifiers defined in~Eq.~(\ref{eq:mem_weight}) with the real normalization factors $a$ and $b$ chosen as in the first line of Eq.~(\ref{eq:ab}), obtained from NLO$+$PS SMEFT events. Positive LO interference is assumed, and results are shown for both inclusive (left panels) and fiducial (right panels) selections based on showered events. In the figure, $\tilde B$ denotes cross sections obtained from NLO$+$PS events using the perturbatively improved classifier, while $\tilde B^{\rm LO}$ corresponds to the LO classifier evaluated on the same NLO$+$PS event sample.

Figure~\ref{fig:NLOvsLO} shows that $\wbsm$ provides a substantially stronger separation between SMEFT and SM events than $\wint$, reflecting the dominance of the BSM-squared contribution over the interference term for the benchmark in~Eq.~(\ref{eq:CWbenchmark}), as summarized in~Table~\ref{tab:ww_xsec_summary}. Another observation is that, for inclusive selections, the NLO-accurate version of $\wbsm$ leads to a modest overall improvement in discrimination power compared to the LO construction. For~fiducial cuts, the overall separation increases for both classifiers, while the LO and NLO predictions become more similar. In this regime, $\wbsm$ performs slightly better at~LO. For~$\wint$, this effect is more pronounced, except in the last bin, where the NLO prediction performs noticeably better. These features can be understood from~Section~\ref{sec:WWproduction} and~Appendix~\ref{app:moredistributions}, where it is shown that the jet-veto cuts in~Table~\ref{tab:fiducial} enhance the relative SMEFT contribution while simultaneously reducing the impact of higher-order~QCD corrections in both SM and SMEFT predictions. The relatively modest impact of NLO effects observed in the present MEM study is therefore not expected to be generic, but rather reflects the specific process and event selection considered here. For processes with larger QCD corrections, such as Higgs production, the improvement from an NLO MEM can be expected to be more pronounced. More generally, the inclusion of NLO corrections provides a more consistent perturbative description and can reduce theoretical uncertainties associated with scale variations and other sources of higher-order corrections. The extent of this reduction, however, is process dependent and should not be inferred directly from the present example. Finally, the good agreement between the LO and NLO results, in particular for~$\wbsm$, indicates that the phase-space partitioning discussed in~Section~\ref{sec:analysis} does not introduce a bias in the construction of the classifiers. Instead, it acts as a stabilizing procedure that preserves the underlying physics information while mitigating numerical cancellations.

\begin{figure}[t!]
\centering
\includegraphics[height=.25\textheight,page=1]{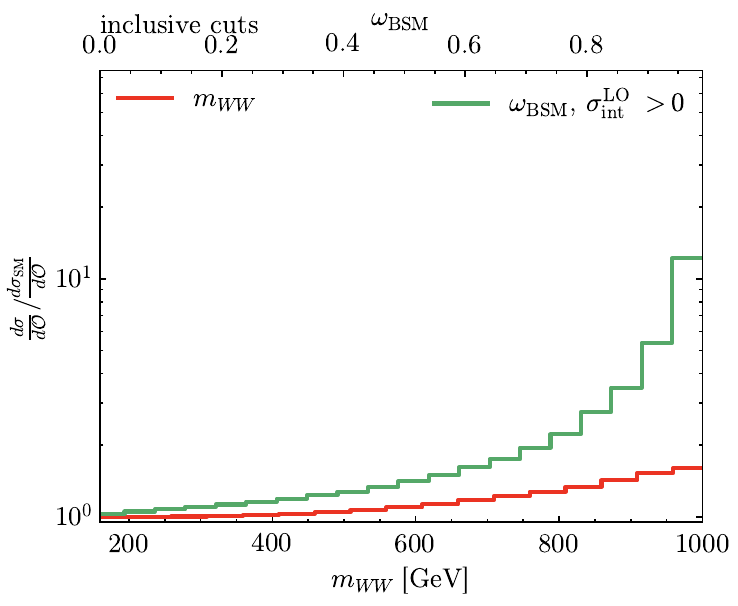} \hfill
\includegraphics[height=.25\textheight,page=1]{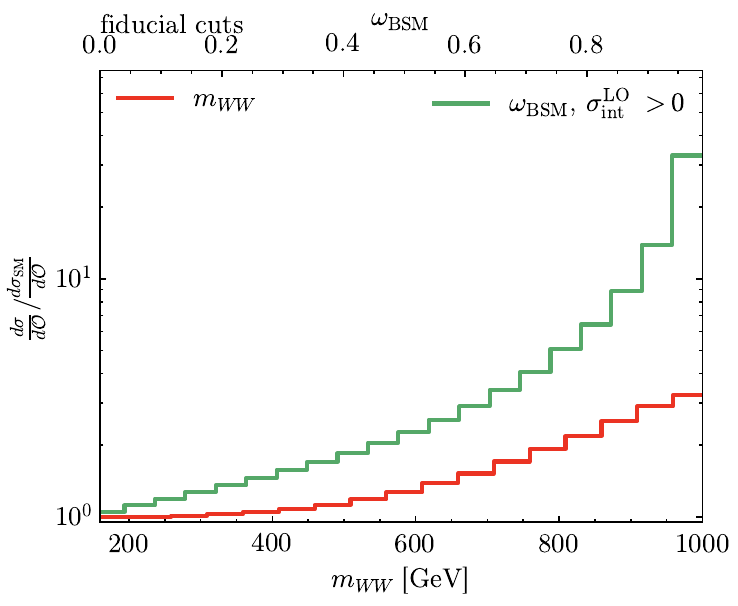}
\vspace{2mm}
\caption{Ratios of the BSM and SM cross sections shown as functions of the invariant mass $m_{WW}$ (bottom axis) of the $W^+ W^-$ system and the classifier~$\wbsm$ (top axis). The~$\wbsm$ results include only events with positive LO interference and are derived from showered events for both the inclusive~(left) and fiducial~(right) event selections. Further details are provided in the main text.}
\label{fig:MEMvsMWW}
\end{figure}

\paragraph{MEM discriminating power:} To illustrate that the classifiers defined in~Eq.~(\ref{eq:mem_weight}) provide greater sensitivity to the underlying dynamics than conventional kinematic observables, Figure~\ref{fig:MEMvsMWW} shows the ratio of BSM to SM cross sections as a function of the invariant mass $m_{WW}$ of the diboson system in $W^+ W^-$ production and the classifier~$\wbsm$. The classifier $\wbsm$ is constructed using only events with a positive LO interference contribution, and the results are obtained from showered events for both the inclusive (left panel) and fiducial (right panel) event selections. The real normalization factors $a$ and $b$ are chosen as in the first line of Eq.~(\ref{eq:ab}) to obtain the depicted results.

It is evident from both panels in~Figure~\ref{fig:MEMvsMWW} that the $\wbsm$ classifier yields improved sensitivity to the~SMEFT contribution for the benchmark value of the Wilson coefficient in~Eq.~(\ref{eq:CWbenchmark}), or equivalently Eq.~(\ref{eq:NlambdaV}). In particular, the $\wbsm$ distributions systematically lie above those based on $m_{WW}$, indicating that this enhancement persists across the entire explored observable space. This improved performance is driven by a significantly stronger separation between signal and background in the region of large $\wbsm$. Quantitatively, for the inclusive selections we obtain a signal-to-background ratio of about $1.6$ at invariant masses $m_{WW} \simeq 1 \, {\rm TeV}$, increasing to approximately $17.2$ at $\wbsm \simeq 1$. For fiducial cuts, the corresponding values are~$3.5$ and~$41.1$, respectively. That the fiducial selections are advantageous for BSM searches within the~SMEFT framework has already been discussed in~Section~\ref{sec:WWproduction}, and Figure~\ref{fig:MEMvsMWW} confirms that this feature also holds for the $\wbsm$ classifier. Notice that Figure~\ref{fig:MEMvsMWW}, together with the left panel of Figure~\ref{fig:MEMbasic}, demonstrates that not only is the signal-to-background separation achieved with $\wbsm$ significantly stronger than that obtained using $m_{WW}$, but also that a non-negligible fraction of the fiducial cross section populates the region of large $\wbsm$ values. As an example, imposing $\wbsm > 0.8$ retains about $6\%$ of the fiducial cross section with positive LO interference, corresponding to $3.8\,{\rm fb}$ out of a total of $63.2\,{\rm fb}$. For the full LHC Run~2 data set of $140\,{\rm fb}^{-1}$, this corresponds to about 530 events, implying a statistical uncertainty of roughly $4\%$ on the event sample.

\begin{figure}[t!]
\centering
\includegraphics[height=.25\textheight,page=2]{plots/obs_for_D1.pdf} \hfill
\includegraphics[height=.25\textheight,page=1]{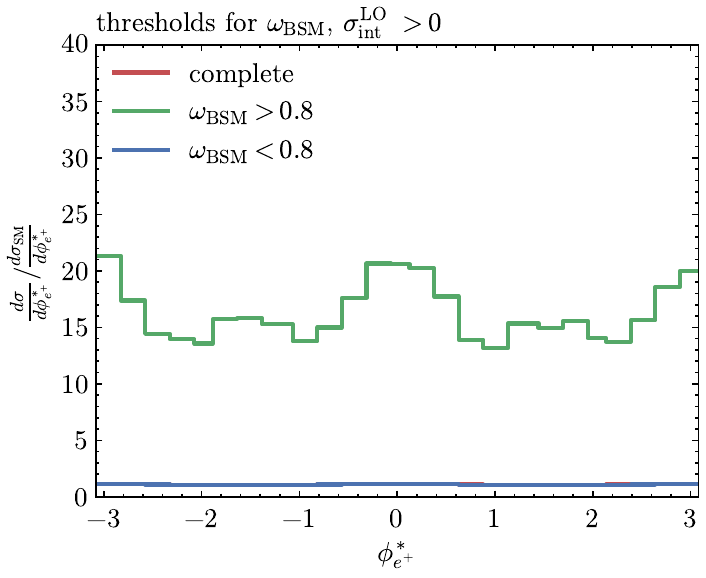}
\vspace{0mm}
\caption{Ratios of the BSM and SM cross sections shown as functions of the invariant mass $m_{WW}$ of the $W^+ W^-$ system (left) and the positron azimuthal decay angle $\phi_{e^+}^\ast$ (right). Predictions are presented for the full event sample, as well as for events satisfying the cuts $\wbsm > 0.8$ and $\wbsm < 0.8$. The results include only events with a positive LO interference contribution and are obtained from NLO$+$PS events under fiducial selection criteria. Additional explanations are provided in the main text.}
\label{fig:MEMtagging}
\end{figure}

\paragraph{MEM as a BSM tagger:} Finally, we illustrate the use of the classifier $\wbsm$ as a BSM~tagger. In this framework, selections on the classifier naturally isolate regions of phase space with enhanced BSM sensitivity, effectively acting as optimized cuts that improve signal-to-background separation. Since BSM-like events populate the region around $\wbsm \simeq 1$, imposing a lower threshold on $\wbsm$ provides a direct way to enrich the sample in BSM-induced events. To demonstrate this feature, we show in~Figure~\ref{fig:MEMtagging} the ratios of the BSM to SM cross sections as functions of the invariant mass $m_{WW}$ of the $W^+ W^-$ system~(left panel) and the positron azimuthal decay angle $\phi_{e^+}^\ast$~(right panel). Results are presented for the full event sample, as well as for events satisfying the selections $\wbsm > 0.8$ and $\wbsm < 0.8$. The predictions are constructed using only events with a positive LO interference contribution, are obtained from NLO$+$PS events under fiducial selections, and the used classifier~(\ref{eq:mem_weight}) employs the values of $a$ and $b$ given in the first line of Eq.~(\ref{eq:ab}).

The left panel in~Figure~\ref{fig:MEMtagging} shows that imposing $\wbsm > 0.8$ yields a BSM-to-SM cross-section ratio that rises significantly more steeply with $m_{WW}$ than in the inclusive sample or in the complementary region with $\wbsm < 0.8$. This indicates that events with a large classifier response are, on average, associated with higher values of $m_{WW}$ than those typical of the SM, and are therefore more BSM-like in their kinematic properties. Interestingly, for $\wbsm > 0.8$ a clear signal-over-background enhancement already appears for $m_{WW} \gtrsim 400 \, {\rm GeV}$, which is a particularly welcome feature with respect to the validity of the EFT expansion, as it reduces the reliance on the extreme high-energy tail of the distribution, where events with $m_{WW} \gtrsim 1 \, {\rm TeV}$ may begin to challenge the applicability of the EFT description for the chosen Wilson coefficient in~Eq.~(\ref{eq:CWbenchmark}).

From the right panel of Figure~\ref{fig:MEMtagging}, one observes that events satisfying $\wbsm > 0.8$ exhibit visible azimuthal correlations in the angle $\phi_{e^+}^\ast$, in contrast to the nearly flat behavior observed in the full sample and in the complementary region with $\wbsm < 0.8$. Figure~\ref{fig:xsec_phi}~shows that the observed dependence on $\phi_{e^+}^\ast$ in the $\wbsm > 0.8$ sample is driven primarily by the enhanced $\inter$ contribution, while both the SM and $\bsm$ components remain approximately flat in the azimuthal angle. This demonstrates that the classifier $\wbsm$ provides a sensitive probe of spin-dependent features in the $W$-boson decays, including angular correlations among the final-state leptons. The above discussion illustrates the underlying mechanism through which the classifier $\wbsm$ acts as an effective BSM tagger. Additional distributions highlighting these features are presented in~Appendix~\ref{app:moretagger}.

\section{Conclusions}
\label{sec:conclusions}

In this article, we have demonstrated that the MEM can be consistently extended to NLO accuracy in QCD by exploiting the \powheg~framework. The key ingredient is the use of the $\Btilde$ function to define the underlying Born kinematics associated with real-emission events, thereby incorporating the hardest QCD radiation while preserving NLO accuracy and normalization of the cross section. This construction enables the definition of event-level probabilities for SM, BSM, and interference contributions, and thus the formulation of likelihood-ratio-inspired classifiers with a transparent probabilistic interpretation at NLO accuracy. In the spirit of~Refs.~\cite{Alwall:2010cq,Campbell:2012cz,Campbell:2012ct,Campbell:2013hz,Martini:2015fsa,Baumeister:2016maz,Kraus:2019qoq,Kraus:2019myc,Martini:2023ylv,Tartarin:2025gbt,Tartarin:2026uoh}, this approach addresses the main technical challenges associated with IR divergences, negative weights, additional radiation, and multi-dimensional phase-space integration.

A central aspect of the framework is the reconstruction of the full partonic phase space from observed final states, including a mapping to the underlying Born configuration and a consistent treatment of radiation and initial-state flavor assignments. This procedure can be implemented as a flexible ``afterburner'' applied to pre-generated {\tt LHE} event samples, allowing for an efficient evaluation of NLO~MEM weights without event regeneration. The~method is robust with respect to reconstruction choices but currently relies on the inversion of the ISR mapping implemented in the \powhegbox, and is therefore mainly applicable to ISR-dominated processes. Extending the formalism to include FSR and more general mappings remains an important direction for future work. Overall, the framework provides a practical and systematically improvable route toward incorporating NLO accuracy into MEM-based~analyses. Given its generality, the method proposed here can be viewed as a first step towards incorporating even higher-order corrections into MEM approaches.
Particularly natural is the extension to next-to-next-to-leading order~(NNLO) effects matched to parton showers via the {\tt MiNNLO} method~\cite{Monni:2019whf}, since in this framework~NNLO contributions are incorporated through suitable modifications of the~$\Btilde$ function, which forms the central building block of our MEM construction.

As a proof of concept, we have applied the method to fully leptonic $W^+W^-$ production in the~SMEFT, focusing on a CP-even dimension-six triple-gauge-boson operator. At~the parton level, our results demonstrate that the NLO MEM efficiently exploits the full kinematic information available in the event, including the spin- and polarization-sensitive correlations encoded in the final-state lepton kinematics. In particular, the angular and azimuthal correlations arising from the $W$-boson decays provide sensitive handles for distinguishing SMEFT effects from the SM prediction. Comparing inclusive and fiducial selections further shows that the MEM retains its discriminating power despite the reduction in event yield induced by fiducial cuts, leading to substantially stronger separation between the SM and SMEFT hypotheses than conventional cut-and-count strategies, especially in regions where inclusive observables exhibit only mild deviations. For the $W^+W^-$ process considered here, the overall discrimination is already largely established at LO, with NLO corrections having a comparatively modest impact. This behavior is specific to the process and event selection and should not be regarded as generic. More broadly, the NLO MEM framework provides a basis for applications to processes with sizable QCD corrections, where higher-order effects may have a more pronounced impact on the discrimination power. A~fully realistic experimental application of the NLO MEM to the fully leptonic $W^+W^-$ channel would additionally require marginalization over the unobserved neutrino degrees of freedom, together with an appropriate treatment of detector and reconstruction~effects.

The classifiers $\wbsm(\Phi)$ and $\wint(\Phi)$, constructed from the full parton-level event kinematics, provide a direct connection between the classifier response and the underlying phase-space structure, allowing detailed differential studies of SMEFT effects. While $\wbsm(\Phi)$ is primarily sensitive to BSM-squared contributions and provides strong SM-BSM discrimination, $\wint(\Phi)$ isolates the interference contribution and is therefore particularly sensitive to sign-changing structures and cancellations across the phase space. Partitioning the phase space according to the sign of the LO interference contribution stabilizes the construction of $\wint(\Phi)$ by reducing cancellations and improving its numerical stability, while leaving $\wbsm(\Phi)$ essentially unaffected. Across the different implementations considered, the MEM classifiers reproduce the corresponding NLO$+$PS predictions with high accuracy, supporting the reliability of the NLO reweighting procedure and the consistency of the LO and NLO formulations. Compared with standard observables such as the invariant mass~$m_{WW}$ of the intermediate $W$ bosons, the classifiers provide enhanced discrimination and more directly exploit the multidimensional kinematic information encoded in the event. In particular, selections based on $\wbsm(\Phi)$ can enhance the sensitivity to BSM contributions without relying exclusively on the high-energy tail, where the validity of the EFT description may become less robust, while also retaining sensitivity to spin and interference effects through the angular correlations in the leptonic final state.

Looking ahead, the NLO MEM provides a promising framework for precision collider studies. Its application to EW, Higgs, and top-quark processes can exploit the multidimensional kinematic information encoded in scattering amplitudes, including polarization and spin correlations as well as SMEFT-induced deviations. Combining the MEM with modern ML techniques may further facilitate the exploration of high-dimensional phase spaces and the extraction of BSM effects. A key step towards realistic experimental applications will be to account for missing transverse momentum and other unobserved degrees of freedom, together with detector effects such as finite resolution and object misidentification. The framework developed here provides a basis for extending NLO MEM techniques to a broader range of collider processes and, ultimately, to precision studies at the LHC and future colliders. In this way, the NLO MEM can serve as a probabilistic framework connecting higher-order theoretical predictions with event-level collider information and enabling systematic studies of SM and BSM effects. 

\acknowledgments{Luc Schnell was supported by the DFG Collaborative Research Centre ``Neutrinos and Dark Matter in Astro- and Particle Physics'' (SFB 1258) as part of project D02. We~thank Tae~(NSBI)~Hyoun Park for helpful discussions and comments on the manuscript. The~Feynman diagrams in this article were generated using {\tt TikZ-Feynman}~\cite{Ellis:2016jkw}.}

\FloatBarrier

\begin{appendix}

\section{Details on ISR reconstruction algorithm}
\label{app:kradReconstructionSelections}

\begin{figure}[t!]
\centering
\includegraphics[height=.35\textheight,page=1]{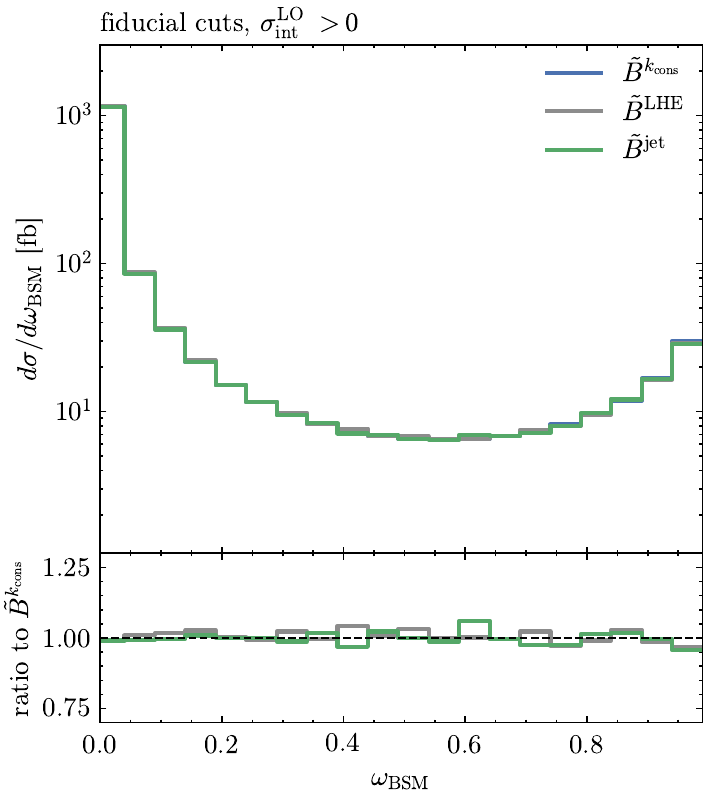} \hfill
\includegraphics[height=.35\textheight,page=5]{plots/hadronic_nlops_dl0p06_fiducial_krad.pdf}
\vspace{2mm}
\caption{Fiducial $W^+ W^-$ NLO+PS SMEFT cross sections as functions of the classifiers~$\wbsm$ and $\wint$. Different prescriptions are employed to reconstruct the radiation momentum $k_{\mathrm{rad}}$ associated with the real phase-space configuration. The~curves labeled~$\btilde^{k_{\mathrm{cons}}}$ correspond to the momentum-conservation procedure introduced in~Section~\ref{sec:implementation} and illustrated in Figure~\ref{fig:mapping_flow_chart}. For $\btilde^{\mathrm{LHE}}$, the unshowered momenta from the {\tt LHE} event are used directly to define the real phase-space point, while in the case of $\btilde^{\mathrm{jet}}$, the hardest reconstructed jet is identified with $k_{\mathrm{rad}}$. Further details are provided in the main~text.}
\label{fig:krad_reconstruction}
\end{figure}

The algorithm discussed in~Section~\ref{sec:implementation} and illustrated in~Figure~\ref{fig:mapping_flow_chart}, used to determine the radiation momentum $k_{\mathrm{rad}}$, relies solely on the four lepton momenta and reconstructs $k_{\mathrm{rad}}$ via momentum conservation. In this appendix, we explore alternative choices and their impact on the observables~$\wbsm$ and~$\wint$, defined in \cref{eq:mem_weight}. Differential distributions in~$\wbsm$ and~$\wint$ are shown in~Figure~\ref{fig:krad_reconstruction}. 

The momentum-conservation-based reconstruction of $k_{\mathrm{rad}}$ is denoted by $\btilde^{k_{\mathrm{cons}}}$ and shown by the blue curve. The gray distribution, labeled $\btilde^{\mathrm{LHE}}$, uses the real phase-space point directly from the {\tt LHE} event file before applying the PS. In this setup, the showered momenta define the fiducial phase space, while the {\tt LHE} momenta are employed in the evaluation of $\wbsm$ and $\wint$. We emphasize that the momenta stored in the {\tt LHE} file are not identical to those used internally by the \powhegbox~during the reweighting procedure --- cf., for example, the black reweighting curve labeled RWGT in~Figure~\ref{fig:MEMbasic}. The latter are modified by the importance-sampling procedure employed in \powheg, whereas the {\tt LHE} momenta originate from a hit-and-miss generation based on a Sudakov form factor. For the green curve, called $\btilde^{\mathrm{jet}}$, jets are clustered with {\tt FastJet} using the anti-$k_t$ algorithm with radius parameter $R=0.4$, and the hardest jet is identified with $k_{\mathrm{rad}}$. Since the resulting phase-space configuration, constructed from this $k_{\mathrm{rad}}$ and the four lepton momenta, does not necessarily correspond to a valid configuration with vanishing total transverse momentum, we apply a mapping analogous to~Eq.~(\ref{eq:boostrestore}) to remove the transverse momentum of the full system. Starting from this configuration, the algorithm illustrated in Figure~\ref{fig:mapping_flow_chart} is then used to reconstruct the underlying Born phase-space point.

As shown in Figure~\ref{fig:krad_reconstruction}, the different prescriptions for determining $k_{\mathrm{rad}}$ have only a minor impact on the classifiers $\wbsm$ and $\wint$. Since the momentum-conservation approach depends exclusively on the four lepton momenta and does not require jet clustering, we adopt it as our default choice. 

\section{Details on initial-state flavor reconstruction}
\label{app:flavorselection}

The selection of the initial-state flavor configuration used in the evaluation of the classifiers $\wbsm$ and $\wint$ is performed via a MC flavor selection based on the probability defined in~Eq.~(\ref{eq:flavorprob}). In this setup, $\tilde{B}_{\sm}(\Phi)$ is used, such that configurations with a larger $\sm$ contribution are preferentially selected. Alternatively, one may use $\tilde{B}_{\bsm}(\Phi)$ or $\tilde{B}_{\mathrm{SMEFT}_{8}}(\Phi)$ in place of $\tilde{B}_{\sm}(\Phi)$, or avoid a stochastic selection altogether by taking a direct sum over all flavor contributions, $\tilde{B}_{\rm sum}(\Phi)$. This appendix compares these different flavor-selection strategies and highlights the advantages of performing a MC flavor selection based on~$\tilde{B}_{\sm}(\Phi)$.

\begin{figure}[t!]
\centering
\includegraphics[height=.45\textheight,page=1]{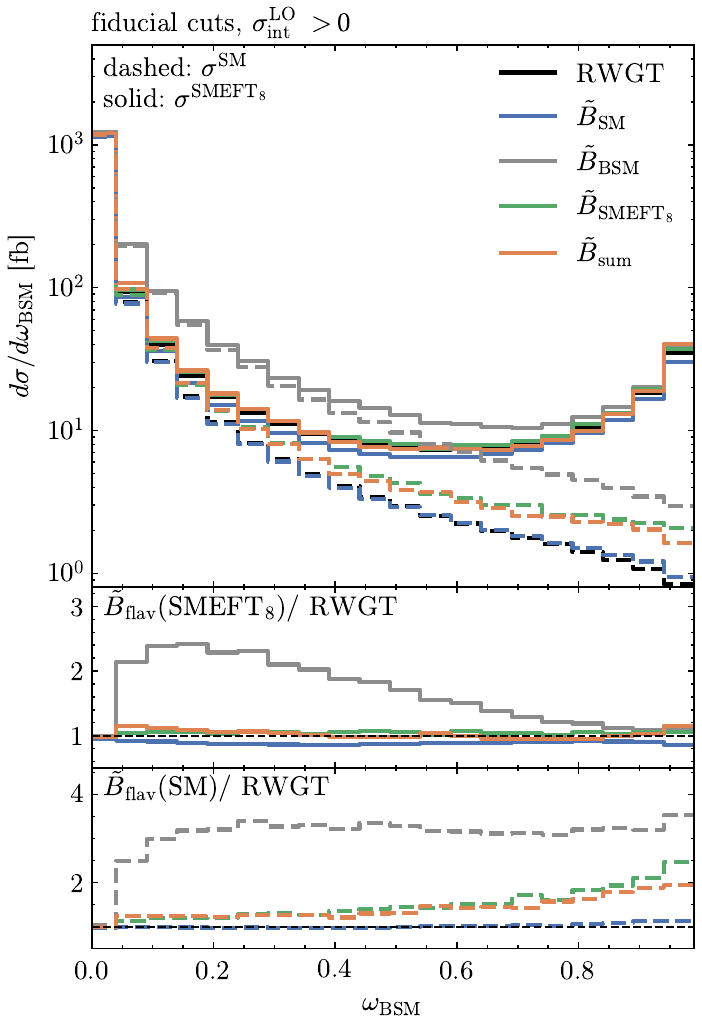} \hfill
\includegraphics[height=.45\textheight,page=5]{plots/hadronic_nlops_dl0p06_fiducial_flav_choice.pdf}
\vspace{2mm}
\caption{
Fiducial $W^+ W^-$ NLO$+$PS SMEFT cross sections as functions of the classifiers~$\wbsm$ and $\wint$, restricted to the phase space with positive LO interference. Different prescriptions for assigning the initial-state flavor configuration are considered. Dashed lines correspond to the $\sm$ cross section, while solid lines represent $\mathrm{SMEFT}_{8}$, i.e.~the sum of the $\sm$, $\bsm$, and $\inter$ contributions. The black curve, labeled RWGT, in which the observables~$\wbsm$ and $\wint$ are obtained via reweighting, is used as a reference for the two ratio plots, which test the behavior of the different flavor-selection models for the $\mathrm{SMEFT}_{8}$~(middle panels) and $\sm$~(bottom panels) cross sections. A MC flavor selection with probabilities defined in~Eq.~\eqref{eq:flavorprob} is used for the blue ($\sm$), gray ($\bsm$), and green ($\mathrm{SMEFT}_{8}$) curves, each corresponding to different components of the $\Btilde$ function (as indicated in parentheses), while the orange curve represents the sum over all flavor contributions. Further details are provided in the main text.}
\label{fig:flav_choice}
\end{figure}

As the main focus of the MEM as defined here, we again consider the differential distributions in $\wbsm$ and $\wint$ defined in~Eq.~(\ref{eq:mem_weight}) with $a$ and $b$ chosen as in the first line of~Eq.~(\ref{eq:ab}). The $\sm$~(dashed) and $\mathrm{SMEFT}_{8}$~(solid), i.e.~the full SM plus BSM result, cross sections are shown in~Figure~\ref{fig:flav_choice} for the different flavor-selection methods. The preferred method should reproduce as closely as possible the differential distributions obtained when the classifiers $\wbsm$ and $\wint$ are computed via the \powheg~reweighting procedure, denoted by RWGT, for both the $\sm$ and $\mathrm{SMEFT}_{8}$ cross sections. In~addition, one aims to maximize the separation between the $\sm$ and $\mathrm{SMEFT}_{8}$ predictions at large values of the observables, where a reduced $\sm$ and an enhanced $\bsm$ contribution is expected.

It is evident from Figure~\ref{fig:flav_choice} that using $\tilde{B}_{\bsm}(\Phi)$~(gray) is not a viable option, since neither the $\sm$ nor the $\mathrm{SMEFT}_{8}$ distributions reproduce the RWGT reference with sufficient accuracy. While both $\tilde{B}_{\mathrm{SMEFT}_{8}}(\Phi)$~(green) and the direct-sum prescription, $\tilde{B}_{\mathrm{sum}}(\Phi)$~(orange), lead to well-behaved $\mathrm{SMEFT}_{8}$ predictions for $\wbsm$ and $\wint$, they fail to provide an equally good description of the corresponding $\sm$ spectra. In the setup considered here, the MC flavor selection based on $\tilde{B}_{\sm}(\Phi)$ therefore emerges as the preferred choice and is adopted as the default throughout this work. We stress, however, that this conclusion is based on an empirical observation for the specific process and observables studied here. In~other~BSM applications, alternative choices may well prove advantageous, and we therefore recommend investigating this issue on a case-by-case basis.

\section{Additional $\bm{W^+ W^-}$ distributions}
\label{app:moredistributions}

In this appendix, we present additional $W^+ W^-$ distributions, analogous to those in~Section~\ref{sec:WWproduction}, for both inclusive and fiducial selections, where the latter are specified in~Table~\ref{tab:fiducial}. These results help clarify the origin of the BSM sensitivity of the classifiers defined in~Eq.~(\ref{eq:mem_weight}) and discussed in~Section~\ref{sec:analysis}.

\begin{figure}[t!]
\centering
\includegraphics[height=.35\textheight,page=3]{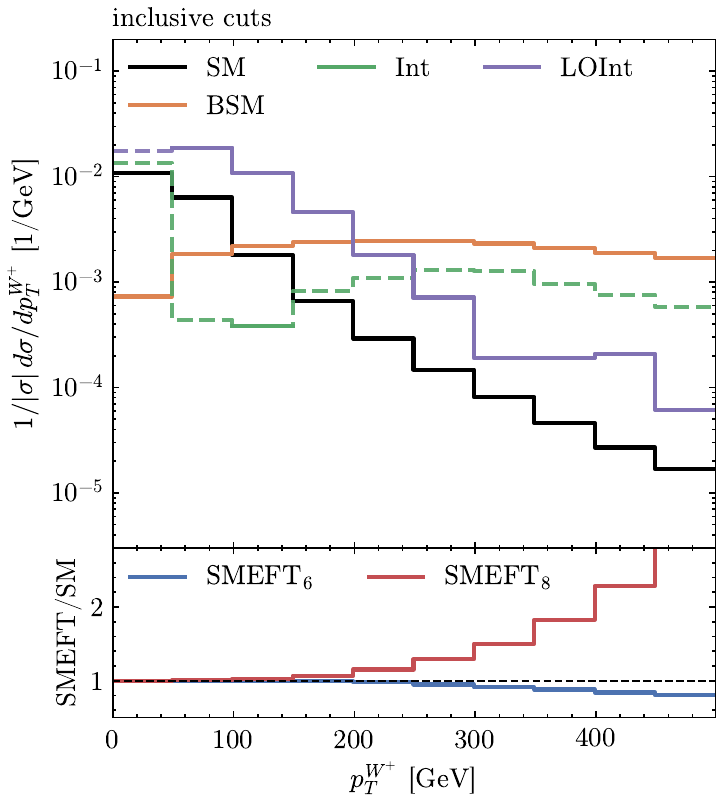} \hfill
\includegraphics[height=.35\textheight,page=3]{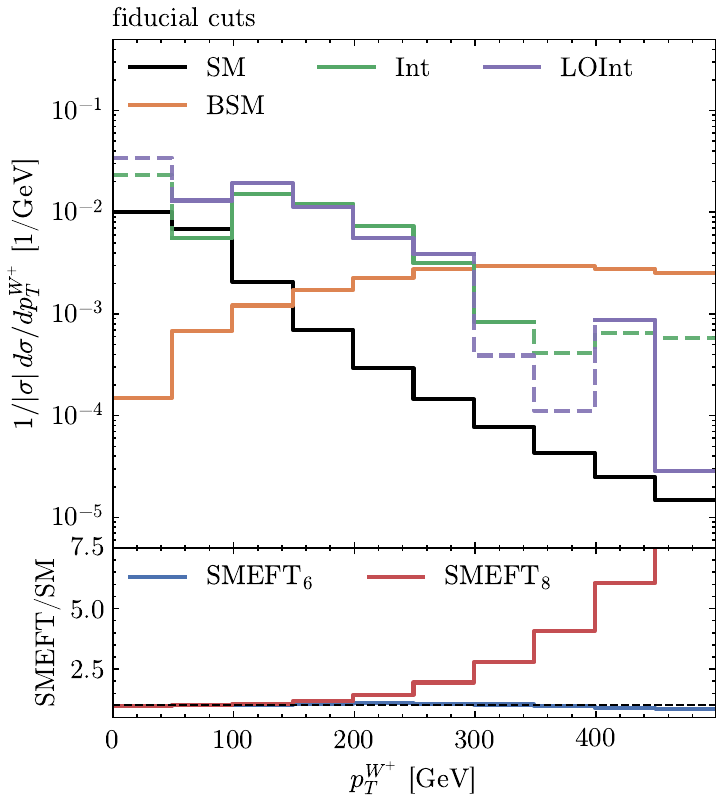}

\vspace{2mm}

\includegraphics[height=.35\textheight,page=4]{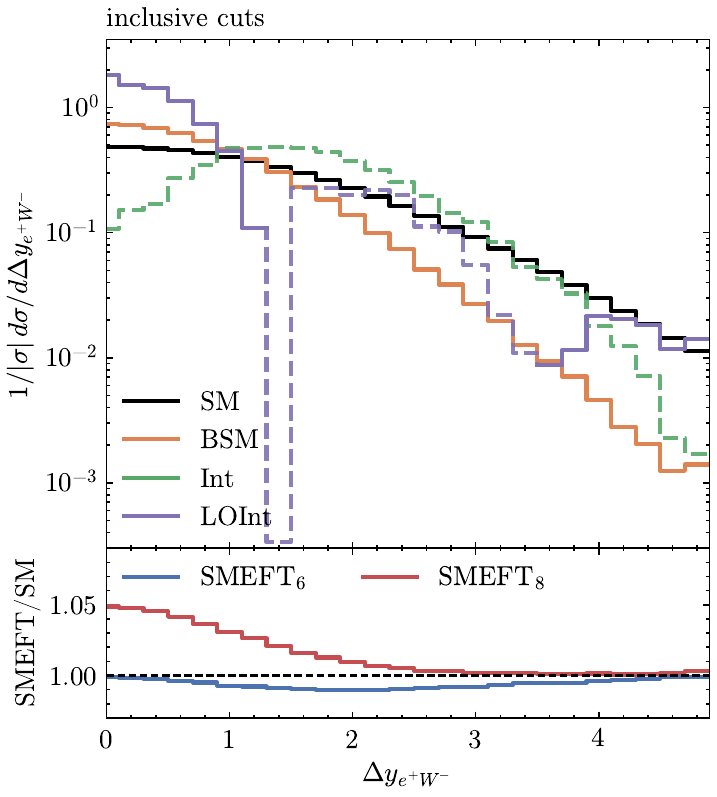} \hfill
\includegraphics[height=.35\textheight,page=4]{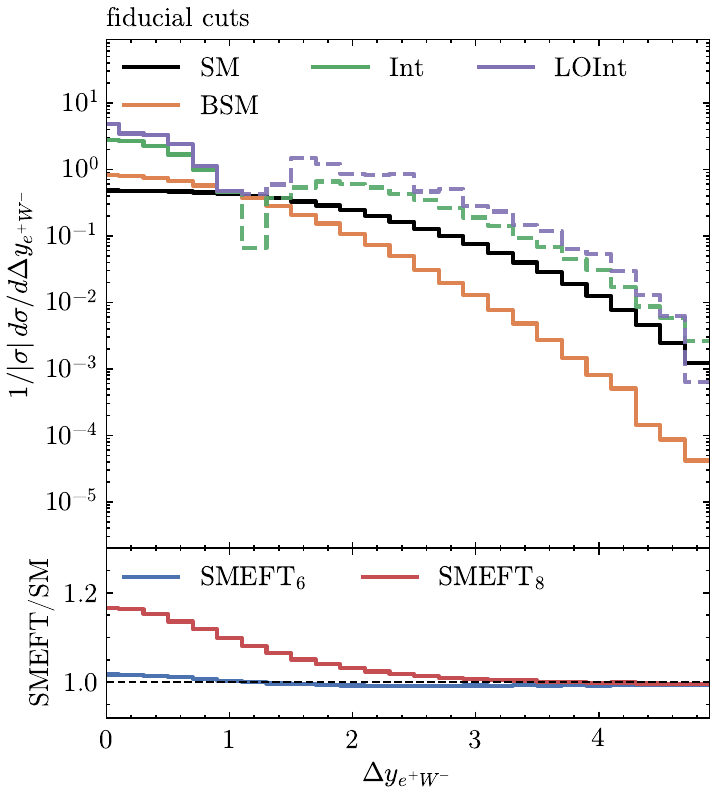}
\vspace{2mm}
\caption{Analogous to Figure~\ref{fig:xsec_mww}, but presenting $p_{T}^{W^+}$ distributions in the upper panels and $\Delta y_{e^+ W^-}$ spectra in the lower panels. Further details are provided in the main text.}
\label{fig:xsec_adddis}
\end{figure}

Our findings are summarized in~Figure~\ref{fig:xsec_adddis}. The upper and lower panels show the predictions for the transverse momentum of the positively charged $W$ boson ($p_{T}^{W^+}$) and the rapidity separation between the positron and the negatively charged $W$ boson ($\Delta y_{e^+ W^-}$), respectively. The left (right) panels correspond to the inclusive (fiducial) selection. In the upper sections, we display the relevant $\sm$, $\bsm$, $\inter$, and $\LOinter$ contributions, normalized to their respective absolute cross sections, while the lower sections present the ratios of the~SMEFT predictions to the SM spectrum. Similar to what was observed in~Figures~\ref{fig:xsec_mww} and \ref{fig:xsec_phi}, the distributions shown here also exhibit a non-trivial modification due to the experimental selections. In the case of the $p_{T}^{W^+}$ spectrum, for instance, the high-$p_{T}^{W^+}$ tail of the $\sm$ distribution is strongly suppressed, whereas the $\bsm$ spectrum tends to be harder. Moreover, the normalized distribution for the $\inter$ contribution exhibits noticeable differences, changing sign at different values of $p_{T}^{W^+}$. For $\Delta y_{e^+ W^-}$, one instead observes that the imposed experimental cuts have only a minor impact on the $\bsm$ contribution, while inducing a sign flip in the normalized $\inter$ spectrum. The normalized distributions can also be substantially altered by evaluating the interference contribution at LO rather than at~NLO. It~is also evident from all panels that the fiducial cuts lead to an enhancement of the relative SMEFT yields.

\section{Impact of phase-space partitioning}
\label{app:interferenceCut}

To mitigate the impact of negative-weight effects in the MEM analysis of~Section~\ref{sec:analysis}, the NLO$+$PS event sample is partitioned according to the sign of the LO interference contribution in $W^+ W^-$ production, thereby separating regions of positive and negative interference. Table~\ref{tab:xsecAmount} summarizes the behavior of the MEM classifiers $\wbsm$ and $\wint$ for NLO$+$PS events in $W^+ W^-$ production, both with and without this interference-sign partitioning. Besides the total fiducial cross sections, the table lists the contributions within the nominal physical ranges of the classifiers, together with the fractions of events that populate regions outside these intervals. The presence of such leakage provides a direct measure of how well the classifiers retain their interpretation as bounded probabilistic observables. The classifiers are calculated with the values of the real normalization factors~$a$ and~$b$ given in~Eq.~(\ref{eq:ab}).

{
\def\arraystretch{1.25}
\centering
\begin{table}[t!]
\centering
\begin{tabular}{|l|c|c|c|c|c|}
\hline
& $\sigma^{\rm LO}_{\inter}$ & total cross section & $\wbsm \in [0,1]$ & $\wbsm < 0$ & $\wbsm > 1$ \\
\hline
\wbsm & $>0$ & $63.11(4) \, {\rm fb}$ & $62.56(4) \, {\rm fb}$ & $0.507(5) \%$ & $0.377(4)\%$ \\
& $<0$ & $88.33(5) \, {\rm fb}$ & $82.34(5) \, {\rm fb}$ & $0.459(4) \%$ & $6.31(1)\%$ \\
& --- & $151.6(2) \, {\rm fb}$ & $100.5(2) \, {\rm fb}$ & $32.20(9)\%$ & $1.53(2)\%$ \\
\hline 
& $\sigma^{\rm LO}_{\inter}$ & total cross section & $\wint \in [0,0.5]$ & $\wint < 0$ & $\wint > 0.5$ \\
\hline
\wint & $>0$ & $63.11(4) \, {\rm fb}$ & $57.76(4) \, {\rm fb}$ & $8.38(2)\%$ & $0.097(2)\%$ \\
& $<0$ & $88.27(5) \, {\rm fb}$ & $81.20(5) \, {\rm fb}$ & $3.24(1)\%$ & $4.77(1)\%$ \\
& --- & $ 151.5(2) \, {\rm fb}$ & $21.19(7) \, {\rm fb}$ & $27.92(8)\%$ & $58.1(1)\%$ \\
\hline
\end{tabular}
\vspace{4mm}
\caption{Cross sections are shown separately for regions with positive and negative LO interference, as well as for the unpartitioned sample, denoted by ``---''. For the classifiers~$\wbsm$ and $\wint$, both the total fiducial cross sections and the contributions within their nominal physical ranges, $[0,1]$ and $[0,0.5]$, respectively, are reported. In addition, the fractions of events with classifier values below $0$ and above $1$ or $0.5$ are provided as a measure of out-of-range behavior. All results are obtained from NLO$+$PS event samples after applying fiducial selection cuts, with statistical uncertainties from MC integration indicated in parentheses. Further details are given in the main text.} 
\label{tab:xsecAmount}
\end{table}
}

The pattern observed in~Table~\ref{tab:xsecAmount} is directly connected to the structure of the~SMEFT interference term. Because the interference contribution changes sign across phase space, its inclusive effect arises from cancellations between positive and negative regions. In an unpartitioned sample, such cancellations can become numerically unstable, propagating into interference-sensitive observables such as~$\wint$ and leading to enhanced fluctuations, including values outside the expected range. This behavior is clearly visible in the table, where about $27\%$ of events yield $\wint < 0$ without partitioning, compared to roughly $8\%$ and $3\%$ in the positive- and negative-interference regions, respectively. Moreover, in the unpartitioned sample the fraction of events with $\wint > 0.5$ reaches about $58\%$, rendering~$\wint$ essentially unusable as a classifier. In contrast, $\wbsm$ remains more stable, with out-of-range fractions of about $33\%$ in the unpartitioned sample and approximately~$1\%$ and~$7\%$ in the positive- and negative-interference regions, respectively, reflecting its dominance by positive-definite contributions. Overall, the table clearly demonstrates that partitioning the phase space according to the sign of the~LO~interference improves the numerical stability and robustness of both classifiers, and we therefore adopt this method in the main body of the article.

The attentive reader might have noticed that, for the classifier $\wbsm$, partitioning the phase space according to the sign of the~LO~interference retains only about $40\%$ of the total fiducial cross section in the region with positive~LO~interference. Throughout Sections~\ref{sec:analysis} and~\ref{sec:MEMperformance}, we nevertheless focus on this subset of events in order to illustrate the properties and performance of the MEM classifiers in a transparent setting. In a realistic experimental analysis, however, one would instead combine $\wbsm$ and $\wint$ for both positive and negative~LO~interference into a single inclusive likelihood ratio discriminating between the SM and the full SM plus BSM hypothesis. Such a construction would constitute a MEM-based analog of the mixture models commonly employed in ML approaches~\cite{Cranmer:2015bka,ATLAS:2025clx,Ghosh:2025fma}. In this way, the full event sample could be exploited without discarding any phase-space regions, thereby maximizing the available statistical sensitivity.

\section{Remarks on ``optimal observables''}
\label{app:cW_dependence}

The classifiers defined in~Eq.~(\ref{eq:mem_weight}) depend on the probability distributions in~Eq.~(\ref{eq:prob_distribution}) and are therefore, in contrast to the ``optimal observables'' defined in~Eq.~(\ref{eq:optimal_mem_weight}), in principle independent of the specific BSM realization. In other words, the classifiers are ``optimal observables'' only for the particular BSM hypotheses used in their construction. Away~from these choices, their performance is not guaranteed to remain optimal, although they typically retain substantial discriminating power due to their direct connection to the underlying MEM structure and its interference pattern with the SM.

\begin{figure}[t!]
\centering
\includegraphics[height=.36\textheight,page=1]{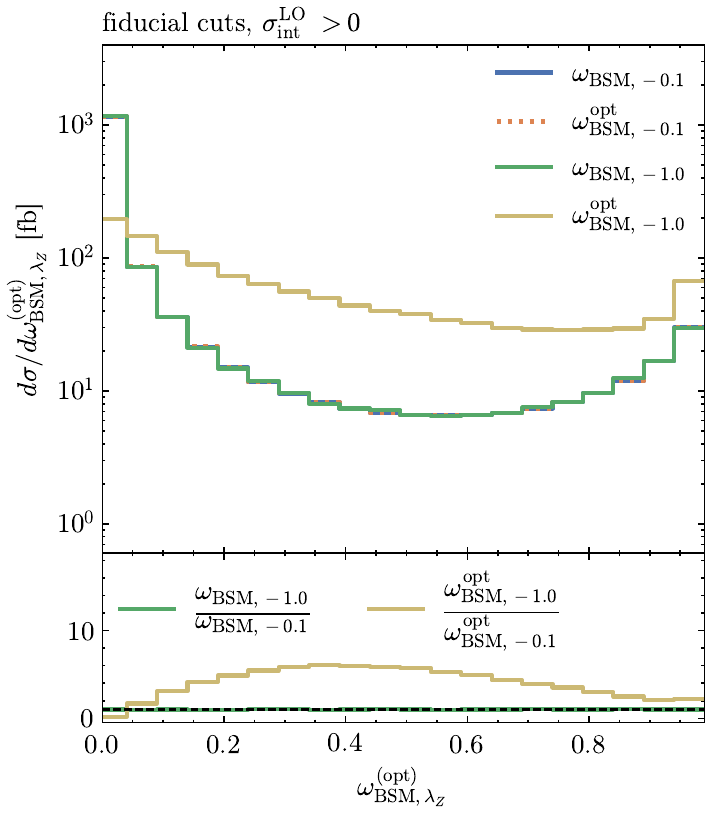} \hfill
\includegraphics[height=.36\textheight,page=3]{plots/cW_hadronic_nlops_dl0p06_fiducial.pdf}
\vspace{2mm}
\caption{NLO$+$PS SMEFT cross sections, as a function of different $\bsm$ (left panel) and $\inter$ (right panel) classifiers. All results correspond to the fiducial cuts summarized in~Table~\ref{tab:fiducial}, positive~LO~interference, and showered events. The~notation $\omega_{\bullet,\lambda_Z}^{({\rm opt})}$ with $\bullet = {\rm BSM}, {\rm Int}$ denotes the classifiers defined in~Eq.~(\ref{eq:mem_weight}) and the~``optimal observables'' of~Eq.~(\ref{eq:optimal_mem_weight}), evaluated for the specified value of $\lambda_Z$. See the main text for additional explanations.}
\label{fig:mem_diffW}
\end{figure}

To illustrate and quantify this feature, Figure~\ref{fig:mem_diffW} shows the NLO$+$PS SMEFT cross sections obtained for the benchmark choice of $\lambda_Z = \lambda_{\gamma}$ specified in Eq.~(\ref{eq:NlambdaV}), displayed as functions of the $\bsm$ (left panel) and $\inter$ (right panel) classifiers and the corresponding ``optimal observables''. All predictions are obtained using the fiducial cuts of~Table~\ref{tab:fiducial}, restricting to events with positive LO interference and employing showered events. The~observables~$\omega_{\bullet,\lambda_Z}$ with $\bullet = \bsm, \inter$ denote the classifiers defined in~Eq.~(\ref{eq:mem_weight}) with the parameters $a$ and $b$ fixed to the values given in the first line of~Eq.~(\ref{eq:ab}), while $\omega_{\bullet,\lambda_Z}^{\rm opt}$ correspond to the observables constructed according to Eq.~(\ref{eq:optimal_mem_weight}). We recall that Eq.~(\ref{eq:mem_weight}) is formally independent of $\lambda_Z$, whereas Eq.~(\ref{eq:optimal_mem_weight}) depends explicitly on the signal hypothesis employed in its construction. From the figure it is evident that the ``optimal observables'' are truly optimal only for $\lambda_Z = -0.1$, since in this case the distributions of $\omega_{\bullet,-0.1}$ and~$\omega_{\bullet,-0.1}^{\rm opt}$ coincide. This is expected because the values of $a$ and $b$ in~Eq.~(\ref{eq:ab}) are fixed precisely such that the two constructions agree for this choice of coupling. For $\lambda_Z = -1.0$, by contrast, only the classifier $\omega_{\bullet,-1.0}$ reproduces the correct spectrum, while $\omega_{\bullet,-1.0}^{\rm opt}$ deviates noticeably from the optimal behavior despite its name. These observations demonstrate that the optimality of the ``optimal observables'' $\omega_{\bullet}^{\rm opt}(\Phi)$ is tied to the specific hypothesis used in their construction and therefore does not persist universally across different parameter choices. The~classifiers~$\omega_{\bullet}(\Phi)$ are instead considerably more robust, making them the preferred choice adopted throughout this work. Notice that constructing the classifiers~$\omega_\bullet(\Phi)$ nevertheless requires fixing the real normalization factors $a$ and $b$. In a realistic application of our NLO MEM method to experimental data, this would entail estimating the approximate sensitivity of a given search to the underlying BSM parameter, such as~$\lambda_Z$, and determining the corresponding values of $a$ and $b$ using~Eq.~(\ref{eq:optimal}) for the chosen signal~hypothesis.

\section{Additional MEM distributions}
\label{app:moreMEM}

\begin{figure}[t!]
\centering
\includegraphics[height=.45\textheight,page=3]{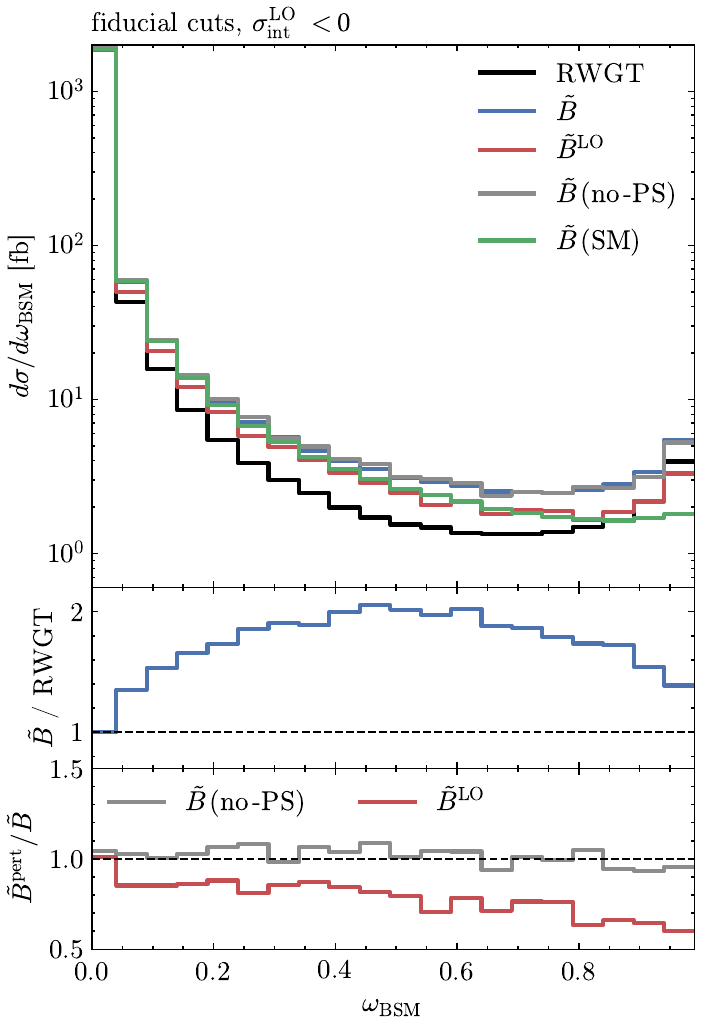} \hfill
\includegraphics[height=.45\textheight,page=7]{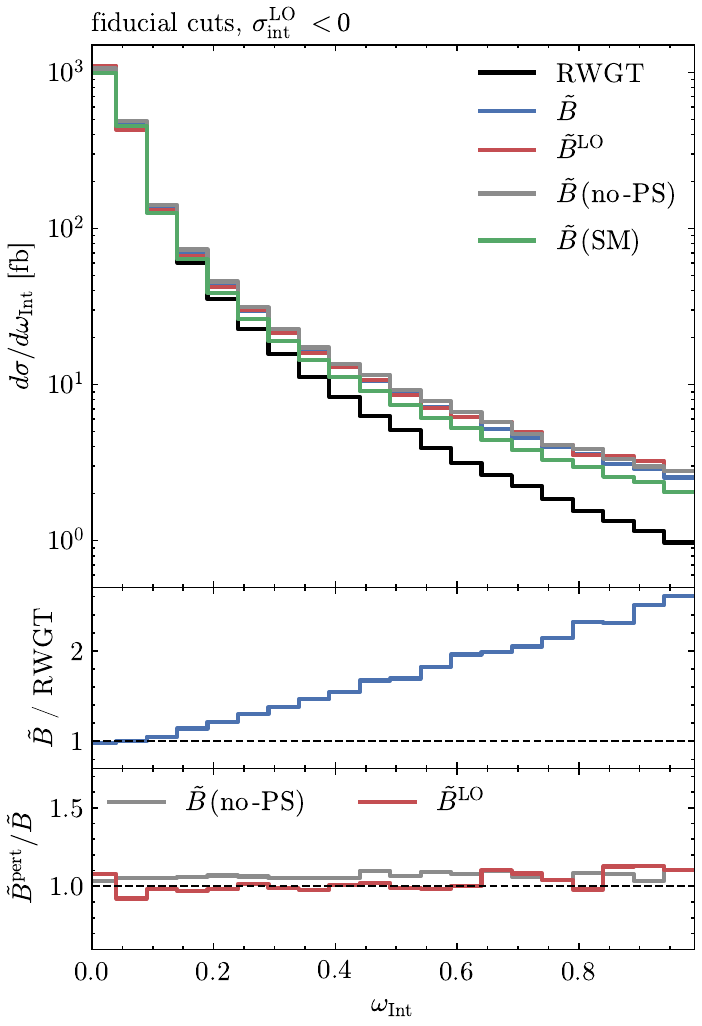}
\vspace{2mm}
\caption{Same as in Figure~\ref{fig:MEMbasic}, but with the requirement of negative LO interference. Consult the main text for further details.} 
\label{fig:moreMEMbasic}
\end{figure}

In~Section~\ref{sec:MEMperformance}, we have illustrated the performance of the MEM observables in the phase-space region with positive LO interference. We now repeat the analysis for the complementary case of negative~LO interference. The corresponding real normalization constants~$a$ and~$b$ are given in the second line of Eq.~(\ref{eq:ab}). Figure~\ref{fig:moreMEMbasic} shows fiducial $W^+ W^-$ cross sections as functions of $\wbsm$ (left) and $\wint$ (right) in this region. Unless stated otherwise, all results are obtained from showered events. The upper panels compare different SMEFT predictions derived using the NLO~MEM ``afterburner'' of Figure~\ref{fig:mem_flow_chart}, including~$\tilde B$~from NLO$+$PS events, the corresponding~LO~prediction $\tilde B^{\mathrm{LO}}$, the unshowered NLO result $\tilde B \, (\noPS)$, and the SM baseline $\tilde B \, (\sm)$. While the differential rates depend on the perturbative setup, the classifier axes are kept identical across all predictions, ensuring a consistent probabilistic interpretation and meaningful comparison. The only exception is~$\tilde B^{{\rm LO}}$, where the LO observable is evaluated on~NLO$+$PS~events.

The first observation from Figure~\ref{fig:moreMEMbasic} is that the classifiers $\wbsm$ and $\wint$ perform less effectively in the negative~LO interference region than in the positive case, as the $\tilde B$ spectra are systematically less well separated from the corresponding $\tilde B \, (\sm)$ distributions. Comparing the classifiers $\wbsm$ and $\wint$, one also finds that while $\wbsm$ still yields a visible separation between BSM- and SM-like events, $\wint$ exhibits very limited discriminating power in the case of negative~LO interference. In addition, the~SMEFT reweighted predictions show noticeably reduced agreement with the full $\tilde B$ results. The quality of the $\tilde B \, (\noPS)$ and $\tilde B^{\rm LO}$ predictions relative to $\tilde B$ is, however, not largely degraded. From the discussion at the beginning of Section~\ref{sec:analysis} and Appendix~\ref{app:interferenceCut}, this reduced performance can be traced to the structure of the $\inter$ contribution in the negative LO interference region. Here, stronger cancellations between positive and negative phase-space contributions reduce the effective signal carried by the interference term, thereby weakening its correlation with the underlying kinematic information. As a consequence, the discrimination power of both $\wbsm$ and, in particular, the interference-sensitive observable~$\wint$ is diminished.

\section{Additional tagger distributions}
\label{app:moretagger}

\begin{figure}[t!]
\centering
\includegraphics[height=.25\textheight,page=3]{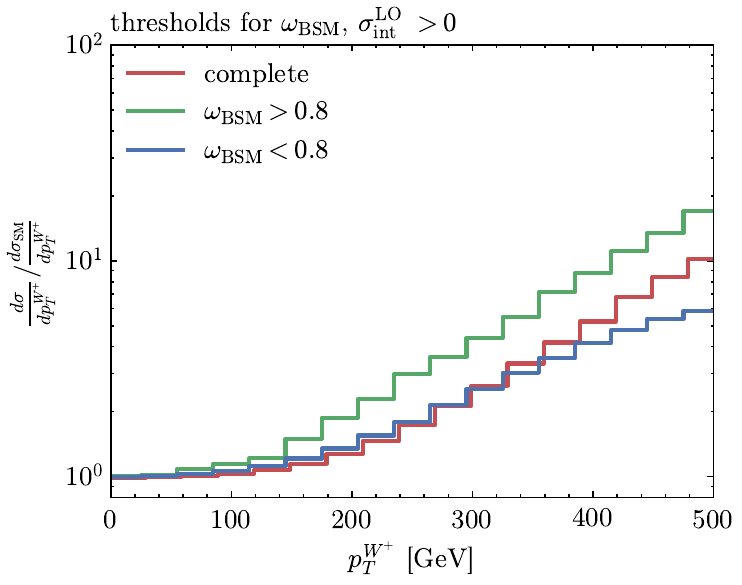} \hfill
\includegraphics[height=.25\textheight,page=4]{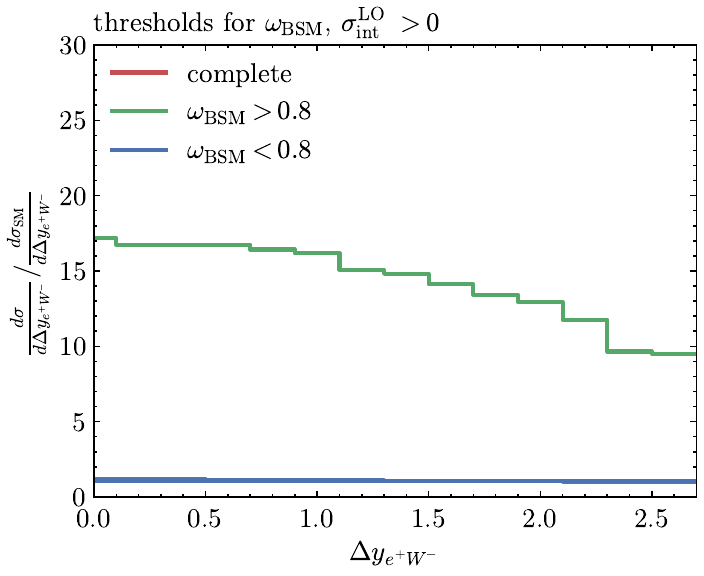}
\vspace{0mm}
\caption{Same as Figure~\ref{fig:MEMtagging} but for $p_T^{W^+}$~(left) and~$\Delta y_{e^+ W^-}$~(right). Further details are given in the main text.}
\label{fig:moreMEMtagging}
\end{figure}

In~Section~\ref{sec:MEMperformance}, we have shown that applying selections on the classifier naturally isolates regions of phase space with enhanced BSM sensitivity, effectively acting as optimized cuts that improve signal-to-background discrimination. In this appendix, we examine how such classifier-based selections affect the $W^+ W^-$ distributions presented in~Appendix~\ref{app:moredistributions}.

Figure~\ref{fig:moreMEMtagging} shows the ratios of BSM to SM cross sections as functions of $p_{T}^{W^+}$~(left panel) and $\Delta y_{e^+ W^-}$~(right panel), for the inclusive event sample as well as for subsets defined by the selections $\wbsm > 0.8$ and $\wbsm < 0.8$, with $a$ and $b$ fixed to the values quoted in the first line of Eq.~(\ref{eq:ab}). The distributions are obtained from NLO$+$PS events after fiducial cuts, retaining only events with a positive LO interference contribution. In~both observables, the requirement $\wbsm > 0.8$ consistently enhances the BSM-to-SM ratio relative to the inclusive result, while the complementary region $\wbsm < 0.8$ tends to suppress it, illustrating the role of the classifier as a BSM-enriching selection. The effect is significantly more pronounced in~$\Delta y_{e^+ W^-}$ than in $p_{T}^{W^+}$, indicating that the former carries a stronger intrinsic sensitivity to the considered SMEFT contributions. Together with the results of Figure~\ref{fig:MEMtagging}, this highlights how $\wbsm$ acts as an effective BSM tagger by exploiting the full matrix-element information and capturing correlations in the event kinematics that go beyond what is accessible through individual observables such as $m_{WW}$, $\phi_{e^+}^\ast$, $p_{T}^{W^+}$, or~$\Delta y_{e^+ W^-}$.
 
\end{appendix}

\FloatBarrier


\begin{thebibliography}{152}%
\makeatletter
\providecommand \@ifxundefined [1]{%
 \@ifx{#1\undefined}
}%
\providecommand \@ifnum [1]{%
 \ifnum #1\expandafter \@firstoftwo
 \else \expandafter \@secondoftwo
 \fi
}%
\providecommand \@ifx [1]{%
 \ifx #1\expandafter \@firstoftwo
 \else \expandafter \@secondoftwo
 \fi
}%
\providecommand \natexlab [1]{#1}%
\providecommand \enquote [1]{``#1''}%
\providecommand \bibnamefont [1]{#1}%
\providecommand \bibfnamefont [1]{#1}%
\providecommand \citenamefont [1]{#1}%
\providecommand \href@noop [0]{\@secondoftwo}%
\providecommand \href [0]{\begingroup \@sanitize@url \@href}%
\providecommand \@href[1]{\@@startlink{#1}\@@href}%
\providecommand \@@href[1]{\endgroup#1\@@endlink}%
\providecommand \@sanitize@url [0]{\catcode `\\12\catcode `\$12\catcode
 `\&12\catcode `\#12\catcode `\^12\catcode `\_12\catcode `\%12\relax}%
\providecommand \@@startlink[1]{}%
\providecommand \@@endlink[0]{}%
\providecommand \url [0]{\begingroup\@sanitize@url \@url }%
\providecommand \@url [1]{\endgroup\@href {#1}{\urlprefix }}%
\providecommand \urlprefix [0]{URL }%
\providecommand \Eprint [0]{\href }%
\providecommand \doibase [0]{http://dx.doi.org/}%
\providecommand \selectlanguage [0]{\@gobble}%
\providecommand \bibinfo [0]{\@secondoftwo}%
\providecommand \bibfield [0]{\@secondoftwo}%
\providecommand \translation [1]{[#1]}%
\providecommand \BibitemOpen [0]{}%
\providecommand \bibitemStop [0]{}%
\providecommand \bibitemNoStop [0]{.\EOS\space}%
\providecommand \EOS [0]{\spacefactor3000\relax}%
\providecommand\BibitemShut [1]{\csname bibitem#1\endcsname}%
\let\auto@bib@innerbib\@empty
\bibitem [{\citenamefont {Kondo}(1988)}]{Kondo:1988yd}%
 \BibitemOpen
 \bibfield {author} {\bibinfo {author} {\bibfnamefont {K.}~\bibnamefont
 {Kondo}},\ }\href {\doibase 10.1143/JPSJ.57.4126} {\bibfield {journal}
 {\bibinfo {journal} {J. Phys. Soc. Jap.}\ }\textbf {\bibinfo {volume}
 {57}},\ \bibinfo {pages} {4126} (\bibinfo {year} {1988})}\BibitemShut
 {NoStop}%
\bibitem [{\citenamefont {Kondo}(1991)}]{Kondo:1991dw}%
 \BibitemOpen
 \bibfield {author} {\bibinfo {author} {\bibfnamefont {K.}~\bibnamefont
 {Kondo}},\ }\href {\doibase 10.1143/JPSJ.60.836} {\bibfield {journal}
 {\bibinfo {journal} {J. Phys. Soc. Jap.}\ }\textbf {\bibinfo {volume}
 {60}},\ \bibinfo {pages} {836} (\bibinfo {year} {1991})}\BibitemShut
 {NoStop}%
\bibitem [{\citenamefont {Dalitz}\ and\ \citenamefont
 {Goldstein}(1992{\natexlab{a}})}]{Dalitz:1991wa}%
 \BibitemOpen
 \bibfield {author} {\bibinfo {author} {\bibfnamefont {R.~H.}\ \bibnamefont
 {Dalitz}}\ and\ \bibinfo {author} {\bibfnamefont {G.~R.}\ \bibnamefont
 {Goldstein}},\ }\href {\doibase 10.1103/PhysRevD.45.1531} {\bibfield
 {journal} {\bibinfo {journal} {Phys. Rev. D}\ }\textbf {\bibinfo {volume}
 {45}},\ \bibinfo {pages} {1531} (\bibinfo {year}
 {1992}{\natexlab{a}})}\BibitemShut {NoStop}%
\bibitem [{\citenamefont {Dalitz}\ and\ \citenamefont
 {Goldstein}(1992{\natexlab{b}})}]{Dalitz:1992np}%
 \BibitemOpen
 \bibfield {author} {\bibinfo {author} {\bibfnamefont {R.~H.}\ \bibnamefont
 {Dalitz}}\ and\ \bibinfo {author} {\bibfnamefont {G.~R.}\ \bibnamefont
 {Goldstein}},\ }\href {\doibase 10.1016/0370-2693(92)91904-N} {\bibfield
 {journal} {\bibinfo {journal} {Phys. Lett. B}\ }\textbf {\bibinfo {volume}
 {287}},\ \bibinfo {pages} {225} (\bibinfo {year}
 {1992}{\natexlab{b}})}\BibitemShut {NoStop}%
\bibitem [{\citenamefont {Kondo}\ \emph {et~al.}(1993)\citenamefont {Kondo},
 \citenamefont {Chikamatsu},\ and\ \citenamefont {Kim}}]{Kondo:1993in}%
 \BibitemOpen
 \bibfield {author} {\bibinfo {author} {\bibfnamefont {K.}~\bibnamefont
 {Kondo}}, \bibinfo {author} {\bibfnamefont {T.}~\bibnamefont {Chikamatsu}}, \
 and\ \bibinfo {author} {\bibfnamefont {S.~H.}\ \bibnamefont {Kim}},\ }\href
 {\doibase 10.1143/JPSJ.62.1177} {\bibfield {journal} {\bibinfo {journal}
 {J. Phys. Soc. Jap.}\ }\textbf {\bibinfo {volume} {62}},\ \bibinfo {pages}
 {1177} (\bibinfo {year} {1993})}\BibitemShut {NoStop}%
\bibitem [{\citenamefont {Fiedler}\ \emph {et~al.}(2010)\citenamefont
 {Fiedler}, \citenamefont {Grohsjean}, \citenamefont {Haefner},\ and\
 \citenamefont {Schieferdecker}}]{Fiedler:2010sg}%
 \BibitemOpen
 \bibfield {author} {\bibinfo {author} {\bibfnamefont {F.}~\bibnamefont
 {Fiedler}}, \bibinfo {author} {\bibfnamefont {A.}~\bibnamefont {Grohsjean}},
 \bibinfo {author} {\bibfnamefont {P.}~\bibnamefont {Haefner}}, \ and\
 \bibinfo {author} {\bibfnamefont {P.}~\bibnamefont {Schieferdecker}},\ }\href
 {\doibase 10.1016/j.nima.2010.09.024} {\bibfield {journal} {\bibinfo
 {journal} {Nucl. Instrum. Meth. A}\ }\textbf {\bibinfo {volume} {624}},\
 \bibinfo {pages} {203} (\bibinfo {year} {2010})},\ \Eprint
 {http://arxiv.org/abs/1003.1316} {arXiv:1003.1316 [hep-ex]}\BibitemShut
 {NoStop}%
\bibitem [{\citenamefont
 {Volobouev}(2011)}]{volobouev2011matrixelementmethodhep}%
 \BibitemOpen
 \bibfield {author} {\bibinfo {author} {\bibfnamefont {I.}~\bibnamefont
 {Volobouev}},\ }\href {https://arxiv.org/abs/1101.2259} {\enquote {\bibinfo
 {title} {Matrix element method in hep: Transfer functions, efficiencies, and
 likelihood normalization},}\ } (\bibinfo {year} {2011}),\ \Eprint
 {http://arxiv.org/abs/1101.2259} {arXiv:1101.2259 [physics.data-an]}\BibitemShut {NoStop}%
\bibitem [{\citenamefont {Gainer}\ \emph {et~al.}(2013)\citenamefont {Gainer},
 \citenamefont {Lykken}, \citenamefont {Matchev}, \citenamefont {Mrenna},\
 and\ \citenamefont {Park}}]{Gainer:2013iya}%
 \BibitemOpen
 \bibfield {author} {\bibinfo {author} {\bibfnamefont {J.~S.}\ \bibnamefont
 {Gainer}}, \bibinfo {author} {\bibfnamefont {J.}~\bibnamefont {Lykken}},
 \bibinfo {author} {\bibfnamefont {K.~T.}\ \bibnamefont {Matchev}}, \bibinfo
 {author} {\bibfnamefont {S.}~\bibnamefont {Mrenna}}, \ and\ \bibinfo {author}
 {\bibfnamefont {M.}~\bibnamefont {Park}},\ }in\ \href@noop {} {\emph
 {\bibinfo {booktitle} {{Snowmass 2013}: {Snowmass on the Mississippi}}}}\
 (\bibinfo {year} {2013})\ \Eprint {http://arxiv.org/abs/1307.3546}
 {arXiv:1307.3546 [hep-ph]}\BibitemShut {NoStop}%
\bibitem [{\citenamefont {Albertsson}\ \emph {et~al.}(2018)\citenamefont
 {Albertsson} \emph {et~al.}}]{Albertsson:2018maf}%
 \BibitemOpen
 \bibfield {author} {\bibinfo {author} {\bibfnamefont {K.}~\bibnamefont
 {Albertsson}} \emph {et~al.},\ }\href {\doibase
 10.1088/1742-6596/1085/2/022008} {\bibfield {journal} {\bibinfo {journal}
 {J. Phys. Conf. Ser.}\ }\textbf {\bibinfo {volume} {1085}},\ \bibinfo {pages}
 {022008} (\bibinfo {year} {2018})},\ \Eprint
 {http://arxiv.org/abs/1807.02876} {arXiv:1807.02876 [physics.comp-ph]}\BibitemShut {NoStop}%
\bibitem [{\citenamefont {Abazov}\ \emph {et~al.}(2005)\citenamefont {Abazov}
 \emph {et~al.}}]{D0:2004lvh}%
 \BibitemOpen
 \bibfield {author} {\bibinfo {author} {\bibfnamefont {V.~M.}\ \bibnamefont
 {Abazov}} \emph {et~al.} (\bibinfo {collaboration} {D0}),\ }\href {\doibase
 10.1016/j.physletb.2005.04.069} {\bibfield {journal} {\bibinfo {journal}
 {Phys. Lett. B}\ }\textbf {\bibinfo {volume} {617}},\ \bibinfo {pages} {1}
 (\bibinfo {year} {2005})},\ \Eprint {http://arxiv.org/abs/hep-ex/0404040}
 {arXiv:hep-ex/0404040}\BibitemShut {NoStop}%
\bibitem [{\citenamefont {Abazov}\ \emph {et~al.}(2004)\citenamefont {Abazov}
 \emph {et~al.}}]{D0:2004rvt}%
 \BibitemOpen
 \bibfield {author} {\bibinfo {author} {\bibfnamefont {V.~M.}\ \bibnamefont
 {Abazov}} \emph {et~al.} (\bibinfo {collaboration} {D0}),\ }\href {\doibase
 10.1038/nature02589} {\bibfield {journal} {\bibinfo {journal} {Nature}\
 }\textbf {\bibinfo {volume} {429}},\ \bibinfo {pages} {638} (\bibinfo {year}
 {2004})},\ \Eprint {http://arxiv.org/abs/hep-ex/0406031}
 {arXiv:hep-ex/0406031}\BibitemShut {NoStop}%
\bibitem [{\citenamefont {Abulencia}\ \emph {et~al.}(2006)\citenamefont
 {Abulencia} \emph {et~al.}}]{CDF:2006nne}%
 \BibitemOpen
 \bibfield {author} {\bibinfo {author} {\bibfnamefont {A.}~\bibnamefont
 {Abulencia}} \emph {et~al.} (\bibinfo {collaboration} {CDF}),\ }\href
 {\doibase 10.1103/PhysRevD.74.032009} {\bibfield {journal} {\bibinfo
 {journal} {Phys. Rev. D}\ }\textbf {\bibinfo {volume} {74}},\ \bibinfo
 {pages} {032009} (\bibinfo {year} {2006})},\ \Eprint
 {http://arxiv.org/abs/hep-ex/0605118} {arXiv:hep-ex/0605118}\BibitemShut
 {NoStop}%
\bibitem [{\citenamefont {Abazov}\ \emph {et~al.}(2006)\citenamefont {Abazov}
 \emph {et~al.}}]{D0:2006fah}%
 \BibitemOpen
 \bibfield {author} {\bibinfo {author} {\bibfnamefont {V.~M.}\ \bibnamefont
 {Abazov}} \emph {et~al.} (\bibinfo {collaboration} {D0}),\ }\href {\doibase
 10.1103/PhysRevD.74.092005} {\bibfield {journal} {\bibinfo {journal} {Phys.
 Rev. D}\ }\textbf {\bibinfo {volume} {74}},\ \bibinfo {pages} {092005}
 (\bibinfo {year} {2006})},\ \Eprint {http://arxiv.org/abs/hep-ex/0609053}
 {arXiv:hep-ex/0609053}\BibitemShut {NoStop}%
\bibitem [{\citenamefont {Aaltonen}\ \emph
 {et~al.}(2009{\natexlab{a}})\citenamefont {Aaltonen} \emph
 {et~al.}}]{CDF:2009uss}%
 \BibitemOpen
 \bibfield {author} {\bibinfo {author} {\bibfnamefont {T.}~\bibnamefont
 {Aaltonen}} \emph {et~al.} (\bibinfo {collaboration} {CDF}),\ }\href
 {\doibase 10.1103/PhysRevLett.103.101802} {\bibfield {journal} {\bibinfo
 {journal} {Phys. Rev. Lett.}\ }\textbf {\bibinfo {volume} {103}},\ \bibinfo
 {pages} {101802} (\bibinfo {year} {2009}{\natexlab{a}})},\ \Eprint
 {http://arxiv.org/abs/0906.5613} {arXiv:0906.5613 [hep-ex]}\BibitemShut
 {NoStop}%
\bibitem [{\citenamefont {Aaltonen}\ \emph
 {et~al.}(2009{\natexlab{b}})\citenamefont {Aaltonen} \emph
 {et~al.}}]{CDF:2009qjw}%
 \BibitemOpen
 \bibfield {author} {\bibinfo {author} {\bibfnamefont {T.}~\bibnamefont
 {Aaltonen}} \emph {et~al.} (\bibinfo {collaboration} {CDF}),\ }\href
 {\doibase 10.1103/PhysRevD.80.071101} {\bibfield {journal} {\bibinfo
 {journal} {Phys. Rev. D}\ }\textbf {\bibinfo {volume} {80}},\ \bibinfo
 {pages} {071101} (\bibinfo {year} {2009}{\natexlab{b}})},\ \Eprint
 {http://arxiv.org/abs/0908.3534} {arXiv:0908.3534 [hep-ex]}\BibitemShut
 {NoStop}%
\bibitem [{\citenamefont {Aaltonen}\ \emph {et~al.}(2010)\citenamefont
 {Aaltonen} \emph {et~al.}}]{CDF:2010zcm}%
 \BibitemOpen
 \bibfield {author} {\bibinfo {author} {\bibfnamefont {T.}~\bibnamefont
 {Aaltonen}} \emph {et~al.} (\bibinfo {collaboration} {CDF}),\ }\href
 {\doibase 10.1103/PhysRevLett.105.252001} {\bibfield {journal} {\bibinfo
 {journal} {Phys. Rev. Lett.}\ }\textbf {\bibinfo {volume} {105}},\ \bibinfo
 {pages} {252001} (\bibinfo {year} {2010})},\ \Eprint
 {http://arxiv.org/abs/1010.4582} {arXiv:1010.4582 [hep-ex]}\BibitemShut
 {NoStop}%
\bibitem [{\citenamefont {Abazov}\ \emph
 {et~al.}(2011{\natexlab{a}})\citenamefont {Abazov} \emph
 {et~al.}}]{D0:2011rkb}%
 \BibitemOpen
 \bibfield {author} {\bibinfo {author} {\bibfnamefont {V.~M.}\ \bibnamefont
 {Abazov}} \emph {et~al.} (\bibinfo {collaboration} {D0}),\ }\href {\doibase
 10.1103/PhysRevLett.107.032001} {\bibfield {journal} {\bibinfo {journal}
 {Phys. Rev. Lett.}\ }\textbf {\bibinfo {volume} {107}},\ \bibinfo {pages}
 {032001} (\bibinfo {year} {2011}{\natexlab{a}})},\ \Eprint
 {http://arxiv.org/abs/1104.5194} {arXiv:1104.5194 [hep-ex]}\BibitemShut
 {NoStop}%
\bibitem [{\citenamefont {Abazov}\ \emph
 {et~al.}(2011{\natexlab{b}})\citenamefont {Abazov} \emph
 {et~al.}}]{D0:2011fla}%
 \BibitemOpen
 \bibfield {author} {\bibinfo {author} {\bibfnamefont {V.~M.}\ \bibnamefont
 {Abazov}} \emph {et~al.} (\bibinfo {collaboration} {D0}),\ }\href {\doibase
 10.1103/PhysRevD.84.032004} {\bibfield {journal} {\bibinfo {journal} {Phys.
 Rev. D}\ }\textbf {\bibinfo {volume} {84}},\ \bibinfo {pages} {032004}
 (\bibinfo {year} {2011}{\natexlab{b}})},\ \Eprint
 {http://arxiv.org/abs/1105.6287} {arXiv:1105.6287 [hep-ex]}\BibitemShut
 {NoStop}%
\bibitem [{\citenamefont {Aaltonen}\ \emph {et~al.}(2011)\citenamefont
 {Aaltonen} \emph {et~al.}}]{CDF:2011yhv}%
 \BibitemOpen
 \bibfield {author} {\bibinfo {author} {\bibfnamefont {T.}~\bibnamefont
 {Aaltonen}} \emph {et~al.} (\bibinfo {collaboration} {CDF}),\ }\href
 {\doibase 10.1103/PhysRevD.84.071105} {\bibfield {journal} {\bibinfo
 {journal} {Phys. Rev. D}\ }\textbf {\bibinfo {volume} {84}},\ \bibinfo
 {pages} {071105} (\bibinfo {year} {2011})},\ \Eprint
 {http://arxiv.org/abs/1108.1601} {arXiv:1108.1601 [hep-ex]}\BibitemShut
 {NoStop}%
\bibitem [{\citenamefont {Aaltonen}\ \emph {et~al.}(2012)\citenamefont
 {Aaltonen} \emph {et~al.}}]{CDF:2011wdm}%
 \BibitemOpen
 \bibfield {author} {\bibinfo {author} {\bibfnamefont {T.}~\bibnamefont
 {Aaltonen}} \emph {et~al.} (\bibinfo {collaboration} {CDF}),\ }\href
 {\doibase 10.1103/PhysRevD.85.072001} {\bibfield {journal} {\bibinfo
 {journal} {Phys. Rev. D}\ }\textbf {\bibinfo {volume} {85}},\ \bibinfo
 {pages} {072001} (\bibinfo {year} {2012})},\ \Eprint
 {http://arxiv.org/abs/1112.4358} {arXiv:1112.4358 [hep-ex]}\BibitemShut
 {NoStop}%
\bibitem [{\citenamefont {Abazov}\ \emph {et~al.}(2016)\citenamefont {Abazov}
 \emph {et~al.}}]{D0:2016ull}%
 \BibitemOpen
 \bibfield {author} {\bibinfo {author} {\bibfnamefont {V.~M.}\ \bibnamefont
 {Abazov}} \emph {et~al.} (\bibinfo {collaboration} {D0}),\ }\href {\doibase
 10.1103/PhysRevD.94.032004} {\bibfield {journal} {\bibinfo {journal} {Phys.
 Rev. D}\ }\textbf {\bibinfo {volume} {94}},\ \bibinfo {pages} {032004}
 (\bibinfo {year} {2016})},\ \Eprint {http://arxiv.org/abs/1606.02814}
 {arXiv:1606.02814 [hep-ex]}\BibitemShut {NoStop}%
\bibitem [{\citenamefont {Chatrchyan}\ \emph {et~al.}(2014)\citenamefont
 {Chatrchyan} \emph {et~al.}}]{CMS:2013fjq}%
 \BibitemOpen
 \bibfield {author} {\bibinfo {author} {\bibfnamefont {S.}~\bibnamefont
 {Chatrchyan}} \emph {et~al.} (\bibinfo {collaboration} {CMS}),\ }\href
 {\doibase 10.1103/PhysRevD.89.092007} {\bibfield {journal} {\bibinfo
 {journal} {Phys. Rev. D}\ }\textbf {\bibinfo {volume} {89}},\ \bibinfo
 {pages} {092007} (\bibinfo {year} {2014})},\ \Eprint
 {http://arxiv.org/abs/1312.5353} {arXiv:1312.5353 [hep-ex]}\BibitemShut
 {NoStop}%
\bibitem [{\citenamefont {Khachatryan}\ \emph {et~al.}(2014)\citenamefont
 {Khachatryan} \emph {et~al.}}]{CMS:2014quz}%
 \BibitemOpen
 \bibfield {author} {\bibinfo {author} {\bibfnamefont {V.}~\bibnamefont
 {Khachatryan}} \emph {et~al.} (\bibinfo {collaboration} {CMS}),\ }\href
 {\doibase 10.1016/j.physletb.2014.06.077} {\bibfield {journal} {\bibinfo
 {journal} {Phys. Lett. B}\ }\textbf {\bibinfo {volume} {736}},\ \bibinfo
 {pages} {64} (\bibinfo {year} {2014})},\ \Eprint
 {http://arxiv.org/abs/1405.3455} {arXiv:1405.3455 [hep-ex]}\BibitemShut
 {NoStop}%
\bibitem [{\citenamefont {Aad}\ \emph {et~al.}(2014)\citenamefont {Aad} \emph
 {et~al.}}]{ATLAS:2014euz}%
 \BibitemOpen
 \bibfield {author} {\bibinfo {author} {\bibfnamefont {G.}~\bibnamefont
 {Aad}} \emph {et~al.} (\bibinfo {collaboration} {ATLAS}),\ }\href {\doibase
 10.1103/PhysRevD.90.052004} {\bibfield {journal} {\bibinfo {journal} {Phys.
 Rev. D}\ }\textbf {\bibinfo {volume} {90}},\ \bibinfo {pages} {052004}
 (\bibinfo {year} {2014})},\ \Eprint {http://arxiv.org/abs/1406.3827}
 {arXiv:1406.3827 [hep-ex]}\BibitemShut {NoStop}%
\bibitem [{\citenamefont {Aad}\ \emph {et~al.}(2015{\natexlab{a}})\citenamefont
 {Aad} \emph {et~al.}}]{ATLAS:2014kct}%
 \BibitemOpen
 \bibfield {author} {\bibinfo {author} {\bibfnamefont {G.}~\bibnamefont
 {Aad}} \emph {et~al.} (\bibinfo {collaboration} {ATLAS}),\ }\href {\doibase
 10.1103/PhysRevD.91.012006} {\bibfield {journal} {\bibinfo {journal} {Phys.
 Rev. D}\ }\textbf {\bibinfo {volume} {91}},\ \bibinfo {pages} {012006}
 (\bibinfo {year} {2015}{\natexlab{a}})},\ \Eprint
 {http://arxiv.org/abs/1408.5191} {arXiv:1408.5191 [hep-ex]}\BibitemShut
 {NoStop}%
\bibitem [{\citenamefont {Khachatryan}\ \emph
 {et~al.}(2015{\natexlab{a}})\citenamefont {Khachatryan} \emph
 {et~al.}}]{CMS:2014nkk}%
 \BibitemOpen
 \bibfield {author} {\bibinfo {author} {\bibfnamefont {V.}~\bibnamefont
 {Khachatryan}} \emph {et~al.} (\bibinfo {collaboration} {CMS}),\ }\href
 {\doibase 10.1103/PhysRevD.92.012004} {\bibfield {journal} {\bibinfo
 {journal} {Phys. Rev. D}\ }\textbf {\bibinfo {volume} {92}},\ \bibinfo
 {pages} {012004} (\bibinfo {year} {2015}{\natexlab{a}})},\ \Eprint
 {http://arxiv.org/abs/1411.3441} {arXiv:1411.3441 [hep-ex]}\BibitemShut
 {NoStop}%
\bibitem [{\citenamefont {Khachatryan}\ \emph
 {et~al.}(2015{\natexlab{b}})\citenamefont {Khachatryan} \emph
 {et~al.}}]{CMS:2015enw}%
 \BibitemOpen
 \bibfield {author} {\bibinfo {author} {\bibfnamefont {V.}~\bibnamefont
 {Khachatryan}} \emph {et~al.} (\bibinfo {collaboration} {CMS}),\ }\href
 {\doibase 10.1140/epjc/s10052-015-3454-1} {\bibfield {journal} {\bibinfo
 {journal} {Eur. Phys. J. C}\ }\textbf {\bibinfo {volume} {75}},\ \bibinfo
 {pages} {251} (\bibinfo {year} {2015}{\natexlab{b}})},\ \Eprint
 {http://arxiv.org/abs/1502.02485} {arXiv:1502.02485 [hep-ex]}\BibitemShut
 {NoStop}%
\bibitem [{\citenamefont {Aad}\ \emph {et~al.}(2015{\natexlab{b}})\citenamefont
 {Aad} \emph {et~al.}}]{ATLAS:2015utn}%
 \BibitemOpen
 \bibfield {author} {\bibinfo {author} {\bibfnamefont {G.}~\bibnamefont
 {Aad}} \emph {et~al.} (\bibinfo {collaboration} {ATLAS}),\ }\href {\doibase
 10.1140/epjc/s10052-015-3543-1} {\bibfield {journal} {\bibinfo {journal}
 {Eur. Phys. J. C}\ }\textbf {\bibinfo {volume} {75}},\ \bibinfo {pages} {349}
 (\bibinfo {year} {2015}{\natexlab{b}})},\ \Eprint
 {http://arxiv.org/abs/1503.05066} {arXiv:1503.05066 [hep-ex]}\BibitemShut
 {NoStop}%
\bibitem [{\citenamefont {Aad}\ \emph {et~al.}(2016)\citenamefont {Aad} \emph
 {et~al.}}]{ATLAS:2015jmq}%
 \BibitemOpen
 \bibfield {author} {\bibinfo {author} {\bibfnamefont {G.}~\bibnamefont
 {Aad}} \emph {et~al.} (\bibinfo {collaboration} {ATLAS}),\ }\href {\doibase
 10.1016/j.physletb.2016.03.017} {\bibfield {journal} {\bibinfo {journal}
 {Phys. Lett. B}\ }\textbf {\bibinfo {volume} {756}},\ \bibinfo {pages} {228}
 (\bibinfo {year} {2016})},\ \Eprint {http://arxiv.org/abs/1511.05980}
 {arXiv:1511.05980 [hep-ex]}\BibitemShut {NoStop}%
\bibitem [{\citenamefont {Khachatryan}\ \emph {et~al.}(2016)\citenamefont
 {Khachatryan} \emph {et~al.}}]{CMS:2015cal}%
 \BibitemOpen
 \bibfield {author} {\bibinfo {author} {\bibfnamefont {V.}~\bibnamefont
 {Khachatryan}} \emph {et~al.} (\bibinfo {collaboration} {CMS}),\ }\href
 {\doibase 10.1016/j.physletb.2016.05.005} {\bibfield {journal} {\bibinfo
 {journal} {Phys. Lett. B}\ }\textbf {\bibinfo {volume} {758}},\ \bibinfo
 {pages} {321} (\bibinfo {year} {2016})},\ \Eprint
 {http://arxiv.org/abs/1511.06170} {arXiv:1511.06170 [hep-ex]}\BibitemShut
 {NoStop}%
\bibitem [{\citenamefont {Sirunyan}\ \emph {et~al.}(2017)\citenamefont
 {Sirunyan} \emph {et~al.}}]{CMS:2017dib}%
 \BibitemOpen
 \bibfield {author} {\bibinfo {author} {\bibfnamefont {A.~M.}\ \bibnamefont
 {Sirunyan}} \emph {et~al.} (\bibinfo {collaboration} {CMS}),\ }\href
 {\doibase 10.1007/JHEP11(2017)047} {\bibfield {journal} {\bibinfo {journal}
 {JHEP}\ }\textbf {\bibinfo {volume} {11}},\ \bibinfo {pages} {047} (\bibinfo
 {year} {2017})},\ \Eprint {http://arxiv.org/abs/1706.09936} {arXiv:1706.09936
 [hep-ex]}\BibitemShut {NoStop}%
\bibitem [{\citenamefont {Sirunyan}\ \emph {et~al.}(2018)\citenamefont
 {Sirunyan} \emph {et~al.}}]{CMS:2018fdh}%
 \BibitemOpen
 \bibfield {author} {\bibinfo {author} {\bibfnamefont {A.~M.}\ \bibnamefont
 {Sirunyan}} \emph {et~al.} (\bibinfo {collaboration} {CMS}),\ }\href
 {\doibase 10.1007/JHEP08(2018)066} {\bibfield {journal} {\bibinfo {journal}
 {JHEP}\ }\textbf {\bibinfo {volume} {08}},\ \bibinfo {pages} {066} (\bibinfo
 {year} {2018})},\ \Eprint {http://arxiv.org/abs/1803.05485} {arXiv:1803.05485
 [hep-ex]}\BibitemShut {NoStop}%
\bibitem [{\citenamefont {Sirunyan}\ \emph {et~al.}(2019)\citenamefont
 {Sirunyan} \emph {et~al.}}]{CMS:2019ekd}%
 \BibitemOpen
 \bibfield {author} {\bibinfo {author} {\bibfnamefont {A.~M.}\ \bibnamefont
 {Sirunyan}} \emph {et~al.} (\bibinfo {collaboration} {CMS}),\ }\href
 {\doibase 10.1103/PhysRevD.99.112003} {\bibfield {journal} {\bibinfo
 {journal} {Phys. Rev. D}\ }\textbf {\bibinfo {volume} {99}},\ \bibinfo
 {pages} {112003} (\bibinfo {year} {2019})},\ \Eprint
 {http://arxiv.org/abs/1901.00174} {arXiv:1901.00174 [hep-ex]}\BibitemShut
 {NoStop}%
\bibitem [{\citenamefont {Sirunyan}\ \emph {et~al.}(2020)\citenamefont
 {Sirunyan} \emph {et~al.}}]{CMS:2020cga}%
 \BibitemOpen
 \bibfield {author} {\bibinfo {author} {\bibfnamefont {A.~M.}\ \bibnamefont
 {Sirunyan}} \emph {et~al.} (\bibinfo {collaboration} {CMS}),\ }\href
 {\doibase 10.1103/PhysRevLett.125.061801} {\bibfield {journal} {\bibinfo
 {journal} {Phys. Rev. Lett.}\ }\textbf {\bibinfo {volume} {125}},\ \bibinfo
 {pages} {061801} (\bibinfo {year} {2020})},\ \Eprint
 {http://arxiv.org/abs/2003.10866} {arXiv:2003.10866 [hep-ex]}\BibitemShut
 {NoStop}%
\bibitem [{\citenamefont {Aad}\ \emph {et~al.}(2025{\natexlab{a}})\citenamefont
 {Aad} \emph {et~al.}}]{ATLAS:2024jry}%
 \BibitemOpen
 \bibfield {author} {\bibinfo {author} {\bibfnamefont {G.}~\bibnamefont
 {Aad}} \emph {et~al.} (\bibinfo {collaboration} {ATLAS}),\ }\href {\doibase
 10.1088/1361-6633/adcd9a} {\bibfield {journal} {\bibinfo {journal} {Rept.
 Prog. Phys.}\ }\textbf {\bibinfo {volume} {88}},\ \bibinfo {pages} {057803}
 (\bibinfo {year} {2025}{\natexlab{a}})},\ \Eprint
 {http://arxiv.org/abs/2412.01548} {arXiv:2412.01548 [hep-ex]}\BibitemShut
 {NoStop}%
\bibitem [{\citenamefont {Aad}\ \emph {et~al.}(2025{\natexlab{b}})\citenamefont
 {Aad} \emph {et~al.}}]{ATLAS:2025clx}%
 \BibitemOpen
 \bibfield {author} {\bibinfo {author} {\bibfnamefont {G.}~\bibnamefont
 {Aad}} \emph {et~al.} (\bibinfo {collaboration} {ATLAS}),\ }\href {\doibase
 10.1088/1361-6633/add370} {\bibfield {journal} {\bibinfo {journal} {Rept.
 Prog. Phys.}\ }\textbf {\bibinfo {volume} {88}},\ \bibinfo {pages} {067801}
 (\bibinfo {year} {2025}{\natexlab{b}})},\ \Eprint
 {http://arxiv.org/abs/2412.01600} {arXiv:2412.01600 [physics.data-an]}\BibitemShut {NoStop}%
\bibitem [{\citenamefont {Gao}\ \emph {et~al.}(2010)\citenamefont {Gao},
 \citenamefont {Gritsan}, \citenamefont {Guo}, \citenamefont {Melnikov},
 \citenamefont {Schulze},\ and\ \citenamefont {Tran}}]{Gao:2010qx}%
 \BibitemOpen
 \bibfield {author} {\bibinfo {author} {\bibfnamefont {Y.}~\bibnamefont
 {Gao}}, \bibinfo {author} {\bibfnamefont {A.~V.}\ \bibnamefont {Gritsan}},
 \bibinfo {author} {\bibfnamefont {Z.}~\bibnamefont {Guo}}, \bibinfo {author}
 {\bibfnamefont {K.}~\bibnamefont {Melnikov}}, \bibinfo {author}
 {\bibfnamefont {M.}~\bibnamefont {Schulze}}, \ and\ \bibinfo {author}
 {\bibfnamefont {N.~V.}\ \bibnamefont {Tran}},\ }\href {\doibase
 10.1103/PhysRevD.81.075022} {\bibfield {journal} {\bibinfo {journal} {Phys.
 Rev. D}\ }\textbf {\bibinfo {volume} {81}},\ \bibinfo {pages} {075022}
 (\bibinfo {year} {2010})},\ \Eprint {http://arxiv.org/abs/1001.3396}
 {arXiv:1001.3396 [hep-ph]}\BibitemShut {NoStop}%
\bibitem [{\citenamefont {Bolognesi}\ \emph {et~al.}(2012)\citenamefont
 {Bolognesi}, \citenamefont {Gao}, \citenamefont {Gritsan}, \citenamefont
 {Melnikov}, \citenamefont {Schulze}, \citenamefont {Tran},\ and\
 \citenamefont {Whitbeck}}]{Bolognesi:2012mm}%
 \BibitemOpen
 \bibfield {author} {\bibinfo {author} {\bibfnamefont {S.}~\bibnamefont
 {Bolognesi}}, \bibinfo {author} {\bibfnamefont {Y.}~\bibnamefont {Gao}},
 \bibinfo {author} {\bibfnamefont {A.~V.}\ \bibnamefont {Gritsan}}, \bibinfo
 {author} {\bibfnamefont {K.}~\bibnamefont {Melnikov}}, \bibinfo {author}
 {\bibfnamefont {M.}~\bibnamefont {Schulze}}, \bibinfo {author} {\bibfnamefont
 {N.~V.}\ \bibnamefont {Tran}}, \ and\ \bibinfo {author} {\bibfnamefont
 {A.}~\bibnamefont {Whitbeck}},\ }\href {\doibase 10.1103/PhysRevD.86.095031}
 {\bibfield {journal} {\bibinfo {journal} {Phys. Rev. D}\ }\textbf {\bibinfo
 {volume} {86}},\ \bibinfo {pages} {095031} (\bibinfo {year} {2012})},\
 \Eprint {http://arxiv.org/abs/1208.4018} {arXiv:1208.4018 [hep-ph]}\BibitemShut {NoStop}%
\bibitem [{\citenamefont {Andersen}\ \emph {et~al.}(2013)\citenamefont
 {Andersen}, \citenamefont {Englert},\ and\ \citenamefont
 {Spannowsky}}]{Andersen:2012kn}%
 \BibitemOpen
 \bibfield {author} {\bibinfo {author} {\bibfnamefont {J.~R.}\ \bibnamefont
 {Andersen}}, \bibinfo {author} {\bibfnamefont {C.}~\bibnamefont {Englert}}, \
 and\ \bibinfo {author} {\bibfnamefont {M.}~\bibnamefont {Spannowsky}},\
 }\href {\doibase 10.1103/PhysRevD.87.015019} {\bibfield {journal} {\bibinfo
 {journal} {Phys. Rev. D}\ }\textbf {\bibinfo {volume} {87}},\ \bibinfo
 {pages} {015019} (\bibinfo {year} {2013})},\ \Eprint
 {http://arxiv.org/abs/1211.3011} {arXiv:1211.3011 [hep-ph]}\BibitemShut
 {NoStop}%
\bibitem [{\citenamefont {Freitas}\ and\ \citenamefont
 {Gainer}(2013)}]{Freitas:2012uk}%
 \BibitemOpen
 \bibfield {author} {\bibinfo {author} {\bibfnamefont {A.}~\bibnamefont
 {Freitas}}\ and\ \bibinfo {author} {\bibfnamefont {J.~S.}\ \bibnamefont
 {Gainer}},\ }\href {\doibase 10.1103/PhysRevD.88.017302} {\bibfield
 {journal} {\bibinfo {journal} {Phys. Rev. D}\ }\textbf {\bibinfo {volume}
 {88}},\ \bibinfo {pages} {017302} (\bibinfo {year} {2013})},\ \Eprint
 {http://arxiv.org/abs/1212.3598} {arXiv:1212.3598 [hep-ph]}\BibitemShut
 {NoStop}%
\bibitem [{\citenamefont {Artoisenet}\ \emph {et~al.}(2013)\citenamefont
 {Artoisenet}, \citenamefont {de~Aquino}, \citenamefont {Maltoni},\ and\
 \citenamefont {Mattelaer}}]{Artoisenet:2013vfa}%
 \BibitemOpen
 \bibfield {author} {\bibinfo {author} {\bibfnamefont {P.}~\bibnamefont
 {Artoisenet}}, \bibinfo {author} {\bibfnamefont {P.}~\bibnamefont
 {de~Aquino}}, \bibinfo {author} {\bibfnamefont {F.}~\bibnamefont {Maltoni}},
 \ and\ \bibinfo {author} {\bibfnamefont {O.}~\bibnamefont {Mattelaer}},\
 }\href {\doibase 10.1103/PhysRevLett.111.091802} {\bibfield {journal}
 {\bibinfo {journal} {Phys. Rev. Lett.}\ }\textbf {\bibinfo {volume} {111}},\
 \bibinfo {pages} {091802} (\bibinfo {year} {2013})},\ \Eprint
 {http://arxiv.org/abs/1304.6414} {arXiv:1304.6414 [hep-ph]}\BibitemShut
 {NoStop}%
\bibitem [{\citenamefont {Anderson}\ \emph {et~al.}(2014)\citenamefont
 {Anderson} \emph {et~al.}}]{Anderson:2013afp}%
 \BibitemOpen
 \bibfield {author} {\bibinfo {author} {\bibfnamefont {I.}~\bibnamefont
 {Anderson}} \emph {et~al.},\ }\href {\doibase 10.1103/PhysRevD.89.035007}
 {\bibfield {journal} {\bibinfo {journal} {Phys. Rev. D}\ }\textbf {\bibinfo
 {volume} {89}},\ \bibinfo {pages} {035007} (\bibinfo {year} {2014})},\
 \Eprint {http://arxiv.org/abs/1309.4819} {arXiv:1309.4819 [hep-ph]}\BibitemShut {NoStop}%
\bibitem [{\citenamefont {Campbell}\ \emph {et~al.}(2014)\citenamefont
 {Campbell}, \citenamefont {Ellis},\ and\ \citenamefont
 {Williams}}]{Campbell:2013una}%
 \BibitemOpen
 \bibfield {author} {\bibinfo {author} {\bibfnamefont {J.~M.}\ \bibnamefont
 {Campbell}}, \bibinfo {author} {\bibfnamefont {R.~K.}\ \bibnamefont {Ellis}},
 \ and\ \bibinfo {author} {\bibfnamefont {C.}~\bibnamefont {Williams}},\
 }\href {\doibase 10.1007/JHEP04(2014)060} {\bibfield {journal} {\bibinfo
 {journal} {JHEP}\ }\textbf {\bibinfo {volume} {04}},\ \bibinfo {pages} {060}
 (\bibinfo {year} {2014})},\ \Eprint {http://arxiv.org/abs/1311.3589}
 {arXiv:1311.3589 [hep-ph]}\BibitemShut {NoStop}%
\bibitem [{\citenamefont {Schouten}\ \emph {et~al.}(2015)\citenamefont
 {Schouten}, \citenamefont {DeAbreu},\ and\ \citenamefont
 {Stelzer}}]{Schouten:2014yza}%
 \BibitemOpen
 \bibfield {author} {\bibinfo {author} {\bibfnamefont {D.}~\bibnamefont
 {Schouten}}, \bibinfo {author} {\bibfnamefont {A.}~\bibnamefont {DeAbreu}}, \
 and\ \bibinfo {author} {\bibfnamefont {B.}~\bibnamefont {Stelzer}},\ }\href
 {\doibase 10.1016/j.cpc.2015.02.020} {\bibfield {journal} {\bibinfo
 {journal} {Comput. Phys. Commun.}\ }\textbf {\bibinfo {volume} {192}},\
 \bibinfo {pages} {54} (\bibinfo {year} {2015})},\ \Eprint
 {http://arxiv.org/abs/1407.7595} {arXiv:1407.7595 [physics.comp-ph]}\BibitemShut {NoStop}%
\bibitem [{\citenamefont {Gritsan}\ \emph {et~al.}(2016)\citenamefont
 {Gritsan}, \citenamefont {R{\"o}ntsch}, \citenamefont {Schulze},\ and\
 \citenamefont {Xiao}}]{Gritsan:2016hjl}%
 \BibitemOpen
 \bibfield {author} {\bibinfo {author} {\bibfnamefont {A.~V.}\ \bibnamefont
 {Gritsan}}, \bibinfo {author} {\bibfnamefont {R.}~\bibnamefont
 {R{\"o}ntsch}}, \bibinfo {author} {\bibfnamefont {M.}~\bibnamefont
 {Schulze}}, \ and\ \bibinfo {author} {\bibfnamefont {M.}~\bibnamefont
 {Xiao}},\ }\href {\doibase 10.1103/PhysRevD.94.055023} {\bibfield {journal}
 {\bibinfo {journal} {Phys. Rev. D}\ }\textbf {\bibinfo {volume} {94}},\
 \bibinfo {pages} {055023} (\bibinfo {year} {2016})},\ \Eprint
 {http://arxiv.org/abs/1606.03107} {arXiv:1606.03107 [hep-ph]}\BibitemShut
 {NoStop}%
\bibitem [{\citenamefont {Elahi}\ and\ \citenamefont
 {Martin}(2017)}]{Elahi:2017ppe}%
 \BibitemOpen
 \bibfield {author} {\bibinfo {author} {\bibfnamefont {F.}~\bibnamefont
 {Elahi}}\ and\ \bibinfo {author} {\bibfnamefont {A.}~\bibnamefont {Martin}},\
 }\href {\doibase 10.1103/PhysRevD.96.015021} {\bibfield {journal} {\bibinfo
 {journal} {Phys. Rev. D}\ }\textbf {\bibinfo {volume} {96}},\ \bibinfo
 {pages} {015021} (\bibinfo {year} {2017})},\ \Eprint
 {http://arxiv.org/abs/1705.02563} {arXiv:1705.02563 [hep-ph]}\BibitemShut
 {NoStop}%
\bibitem [{\citenamefont {Betancur}\ \emph {et~al.}(2019)\citenamefont
 {Betancur}, \citenamefont {Debnath}, \citenamefont {Gainer}, \citenamefont
 {Matchev},\ and\ \citenamefont {Shyamsundar}}]{Betancur:2017kqe}%
 \BibitemOpen
 \bibfield {author} {\bibinfo {author} {\bibfnamefont {A.}~\bibnamefont
 {Betancur}}, \bibinfo {author} {\bibfnamefont {D.}~\bibnamefont {Debnath}},
 \bibinfo {author} {\bibfnamefont {J.~S.}\ \bibnamefont {Gainer}}, \bibinfo
 {author} {\bibfnamefont {K.~T.}\ \bibnamefont {Matchev}}, \ and\ \bibinfo
 {author} {\bibfnamefont {P.}~\bibnamefont {Shyamsundar}},\ }\href {\doibase
 10.1103/PhysRevD.99.116007} {\bibfield {journal} {\bibinfo {journal} {Phys.
 Rev. D}\ }\textbf {\bibinfo {volume} {99}},\ \bibinfo {pages} {116007}
 (\bibinfo {year} {2019})},\ \Eprint {http://arxiv.org/abs/1708.07641}
 {arXiv:1708.07641 [hep-ph]}\BibitemShut {NoStop}%
\bibitem [{\citenamefont {Gritsan}\ \emph {et~al.}(2020)\citenamefont
 {Gritsan}, \citenamefont {Roskes}, \citenamefont {Sarica}, \citenamefont
 {Schulze}, \citenamefont {Xiao},\ and\ \citenamefont
 {Zhou}}]{Gritsan:2020pib}%
 \BibitemOpen
 \bibfield {author} {\bibinfo {author} {\bibfnamefont {A.~V.}\ \bibnamefont
 {Gritsan}}, \bibinfo {author} {\bibfnamefont {J.}~\bibnamefont {Roskes}},
 \bibinfo {author} {\bibfnamefont {U.}~\bibnamefont {Sarica}}, \bibinfo
 {author} {\bibfnamefont {M.}~\bibnamefont {Schulze}}, \bibinfo {author}
 {\bibfnamefont {M.}~\bibnamefont {Xiao}}, \ and\ \bibinfo {author}
 {\bibfnamefont {Y.}~\bibnamefont {Zhou}},\ }\href {\doibase
 10.1103/PhysRevD.102.056022} {\bibfield {journal} {\bibinfo {journal}
 {Phys. Rev. D}\ }\textbf {\bibinfo {volume} {102}},\ \bibinfo {pages}
 {056022} (\bibinfo {year} {2020})},\ \Eprint
 {http://arxiv.org/abs/2002.09888} {arXiv:2002.09888 [hep-ph]}\BibitemShut
 {NoStop}%
\bibitem [{\citenamefont {Bahl}\ and\ \citenamefont
 {Brass}(2022)}]{Bahl:2021dnc}%
 \BibitemOpen
 \bibfield {author} {\bibinfo {author} {\bibfnamefont {H.}~\bibnamefont
 {Bahl}}\ and\ \bibinfo {author} {\bibfnamefont {S.}~\bibnamefont {Brass}},\
 }\href {\doibase 10.1007/JHEP03(2022)017} {\bibfield {journal} {\bibinfo
 {journal} {JHEP}\ }\textbf {\bibinfo {volume} {03}},\ \bibinfo {pages} {017}
 (\bibinfo {year} {2022})},\ \Eprint {http://arxiv.org/abs/2110.10177}
 {arXiv:2110.10177 [hep-ph]}\BibitemShut {NoStop}%
\bibitem [{\citenamefont {Haisch}\ and\ \citenamefont
 {Koole}(2022{\natexlab{a}})}]{Haisch:2021hvy}%
 \BibitemOpen
 \bibfield {author} {\bibinfo {author} {\bibfnamefont {U.}~\bibnamefont
 {Haisch}}\ and\ \bibinfo {author} {\bibfnamefont {G.}~\bibnamefont {Koole}},\
 }\href {\doibase 10.1007/JHEP02(2022)030} {\bibfield {journal} {\bibinfo
 {journal} {JHEP}\ }\textbf {\bibinfo {volume} {02}},\ \bibinfo {pages} {030}
 (\bibinfo {year} {2022}{\natexlab{a}})},\ \Eprint
 {http://arxiv.org/abs/2111.12589} {arXiv:2111.12589 [hep-ph]}\BibitemShut
 {NoStop}%
\bibitem [{\citenamefont {Haisch}\ and\ \citenamefont
 {Koole}(2022{\natexlab{b}})}]{Haisch:2022rkm}%
 \BibitemOpen
 \bibfield {author} {\bibinfo {author} {\bibfnamefont {U.}~\bibnamefont
 {Haisch}}\ and\ \bibinfo {author} {\bibfnamefont {G.}~\bibnamefont {Koole}},\
 }\href {\doibase 10.1007/JHEP04(2022)166} {\bibfield {journal} {\bibinfo
 {journal} {JHEP}\ }\textbf {\bibinfo {volume} {04}},\ \bibinfo {pages} {166}
 (\bibinfo {year} {2022}{\natexlab{b}})},\ \Eprint
 {http://arxiv.org/abs/2201.09711} {arXiv:2201.09711 [hep-ph]}\BibitemShut
 {NoStop}%
\bibitem [{\citenamefont {Butter}\ \emph {et~al.}(2023)\citenamefont {Butter},
 \citenamefont {Heimel}, \citenamefont {Martini}, \citenamefont {Peitzsch},\
 and\ \citenamefont {Plehn}}]{Butter:2022vkj}%
 \BibitemOpen
 \bibfield {author} {\bibinfo {author} {\bibfnamefont {A.}~\bibnamefont
 {Butter}}, \bibinfo {author} {\bibfnamefont {T.}~\bibnamefont {Heimel}},
 \bibinfo {author} {\bibfnamefont {T.}~\bibnamefont {Martini}}, \bibinfo
 {author} {\bibfnamefont {S.}~\bibnamefont {Peitzsch}}, \ and\ \bibinfo
 {author} {\bibfnamefont {T.}~\bibnamefont {Plehn}},\ }\href {\doibase
 10.21468/SciPostPhys.15.3.094} {\bibfield {journal} {\bibinfo {journal}
 {SciPost Phys.}\ }\textbf {\bibinfo {volume} {15}},\ \bibinfo {pages} {094}
 (\bibinfo {year} {2023})},\ \Eprint {http://arxiv.org/abs/2210.00019}
 {arXiv:2210.00019 [hep-ph]}\BibitemShut {NoStop}%
\bibitem [{\citenamefont {Heimel}\ \emph {et~al.}(2024)\citenamefont {Heimel},
 \citenamefont {Huetsch}, \citenamefont {Winterhalder}, \citenamefont
 {Plehn},\ and\ \citenamefont {Butter}}]{Heimel:2023mvw}%
 \BibitemOpen
 \bibfield {author} {\bibinfo {author} {\bibfnamefont {T.}~\bibnamefont
 {Heimel}}, \bibinfo {author} {\bibfnamefont {N.}~\bibnamefont {Huetsch}},
 \bibinfo {author} {\bibfnamefont {R.}~\bibnamefont {Winterhalder}}, \bibinfo
 {author} {\bibfnamefont {T.}~\bibnamefont {Plehn}}, \ and\ \bibinfo {author}
 {\bibfnamefont {A.}~\bibnamefont {Butter}},\ }\href {\doibase
 10.21468/SciPostPhys.17.5.129} {\bibfield {journal} {\bibinfo {journal}
 {SciPost Phys.}\ }\textbf {\bibinfo {volume} {17}},\ \bibinfo {pages} {129}
 (\bibinfo {year} {2024})},\ \Eprint {http://arxiv.org/abs/2310.07752}
 {arXiv:2310.07752 [hep-ph]}\BibitemShut {NoStop}%
\bibitem [{\citenamefont {Balzani}\ \emph {et~al.}(2023)\citenamefont
 {Balzani}, \citenamefont {Gr{\"o}ber},\ and\ \citenamefont
 {Vitti}}]{Balzani:2023jas}%
 \BibitemOpen
 \bibfield {author} {\bibinfo {author} {\bibfnamefont {E.}~\bibnamefont
 {Balzani}}, \bibinfo {author} {\bibfnamefont {R.}~\bibnamefont {Gr{\"o}ber}},
 \ and\ \bibinfo {author} {\bibfnamefont {M.}~\bibnamefont {Vitti}},\ }\href
 {\doibase 10.1007/JHEP10(2023)027} {\bibfield {journal} {\bibinfo {journal}
 {JHEP}\ }\textbf {\bibinfo {volume} {10}},\ \bibinfo {pages} {027} (\bibinfo
 {year} {2023})},\ \Eprint {http://arxiv.org/abs/2304.09772} {arXiv:2304.09772
 [hep-ph]}\BibitemShut {NoStop}%
\bibitem [{\citenamefont {Haisch}\ \emph {et~al.}(2024)\citenamefont {Haisch},
 \citenamefont {Ruhdorfer}, \citenamefont {Schmid},\ and\ \citenamefont
 {Weiler}}]{Haisch:2023aiz}%
 \BibitemOpen
 \bibfield {author} {\bibinfo {author} {\bibfnamefont {U.}~\bibnamefont
 {Haisch}}, \bibinfo {author} {\bibfnamefont {M.}~\bibnamefont {Ruhdorfer}},
 \bibinfo {author} {\bibfnamefont {K.}~\bibnamefont {Schmid}}, \ and\ \bibinfo
 {author} {\bibfnamefont {A.}~\bibnamefont {Weiler}},\ }\href {\doibase
 10.21468/SciPostPhys.16.4.112} {\bibfield {journal} {\bibinfo {journal}
 {SciPost Phys.}\ }\textbf {\bibinfo {volume} {16}},\ \bibinfo {pages} {112}
 (\bibinfo {year} {2024})},\ \Eprint {http://arxiv.org/abs/2311.03995}
 {arXiv:2311.03995 [hep-ph]}\BibitemShut {NoStop}%
\bibitem [{\citenamefont {Mastandrea}\ \emph {et~al.}(2024)\citenamefont
 {Mastandrea}, \citenamefont {Nachman},\ and\ \citenamefont
 {Plehn}}]{Mastandrea:2024irf}%
 \BibitemOpen
 \bibfield {author} {\bibinfo {author} {\bibfnamefont {R.}~\bibnamefont
 {Mastandrea}}, \bibinfo {author} {\bibfnamefont {B.}~\bibnamefont {Nachman}},
 \ and\ \bibinfo {author} {\bibfnamefont {T.}~\bibnamefont {Plehn}},\ }\href
 {\doibase 10.1103/PhysRevD.110.056004} {\bibfield {journal} {\bibinfo
 {journal} {Phys. Rev. D}\ }\textbf {\bibinfo {volume} {110}},\ \bibinfo
 {pages} {056004} (\bibinfo {year} {2024})},\ \Eprint
 {http://arxiv.org/abs/2405.15847} {arXiv:2405.15847 [hep-ph]}\BibitemShut
 {NoStop}%
\bibitem [{\citenamefont {Ghosh}\ \emph {et~al.}(2026)\citenamefont {Ghosh},
 \citenamefont {Griese}, \citenamefont {Haisch},\ and\ \citenamefont
 {Park}}]{Ghosh:2025fma}%
 \BibitemOpen
 \bibfield {author} {\bibinfo {author} {\bibfnamefont {A.}~\bibnamefont
 {Ghosh}}, \bibinfo {author} {\bibfnamefont {M.}~\bibnamefont {Griese}},
 \bibinfo {author} {\bibfnamefont {U.}~\bibnamefont {Haisch}}, \ and\ \bibinfo
 {author} {\bibfnamefont {T.~H.}\ \bibnamefont {Park}},\ }\href {\doibase
 10.1140/epjc/s10052-026-15605-3} {\bibfield {journal} {\bibinfo {journal}
 {Eur. Phys. J. C}\ }\textbf {\bibinfo {volume} {86}},\ \bibinfo {pages} {415}
 (\bibinfo {year} {2026})},\ \Eprint {http://arxiv.org/abs/2507.02032}
 {arXiv:2507.02032 [hep-ph]}\BibitemShut {NoStop}%
\bibitem [{\citenamefont {Silva}\ \emph {et~al.}(2026)\citenamefont {Silva},
 \citenamefont {Barru{\'e}}, \citenamefont {Ochoa},\ and\ \citenamefont
 {Conde~Mu{\'\i}{\~n}o}}]{Silva:2025hzo}%
 \BibitemOpen
 \bibfield {author} {\bibinfo {author} {\bibfnamefont {M.}~\bibnamefont
 {Silva}}, \bibinfo {author} {\bibfnamefont {R.}~\bibnamefont {Barru{\'e}}},
 \bibinfo {author} {\bibfnamefont {I.}~\bibnamefont {Ochoa}}, \ and\ \bibinfo
 {author} {\bibfnamefont {P.}~\bibnamefont {Conde~Mu{\'\i}{\~n}o}},\ }\href
 {\doibase 10.1103/29xq-b66r} {\bibfield {journal} {\bibinfo {journal}
 {Phys. Rev. D}\ }\textbf {\bibinfo {volume} {113}},\ \bibinfo {pages}
 {056011} (\bibinfo {year} {2026})},\ \Eprint
 {http://arxiv.org/abs/2509.03307} {arXiv:2509.03307 [hep-ph]}\BibitemShut
 {NoStop}%
\bibitem [{\citenamefont {Bahl}\ \emph {et~al.}(2025)\citenamefont {Bahl},
 \citenamefont {Plehn},\ and\ \citenamefont {Schmal}}]{Bahl:2025mib}%
 \BibitemOpen
 \bibfield {author} {\bibinfo {author} {\bibfnamefont {H.}~\bibnamefont
 {Bahl}}, \bibinfo {author} {\bibfnamefont {T.}~\bibnamefont {Plehn}}, \ and\
 \bibinfo {author} {\bibfnamefont {N.}~\bibnamefont {Schmal}},\ }\href@noop {}
 {\ (\bibinfo {year} {2025})},\ \Eprint {http://arxiv.org/abs/2509.05409}
 {arXiv:2509.05409 [hep-ph]}\BibitemShut {NoStop}%
\bibitem [{\citenamefont {Artoisenet}\ and\ \citenamefont
 {Mattelaer}(2008)}]{Artoisenet:2008zz}%
 \BibitemOpen
 \bibfield {author} {\bibinfo {author} {\bibfnamefont {P.}~\bibnamefont
 {Artoisenet}}\ and\ \bibinfo {author} {\bibfnamefont {O.}~\bibnamefont
 {Mattelaer}},\ }\href {\doibase 10.22323/1.073.0025} {\bibfield {journal}
 {\bibinfo {journal} {PoS}\ }\textbf {\bibinfo {volume} {CHARGED2008}},\
 \bibinfo {pages} {025} (\bibinfo {year} {2008})}\BibitemShut {NoStop}%
\bibitem [{\citenamefont {Artoisenet}\ \emph {et~al.}(2010)\citenamefont
 {Artoisenet}, \citenamefont {Lemaitre}, \citenamefont {Maltoni},\ and\
 \citenamefont {Mattelaer}}]{Artoisenet:2010cn}%
 \BibitemOpen
 \bibfield {author} {\bibinfo {author} {\bibfnamefont {P.}~\bibnamefont
 {Artoisenet}}, \bibinfo {author} {\bibfnamefont {V.}~\bibnamefont
 {Lemaitre}}, \bibinfo {author} {\bibfnamefont {F.}~\bibnamefont {Maltoni}}, \
 and\ \bibinfo {author} {\bibfnamefont {O.}~\bibnamefont {Mattelaer}},\ }\href
 {\doibase 10.1007/JHEP12(2010)068} {\bibfield {journal} {\bibinfo {journal}
 {JHEP}\ }\textbf {\bibinfo {volume} {12}},\ \bibinfo {pages} {068} (\bibinfo
 {year} {2010})},\ \Eprint {http://arxiv.org/abs/1007.3300} {arXiv:1007.3300
 [hep-ph]}\BibitemShut {NoStop}%
\bibitem [{\citenamefont {Brochet}\ \emph {et~al.}(2019)\citenamefont
 {Brochet}, \citenamefont {Delaere}, \citenamefont {Fran{\c{c}}ois},
 \citenamefont {Lema{\^\i}tre}, \citenamefont {Mertens}, \citenamefont
 {Saggio}, \citenamefont {Vidal~Marono},\ and\ \citenamefont
 {Wertz}}]{Brochet:2018pqf}%
 \BibitemOpen
 \bibfield {author} {\bibinfo {author} {\bibfnamefont {S.}~\bibnamefont
 {Brochet}}, \bibinfo {author} {\bibfnamefont {C.}~\bibnamefont {Delaere}},
 \bibinfo {author} {\bibfnamefont {B.}~\bibnamefont {Fran{\c{c}}ois}},
 \bibinfo {author} {\bibfnamefont {V.}~\bibnamefont {Lema{\^\i}tre}}, \bibinfo
 {author} {\bibfnamefont {A.}~\bibnamefont {Mertens}}, \bibinfo {author}
 {\bibfnamefont {A.}~\bibnamefont {Saggio}}, \bibinfo {author} {\bibfnamefont
 {M.}~\bibnamefont {Vidal~Marono}}, \ and\ \bibinfo {author} {\bibfnamefont
 {S.}~\bibnamefont {Wertz}},\ }\href {\doibase 10.1140/epjc/s10052-019-6635-5}
 {\bibfield {journal} {\bibinfo {journal} {Eur. Phys. J. C}\ }\textbf
 {\bibinfo {volume} {79}},\ \bibinfo {pages} {126} (\bibinfo {year} {2019})},\
 \Eprint {http://arxiv.org/abs/1805.08555} {arXiv:1805.08555 [hep-ph]}\BibitemShut {NoStop}%
\bibitem [{\citenamefont {Alwall}\ \emph {et~al.}(2011)\citenamefont {Alwall},
 \citenamefont {Freitas},\ and\ \citenamefont {Mattelaer}}]{Alwall:2010cq}%
 \BibitemOpen
 \bibfield {author} {\bibinfo {author} {\bibfnamefont {J.}~\bibnamefont
 {Alwall}}, \bibinfo {author} {\bibfnamefont {A.}~\bibnamefont {Freitas}}, \
 and\ \bibinfo {author} {\bibfnamefont {O.}~\bibnamefont {Mattelaer}},\ }\href
 {\doibase 10.1103/PhysRevD.83.074010} {\bibfield {journal} {\bibinfo
 {journal} {Phys. Rev. D}\ }\textbf {\bibinfo {volume} {83}},\ \bibinfo
 {pages} {074010} (\bibinfo {year} {2011})},\ \Eprint
 {http://arxiv.org/abs/1010.2263} {arXiv:1010.2263 [hep-ph]}\BibitemShut
 {NoStop}%
\bibitem [{\citenamefont {Campbell}\ \emph
 {et~al.}(2012{\natexlab{a}})\citenamefont {Campbell}, \citenamefont {Giele},\
 and\ \citenamefont {Williams}}]{Campbell:2012cz}%
 \BibitemOpen
 \bibfield {author} {\bibinfo {author} {\bibfnamefont {J.~M.}\ \bibnamefont
 {Campbell}}, \bibinfo {author} {\bibfnamefont {W.~T.}\ \bibnamefont {Giele}},
 \ and\ \bibinfo {author} {\bibfnamefont {C.}~\bibnamefont {Williams}},\
 }\href {\doibase 10.1007/JHEP11(2012)043} {\bibfield {journal} {\bibinfo
 {journal} {JHEP}\ }\textbf {\bibinfo {volume} {11}},\ \bibinfo {pages} {043}
 (\bibinfo {year} {2012}{\natexlab{a}})},\ \Eprint
 {http://arxiv.org/abs/1204.4424} {arXiv:1204.4424 [hep-ph]}\BibitemShut
 {NoStop}%
\bibitem [{\citenamefont {Campbell}\ \emph
 {et~al.}(2012{\natexlab{b}})\citenamefont {Campbell}, \citenamefont {Giele},\
 and\ \citenamefont {Williams}}]{Campbell:2012ct}%
 \BibitemOpen
 \bibfield {author} {\bibinfo {author} {\bibfnamefont {J.~M.}\ \bibnamefont
 {Campbell}}, \bibinfo {author} {\bibfnamefont {W.~T.}\ \bibnamefont {Giele}},
 \ and\ \bibinfo {author} {\bibfnamefont {C.}~\bibnamefont {Williams}},\ }in\
 \href@noop {} {\emph {\bibinfo {booktitle} {{47th Rencontres de Moriond on
 QCD and High Energy Interactions}}}}\ (\bibinfo {year} {2012})\ pp.\ \bibinfo
 {pages} {319--322},\ \Eprint {http://arxiv.org/abs/1205.3434}
 {arXiv:1205.3434 [hep-ph]}\BibitemShut {NoStop}%
\bibitem [{\citenamefont {Campbell}\ \emph {et~al.}(2013)\citenamefont
 {Campbell}, \citenamefont {Ellis}, \citenamefont {Giele},\ and\ \citenamefont
 {Williams}}]{Campbell:2013hz}%
 \BibitemOpen
 \bibfield {author} {\bibinfo {author} {\bibfnamefont {J.~M.}\ \bibnamefont
 {Campbell}}, \bibinfo {author} {\bibfnamefont {R.~K.}\ \bibnamefont {Ellis}},
 \bibinfo {author} {\bibfnamefont {W.~T.}\ \bibnamefont {Giele}}, \ and\
 \bibinfo {author} {\bibfnamefont {C.}~\bibnamefont {Williams}},\ }\href
 {\doibase 10.1103/PhysRevD.87.073005} {\bibfield {journal} {\bibinfo
 {journal} {Phys. Rev. D}\ }\textbf {\bibinfo {volume} {87}},\ \bibinfo
 {pages} {073005} (\bibinfo {year} {2013})},\ \Eprint
 {http://arxiv.org/abs/1301.7086} {arXiv:1301.7086 [hep-ph]}\BibitemShut
 {NoStop}%
\bibitem [{\citenamefont {Martini}\ and\ \citenamefont
 {Uwer}(2015)}]{Martini:2015fsa}%
 \BibitemOpen
 \bibfield {author} {\bibinfo {author} {\bibfnamefont {T.}~\bibnamefont
 {Martini}}\ and\ \bibinfo {author} {\bibfnamefont {P.}~\bibnamefont {Uwer}},\
 }\href {\doibase 10.1007/JHEP09(2015)083} {\bibfield {journal} {\bibinfo
 {journal} {JHEP}\ }\textbf {\bibinfo {volume} {09}},\ \bibinfo {pages} {083}
 (\bibinfo {year} {2015})},\ \Eprint {http://arxiv.org/abs/1506.08798}
 {arXiv:1506.08798 [hep-ph]}\BibitemShut {NoStop}%
\bibitem [{\citenamefont {Baumeister}\ and\ \citenamefont
 {Weinzierl}(2017)}]{Baumeister:2016maz}%
 \BibitemOpen
 \bibfield {author} {\bibinfo {author} {\bibfnamefont {R.}~\bibnamefont
 {Baumeister}}\ and\ \bibinfo {author} {\bibfnamefont {S.}~\bibnamefont
 {Weinzierl}},\ }\href {\doibase 10.1103/PhysRevD.95.036019} {\bibfield
 {journal} {\bibinfo {journal} {Phys. Rev. D}\ }\textbf {\bibinfo {volume}
 {95}},\ \bibinfo {pages} {036019} (\bibinfo {year} {2017})},\ \Eprint
 {http://arxiv.org/abs/1612.07252} {arXiv:1612.07252 [hep-ph]}\BibitemShut
 {NoStop}%
\bibitem [{\citenamefont {Kraus}\ \emph
 {et~al.}(2019{\natexlab{a}})\citenamefont {Kraus}, \citenamefont {Martini},\
 and\ \citenamefont {Uwer}}]{Kraus:2019qoq}%
 \BibitemOpen
 \bibfield {author} {\bibinfo {author} {\bibfnamefont {M.}~\bibnamefont
 {Kraus}}, \bibinfo {author} {\bibfnamefont {T.}~\bibnamefont {Martini}}, \
 and\ \bibinfo {author} {\bibfnamefont {P.}~\bibnamefont {Uwer}},\ }\href
 {\doibase 10.1103/PhysRevD.100.076010} {\bibfield {journal} {\bibinfo
 {journal} {Phys. Rev. D}\ }\textbf {\bibinfo {volume} {100}},\ \bibinfo
 {pages} {076010} (\bibinfo {year} {2019}{\natexlab{a}})},\ \Eprint
 {http://arxiv.org/abs/1901.08008} {arXiv:1901.08008 [hep-ph]}\BibitemShut
 {NoStop}%
\bibitem [{\citenamefont {Kraus}\ \emph
 {et~al.}(2019{\natexlab{b}})\citenamefont {Kraus}, \citenamefont {Martini},
 \citenamefont {Peitzsch},\ and\ \citenamefont {Uwer}}]{Kraus:2019myc}%
 \BibitemOpen
 \bibfield {author} {\bibinfo {author} {\bibfnamefont {M.}~\bibnamefont
 {Kraus}}, \bibinfo {author} {\bibfnamefont {T.}~\bibnamefont {Martini}},
 \bibinfo {author} {\bibfnamefont {S.}~\bibnamefont {Peitzsch}}, \ and\
 \bibinfo {author} {\bibfnamefont {P.}~\bibnamefont {Uwer}},\ }\href@noop {}
 {\ (\bibinfo {year} {2019}{\natexlab{b}})},\ \Eprint
 {http://arxiv.org/abs/1908.09100} {arXiv:1908.09100 [hep-ph]}\BibitemShut
 {NoStop}%
\bibitem [{\citenamefont {Martini}\ \emph {et~al.}(2023)\citenamefont
 {Martini}, \citenamefont {Nuraliyev},\ and\ \citenamefont
 {Uwer}}]{Martini:2023ylv}%
 \BibitemOpen
 \bibfield {author} {\bibinfo {author} {\bibfnamefont {T.}~\bibnamefont
 {Martini}}, \bibinfo {author} {\bibfnamefont {T.}~\bibnamefont {Nuraliyev}},
 \ and\ \bibinfo {author} {\bibfnamefont {P.}~\bibnamefont {Uwer}},\ }\href
 {\doibase 10.1103/PhysRevD.107.076013} {\bibfield {journal} {\bibinfo
 {journal} {Phys. Rev. D}\ }\textbf {\bibinfo {volume} {107}},\ \bibinfo
 {pages} {076013} (\bibinfo {year} {2023})},\ \Eprint
 {http://arxiv.org/abs/2301.03280} {arXiv:2301.03280 [hep-ph]}\BibitemShut
 {NoStop}%
\bibitem [{\citenamefont {Tartarin}(2025)}]{Tartarin:2025gbt}%
 \BibitemOpen
 \bibfield {author} {\bibinfo {author} {\bibfnamefont {M.}~\bibnamefont
 {Tartarin}},\ }\emph {\bibinfo {title} {\href{https://theses.hal.science/tel-05441500}{Contribution to the study of the
 Higgs boson self-coupling in the $b \bar b \gamma \gamma$ channel using the
 matrix element method at NLO with the ATLAS experiment at LHC, CERN}}},\
 \href@noop {} {Ph.D. thesis} (\bibinfo {year} {2025})\BibitemShut {NoStop}%
\bibitem [{\citenamefont {Tartarin}\ and\ \citenamefont
 {Stark}(2026)}]{Tartarin:2026uoh}%
 \BibitemOpen
 \bibfield {author} {\bibinfo {author} {\bibfnamefont {M.}~\bibnamefont
 {Tartarin}}\ and\ \bibinfo {author} {\bibfnamefont {J.}~\bibnamefont
 {Stark}},\ }\href {\doibase 10.22323/1.485.0363} {\bibfield {journal}
 {\bibinfo {journal} {PoS}\ }\textbf {\bibinfo {volume} {EPS-HEP2025}},\
 \bibinfo {pages} {363} (\bibinfo {year} {2026})},\ \Eprint
 {http://arxiv.org/abs/2602.02303} {arXiv:2602.02303 [hep-ph]}\BibitemShut
 {NoStop}%
\bibitem [{\citenamefont {Nason}(2004)}]{Nason:2004rx}%
 \BibitemOpen
 \bibfield {author} {\bibinfo {author} {\bibfnamefont {P.}~\bibnamefont
 {Nason}},\ }\href {\doibase 10.1088/1126-6708/2004/11/040} {\bibfield
 {journal} {\bibinfo {journal} {JHEP}\ }\textbf {\bibinfo {volume} {11}},\
 \bibinfo {pages} {040} (\bibinfo {year} {2004})},\ \Eprint
 {http://arxiv.org/abs/hep-ph/0409146} {arXiv:hep-ph/0409146}\BibitemShut
 {NoStop}%
\bibitem [{\citenamefont {Frixione}\ \emph {et~al.}(2007)\citenamefont
 {Frixione}, \citenamefont {Nason},\ and\ \citenamefont
 {Oleari}}]{Frixione:2007vw}%
 \BibitemOpen
 \bibfield {author} {\bibinfo {author} {\bibfnamefont {S.}~\bibnamefont
 {Frixione}}, \bibinfo {author} {\bibfnamefont {P.}~\bibnamefont {Nason}}, \
 and\ \bibinfo {author} {\bibfnamefont {C.}~\bibnamefont {Oleari}},\ }\href
 {\doibase 10.1088/1126-6708/2007/11/070} {\bibfield {journal} {\bibinfo
 {journal} {JHEP}\ }\textbf {\bibinfo {volume} {11}},\ \bibinfo {pages} {070}
 (\bibinfo {year} {2007})},\ \Eprint {http://arxiv.org/abs/0709.2092}
 {arXiv:0709.2092 [hep-ph]}\BibitemShut {NoStop}%
\bibitem [{\citenamefont {Alioli}\ \emph {et~al.}(2010)\citenamefont {Alioli},
 \citenamefont {Nason}, \citenamefont {Oleari},\ and\ \citenamefont
 {Re}}]{Alioli:2010xd}%
 \BibitemOpen
 \bibfield {author} {\bibinfo {author} {\bibfnamefont {S.}~\bibnamefont
 {Alioli}}, \bibinfo {author} {\bibfnamefont {P.}~\bibnamefont {Nason}},
 \bibinfo {author} {\bibfnamefont {C.}~\bibnamefont {Oleari}}, \ and\ \bibinfo
 {author} {\bibfnamefont {E.}~\bibnamefont {Re}},\ }\href {\doibase
 10.1007/JHEP06(2010)043} {\bibfield {journal} {\bibinfo {journal} {JHEP}\
 }\textbf {\bibinfo {volume} {06}},\ \bibinfo {pages} {043} (\bibinfo {year}
 {2010})},\ \Eprint {http://arxiv.org/abs/1002.2581} {arXiv:1002.2581
 [hep-ph]}\BibitemShut {NoStop}%
\bibitem [{\citenamefont {Je\v{z}o}\ and\ \citenamefont
 {Nason}(2015)}]{Jezo:2015aia}%
 \BibitemOpen
 \bibfield {author} {\bibinfo {author} {\bibfnamefont {T.}~\bibnamefont
 {Je\v{z}o}}\ and\ \bibinfo {author} {\bibfnamefont {P.}~\bibnamefont
 {Nason}},\ }\href {\doibase 10.1007/JHEP12(2015)065} {\bibfield {journal}
 {\bibinfo {journal} {JHEP}\ }\textbf {\bibinfo {volume} {12}},\ \bibinfo
 {pages} {065} (\bibinfo {year} {2015})},\ \Eprint
 {http://arxiv.org/abs/1509.09071} {arXiv:1509.09071 [hep-ph]}\BibitemShut
 {NoStop}%
\bibitem [{\citenamefont {Buchm{\"u}ller}\ and\ \citenamefont
 {Wyler}(1986)}]{Buchmuller:1985jz}%
 \BibitemOpen
 \bibfield {author} {\bibinfo {author} {\bibfnamefont {W.}~\bibnamefont
 {Buchm{\"u}ller}}\ and\ \bibinfo {author} {\bibfnamefont {D.}~\bibnamefont
 {Wyler}},\ }\href {\doibase 10.1016/0550-3213(86)90262-2} {\bibfield
 {journal} {\bibinfo {journal} {Nucl. Phys. B}\ }\textbf {\bibinfo {volume}
 {268}},\ \bibinfo {pages} {621} (\bibinfo {year} {1986})}\BibitemShut
 {NoStop}%
\bibitem [{\citenamefont {Grzadkowski}\ \emph {et~al.}(2010)\citenamefont
 {Grzadkowski}, \citenamefont {Iskrzynski}, \citenamefont {Misiak},\ and\
 \citenamefont {Rosiek}}]{Grzadkowski:2010es}%
 \BibitemOpen
 \bibfield {author} {\bibinfo {author} {\bibfnamefont {B.}~\bibnamefont
 {Grzadkowski}}, \bibinfo {author} {\bibfnamefont {M.}~\bibnamefont
 {Iskrzynski}}, \bibinfo {author} {\bibfnamefont {M.}~\bibnamefont {Misiak}},
 \ and\ \bibinfo {author} {\bibfnamefont {J.}~\bibnamefont {Rosiek}},\ }\href
 {\doibase 10.1007/JHEP10(2010)085} {\bibfield {journal} {\bibinfo {journal}
 {JHEP}\ }\textbf {\bibinfo {volume} {10}},\ \bibinfo {pages} {085} (\bibinfo
 {year} {2010})},\ \Eprint {http://arxiv.org/abs/1008.4884} {arXiv:1008.4884
 [hep-ph]}\BibitemShut {NoStop}%
\bibitem [{\citenamefont {Brivio}\ and\ \citenamefont
 {Trott}(2019)}]{Brivio:2017vri}%
 \BibitemOpen
 \bibfield {author} {\bibinfo {author} {\bibfnamefont {I.}~\bibnamefont
 {Brivio}}\ and\ \bibinfo {author} {\bibfnamefont {M.}~\bibnamefont {Trott}},\
 }\href {\doibase 10.1016/j.physrep.2018.11.002} {\bibfield {journal}
 {\bibinfo {journal} {Phys. Rept.}\ }\textbf {\bibinfo {volume} {793}},\
 \bibinfo {pages} {1} (\bibinfo {year} {2019})},\ \Eprint
 {http://arxiv.org/abs/1706.08945} {arXiv:1706.08945 [hep-ph]}\BibitemShut
 {NoStop}%
\bibitem [{\citenamefont {Isidori}\ \emph {et~al.}(2024)\citenamefont
 {Isidori}, \citenamefont {Wilsch},\ and\ \citenamefont
 {Wyler}}]{Isidori:2023pyp}%
 \BibitemOpen
 \bibfield {author} {\bibinfo {author} {\bibfnamefont {G.}~\bibnamefont
 {Isidori}}, \bibinfo {author} {\bibfnamefont {F.}~\bibnamefont {Wilsch}}, \
 and\ \bibinfo {author} {\bibfnamefont {D.}~\bibnamefont {Wyler}},\ }\href
 {\doibase 10.1103/RevModPhys.96.015006} {\bibfield {journal} {\bibinfo
 {journal} {Rev. Mod. Phys.}\ }\textbf {\bibinfo {volume} {96}},\ \bibinfo
 {pages} {015006} (\bibinfo {year} {2024})},\ \Eprint
 {http://arxiv.org/abs/2303.16922} {arXiv:2303.16922 [hep-ph]}\BibitemShut
 {NoStop}%
\bibitem [{\citenamefont {Frixione}\ \emph {et~al.}(1996)\citenamefont
 {Frixione}, \citenamefont {Kunszt},\ and\ \citenamefont
 {Signer}}]{Frixione:1995ms}%
 \BibitemOpen
 \bibfield {author} {\bibinfo {author} {\bibfnamefont {S.}~\bibnamefont
 {Frixione}}, \bibinfo {author} {\bibfnamefont {Z.}~\bibnamefont {Kunszt}}, \
 and\ \bibinfo {author} {\bibfnamefont {A.}~\bibnamefont {Signer}},\ }\href
 {\doibase 10.1016/0550-3213(96)00110-1} {\bibfield {journal} {\bibinfo
 {journal} {Nucl. Phys. B}\ }\textbf {\bibinfo {volume} {467}},\ \bibinfo
 {pages} {399} (\bibinfo {year} {1996})},\ \Eprint
 {http://arxiv.org/abs/hep-ph/9512328} {arXiv:hep-ph/9512328}\BibitemShut
 {NoStop}%
\bibitem [{\citenamefont {Neyman}\ and\ \citenamefont
 {Pearson}(1933)}]{Neyman:1933wgr}%
 \BibitemOpen
 \bibfield {author} {\bibinfo {author} {\bibfnamefont {J.}~\bibnamefont
 {Neyman}}\ and\ \bibinfo {author} {\bibfnamefont {E.~S.}\ \bibnamefont
 {Pearson}},\ }\href {\doibase 10.1098/rsta.1933.0009} {\bibfield {journal}
 {\bibinfo {journal} {Phil. Trans. Roy. Soc. Lond. A}\ }\textbf {\bibinfo
 {volume} {231}},\ \bibinfo {pages} {289} (\bibinfo {year}
 {1933})}\BibitemShut {NoStop}%
\bibitem [{\citenamefont {Barlow}(1987)}]{Barlow:1986ek}%
 \BibitemOpen
 \bibfield {author} {\bibinfo {author} {\bibfnamefont {R.~J.}\ \bibnamefont
 {Barlow}},\ }\href {\doibase 10.1016/0021-9991(87)90078-7} {\bibfield
 {journal} {\bibinfo {journal} {J. Comput. Phys.}\ }\textbf {\bibinfo
 {volume} {72}},\ \bibinfo {pages} {202} (\bibinfo {year} {1987})}\BibitemShut
 {NoStop}%
\bibitem [{\citenamefont {Pivk}\ and\ \citenamefont
 {Le~Diberder}(2005)}]{Pivk:2004ty}%
 \BibitemOpen
 \bibfield {author} {\bibinfo {author} {\bibfnamefont {M.}~\bibnamefont
 {Pivk}}\ and\ \bibinfo {author} {\bibfnamefont {F.~R.}\ \bibnamefont
 {Le~Diberder}},\ }\href {\doibase 10.1016/j.nima.2005.08.106} {\bibfield
 {journal} {\bibinfo {journal} {Nucl. Instrum. Meth. A}\ }\textbf {\bibinfo
 {volume} {555}},\ \bibinfo {pages} {356} (\bibinfo {year} {2005})},\ \Eprint
 {http://arxiv.org/abs/physics/0402083} {arXiv:physics/0402083}\BibitemShut
 {NoStop}%
\bibitem [{\citenamefont {Baldi}\ \emph {et~al.}(2014)\citenamefont {Baldi},
 \citenamefont {Sadowski},\ and\ \citenamefont {Whiteson}}]{Baldi:2014kfa}%
 \BibitemOpen
 \bibfield {author} {\bibinfo {author} {\bibfnamefont {P.}~\bibnamefont
 {Baldi}}, \bibinfo {author} {\bibfnamefont {P.}~\bibnamefont {Sadowski}}, \
 and\ \bibinfo {author} {\bibfnamefont {D.}~\bibnamefont {Whiteson}},\ }\href
 {\doibase 10.1038/ncomms5308} {\bibfield {journal} {\bibinfo {journal}
 {Nature Commun.}\ }\textbf {\bibinfo {volume} {5}},\ \bibinfo {pages} {4308}
 (\bibinfo {year} {2014})},\ \Eprint {http://arxiv.org/abs/1402.4735}
 {arXiv:1402.4735 [hep-ph]}\BibitemShut {NoStop}%
\bibitem [{\citenamefont {Cranmer}\ \emph {et~al.}(2015)\citenamefont
 {Cranmer}, \citenamefont {Pavez},\ and\ \citenamefont
 {Louppe}}]{Cranmer:2015bka}%
 \BibitemOpen
 \bibfield {author} {\bibinfo {author} {\bibfnamefont {K.}~\bibnamefont
 {Cranmer}}, \bibinfo {author} {\bibfnamefont {J.}~\bibnamefont {Pavez}}, \
 and\ \bibinfo {author} {\bibfnamefont {G.}~\bibnamefont {Louppe}},\
 }\href@noop {} {\ (\bibinfo {year} {2015})},\ \Eprint
 {http://arxiv.org/abs/1506.02169} {arXiv:1506.02169 [stat.AP]}\BibitemShut
 {NoStop}%
\bibitem [{\citenamefont {Brehmer}\ \emph
 {et~al.}(2018{\natexlab{a}})\citenamefont {Brehmer}, \citenamefont {Cranmer},
 \citenamefont {Louppe},\ and\ \citenamefont {Pavez}}]{Brehmer:2018kdj}%
 \BibitemOpen
 \bibfield {author} {\bibinfo {author} {\bibfnamefont {J.}~\bibnamefont
 {Brehmer}}, \bibinfo {author} {\bibfnamefont {K.}~\bibnamefont {Cranmer}},
 \bibinfo {author} {\bibfnamefont {G.}~\bibnamefont {Louppe}}, \ and\ \bibinfo
 {author} {\bibfnamefont {J.}~\bibnamefont {Pavez}},\ }\href {\doibase
 10.1103/PhysRevLett.121.111801} {\bibfield {journal} {\bibinfo {journal}
 {Phys. Rev. Lett.}\ }\textbf {\bibinfo {volume} {121}},\ \bibinfo {pages}
 {111801} (\bibinfo {year} {2018}{\natexlab{a}})},\ \Eprint
 {http://arxiv.org/abs/1805.00013} {arXiv:1805.00013 [hep-ph]}\BibitemShut
 {NoStop}%
\bibitem [{\citenamefont {Brehmer}\ \emph
 {et~al.}(2018{\natexlab{b}})\citenamefont {Brehmer}, \citenamefont {Cranmer},
 \citenamefont {Louppe},\ and\ \citenamefont {Pavez}}]{Brehmer:2018eca}%
 \BibitemOpen
 \bibfield {author} {\bibinfo {author} {\bibfnamefont {J.}~\bibnamefont
 {Brehmer}}, \bibinfo {author} {\bibfnamefont {K.}~\bibnamefont {Cranmer}},
 \bibinfo {author} {\bibfnamefont {G.}~\bibnamefont {Louppe}}, \ and\ \bibinfo
 {author} {\bibfnamefont {J.}~\bibnamefont {Pavez}},\ }\href {\doibase
 10.1103/PhysRevD.98.052004} {\bibfield {journal} {\bibinfo {journal} {Phys.
 Rev. D}\ }\textbf {\bibinfo {volume} {98}},\ \bibinfo {pages} {052004}
 (\bibinfo {year} {2018}{\natexlab{b}})},\ \Eprint
 {http://arxiv.org/abs/1805.00020} {arXiv:1805.00020 [hep-ph]}\BibitemShut
 {NoStop}%
\bibitem [{\citenamefont {Brehmer}\ \emph
 {et~al.}(2020{\natexlab{a}})\citenamefont {Brehmer}, \citenamefont {Louppe},
 \citenamefont {Pavez},\ and\ \citenamefont {Cranmer}}]{Brehmer:2018hga}%
 \BibitemOpen
 \bibfield {author} {\bibinfo {author} {\bibfnamefont {J.}~\bibnamefont
 {Brehmer}}, \bibinfo {author} {\bibfnamefont {G.}~\bibnamefont {Louppe}},
 \bibinfo {author} {\bibfnamefont {J.}~\bibnamefont {Pavez}}, \ and\ \bibinfo
 {author} {\bibfnamefont {K.}~\bibnamefont {Cranmer}},\ }\href {\doibase
 10.1073/pnas.1915980117} {\bibfield {journal} {\bibinfo {journal} {Proc.
 Nat. Acad. Sci.}\ }\textbf {\bibinfo {volume} {117}},\ \bibinfo {pages}
 {5242} (\bibinfo {year} {2020}{\natexlab{a}})},\ \Eprint
 {http://arxiv.org/abs/1805.12244} {arXiv:1805.12244 [stat.ML]}\BibitemShut
 {NoStop}%
\bibitem [{\citenamefont {Brehmer}\ \emph
 {et~al.}(2020{\natexlab{b}})\citenamefont {Brehmer}, \citenamefont {Kling},
 \citenamefont {Espejo},\ and\ \citenamefont {Cranmer}}]{Brehmer:2019xox}%
 \BibitemOpen
 \bibfield {author} {\bibinfo {author} {\bibfnamefont {J.}~\bibnamefont
 {Brehmer}}, \bibinfo {author} {\bibfnamefont {F.}~\bibnamefont {Kling}},
 \bibinfo {author} {\bibfnamefont {I.}~\bibnamefont {Espejo}}, \ and\ \bibinfo
 {author} {\bibfnamefont {K.}~\bibnamefont {Cranmer}},\ }\href {\doibase
 10.1007/s41781-020-0035-2} {\bibfield {journal} {\bibinfo {journal}
 {Comput. Softw. Big Sci.}\ }\textbf {\bibinfo {volume} {4}},\ \bibinfo
 {pages} {3} (\bibinfo {year} {2020}{\natexlab{b}})},\ \Eprint
 {http://arxiv.org/abs/1907.10621} {arXiv:1907.10621 [hep-ph]}\BibitemShut
 {NoStop}%
\bibitem [{\citenamefont {Cranmer}\ \emph {et~al.}(2020)\citenamefont
 {Cranmer}, \citenamefont {Brehmer},\ and\ \citenamefont
 {Louppe}}]{Cranmer_2020}%
 \BibitemOpen
 \bibfield {author} {\bibinfo {author} {\bibfnamefont {K.}~\bibnamefont
 {Cranmer}}, \bibinfo {author} {\bibfnamefont {J.}~\bibnamefont {Brehmer}}, \
 and\ \bibinfo {author} {\bibfnamefont {G.}~\bibnamefont {Louppe}},\ }\href
 {\doibase 10.1073/pnas.1912789117} {\bibfield {journal} {\bibinfo {journal}
 {Proceedings of the National Academy of Sciences}\ }\textbf {\bibinfo
 {volume} {117}},\ \bibinfo {pages} {30055–30062} (\bibinfo {year}
 {2020})}\BibitemShut {NoStop}%
\bibitem [{\citenamefont {van Beekveld}\ \emph {et~al.}(2025)\citenamefont {van
 Beekveld}, \citenamefont {Ferrario~Ravasio}, \citenamefont {Helliwell},
 \citenamefont {Karlberg}, \citenamefont {Salam}, \citenamefont {Scyboz},
 \citenamefont {Soto-Ontoso}, \citenamefont {Soyez},\ and\ \citenamefont
 {Zanoli}}]{vanBeekveld:2025lpz}%
 \BibitemOpen
 \bibfield {author} {\bibinfo {author} {\bibfnamefont {M.}~\bibnamefont {van
 Beekveld}}, \bibinfo {author} {\bibfnamefont {S.}~\bibnamefont
 {Ferrario~Ravasio}}, \bibinfo {author} {\bibfnamefont {J.}~\bibnamefont
 {Helliwell}}, \bibinfo {author} {\bibfnamefont {A.}~\bibnamefont {Karlberg}},
 \bibinfo {author} {\bibfnamefont {G.~P.}\ \bibnamefont {Salam}}, \bibinfo
 {author} {\bibfnamefont {L.}~\bibnamefont {Scyboz}}, \bibinfo {author}
 {\bibfnamefont {A.}~\bibnamefont {Soto-Ontoso}}, \bibinfo {author}
 {\bibfnamefont {G.}~\bibnamefont {Soyez}}, \ and\ \bibinfo {author}
 {\bibfnamefont {S.}~\bibnamefont {Zanoli}},\ }\href {\doibase
 10.1007/JHEP10(2025)038} {\bibfield {journal} {\bibinfo {journal} {JHEP}\
 }\textbf {\bibinfo {volume} {10}},\ \bibinfo {pages} {038} (\bibinfo {year}
 {2025})},\ \Eprint {http://arxiv.org/abs/2504.05377} {arXiv:2504.05377
 [hep-ph]}\BibitemShut {NoStop}%
\bibitem [{\citenamefont {Nason}(2007)}]{Nason:2007vt}%
 \BibitemOpen
 \bibfield {author} {\bibinfo {author} {\bibfnamefont {P.}~\bibnamefont
 {Nason}},\ }\href@noop {} {\ (\bibinfo {year} {2007})},\ \Eprint
 {http://arxiv.org/abs/0709.2085} {arXiv:0709.2085 [hep-ph]}\BibitemShut
 {NoStop}%
\bibitem [{\citenamefont {Jadach}\ \emph {et~al.}(2015)\citenamefont {Jadach},
 \citenamefont {P{\l}aczek}, \citenamefont {Sapeta}, \citenamefont
 {Si{\'o}dmok},\ and\ \citenamefont {Skrzypek}}]{Jadach:2015mza}%
 \BibitemOpen
 \bibfield {author} {\bibinfo {author} {\bibfnamefont {S.}~\bibnamefont
 {Jadach}}, \bibinfo {author} {\bibfnamefont {W.}~\bibnamefont {P{\l}aczek}},
 \bibinfo {author} {\bibfnamefont {S.}~\bibnamefont {Sapeta}}, \bibinfo
 {author} {\bibfnamefont {A.}~\bibnamefont {Si{\'o}dmok}}, \ and\ \bibinfo
 {author} {\bibfnamefont {M.}~\bibnamefont {Skrzypek}},\ }\href {\doibase
 10.1007/JHEP10(2015)052} {\bibfield {journal} {\bibinfo {journal} {JHEP}\
 }\textbf {\bibinfo {volume} {10}},\ \bibinfo {pages} {052} (\bibinfo {year}
 {2015})},\ \Eprint {http://arxiv.org/abs/1503.06849} {arXiv:1503.06849
 [hep-ph]}\BibitemShut {NoStop}%
\bibitem [{\citenamefont {Jadach}\ \emph {et~al.}(2017)\citenamefont {Jadach},
 \citenamefont {Nail}, \citenamefont {P{\l}aczek}, \citenamefont {Sapeta},
 \citenamefont {Siodmok},\ and\ \citenamefont {Skrzypek}}]{Jadach:2016qti}%
 \BibitemOpen
 \bibfield {author} {\bibinfo {author} {\bibfnamefont {S.}~\bibnamefont
 {Jadach}}, \bibinfo {author} {\bibfnamefont {G.}~\bibnamefont {Nail}},
 \bibinfo {author} {\bibfnamefont {W.}~\bibnamefont {P{\l}aczek}}, \bibinfo
 {author} {\bibfnamefont {S.}~\bibnamefont {Sapeta}}, \bibinfo {author}
 {\bibfnamefont {A.}~\bibnamefont {Siodmok}}, \ and\ \bibinfo {author}
 {\bibfnamefont {M.}~\bibnamefont {Skrzypek}},\ }\href {\doibase
 10.1140/epjc/s10052-017-4733-9} {\bibfield {journal} {\bibinfo {journal}
 {Eur. Phys. J. C}\ }\textbf {\bibinfo {volume} {77}},\ \bibinfo {pages} {164}
 (\bibinfo {year} {2017})},\ \Eprint {http://arxiv.org/abs/1607.06799}
 {arXiv:1607.06799 [hep-ph]}\BibitemShut {NoStop}%
\bibitem [{\citenamefont {Frederix}\ \emph {et~al.}(2020)\citenamefont
 {Frederix}, \citenamefont {Frixione}, \citenamefont {Prestel},\ and\
 \citenamefont {Torrielli}}]{Frederix:2020trv}%
 \BibitemOpen
 \bibfield {author} {\bibinfo {author} {\bibfnamefont {R.}~\bibnamefont
 {Frederix}}, \bibinfo {author} {\bibfnamefont {S.}~\bibnamefont {Frixione}},
 \bibinfo {author} {\bibfnamefont {S.}~\bibnamefont {Prestel}}, \ and\
 \bibinfo {author} {\bibfnamefont {P.}~\bibnamefont {Torrielli}},\ }\href
 {\doibase 10.1007/JHEP07(2020)238} {\bibfield {journal} {\bibinfo {journal}
 {JHEP}\ }\textbf {\bibinfo {volume} {07}},\ \bibinfo {pages} {238} (\bibinfo
 {year} {2020})},\ \Eprint {http://arxiv.org/abs/2002.12716} {arXiv:2002.12716
 [hep-ph]}\BibitemShut {NoStop}%
\bibitem [{\citenamefont {Nachman}\ and\ \citenamefont
 {Thaler}(2020)}]{Nachman:2020fff}%
 \BibitemOpen
 \bibfield {author} {\bibinfo {author} {\bibfnamefont {B.}~\bibnamefont
 {Nachman}}\ and\ \bibinfo {author} {\bibfnamefont {J.}~\bibnamefont
 {Thaler}},\ }\href {\doibase 10.1103/PhysRevD.102.076004} {\bibfield
 {journal} {\bibinfo {journal} {Phys. Rev. D}\ }\textbf {\bibinfo {volume}
 {102}},\ \bibinfo {pages} {076004} (\bibinfo {year} {2020})},\ \Eprint
 {http://arxiv.org/abs/2007.11586} {arXiv:2007.11586 [hep-ph]}\BibitemShut
 {NoStop}%
\bibitem [{\citenamefont {Andersen}\ and\ \citenamefont
 {Maier}(2022)}]{Andersen:2021mvw}%
 \BibitemOpen
 \bibfield {author} {\bibinfo {author} {\bibfnamefont {J.~R.}\ \bibnamefont
 {Andersen}}\ and\ \bibinfo {author} {\bibfnamefont {A.}~\bibnamefont
 {Maier}},\ }\href {\doibase 10.1140/epjc/s10052-022-10372-3} {\bibfield
 {journal} {\bibinfo {journal} {Eur. Phys. J. C}\ }\textbf {\bibinfo {volume}
 {82}},\ \bibinfo {pages} {433} (\bibinfo {year} {2022})},\ \Eprint
 {http://arxiv.org/abs/2109.07851} {arXiv:2109.07851 [hep-ph]}\BibitemShut
 {NoStop}%
\bibitem [{\citenamefont {Danziger}\ \emph {et~al.}(2021)\citenamefont
 {Danziger}, \citenamefont {H{\"o}che},\ and\ \citenamefont
 {Siegert}}]{Danziger:2021xvr}%
 \BibitemOpen
 \bibfield {author} {\bibinfo {author} {\bibfnamefont {K.}~\bibnamefont
 {Danziger}}, \bibinfo {author} {\bibfnamefont {S.}~\bibnamefont {H{\"o}che}},
 \ and\ \bibinfo {author} {\bibfnamefont {F.}~\bibnamefont {Siegert}},\
 }\href@noop {} {\ (\bibinfo {year} {2021})},\ \Eprint
 {http://arxiv.org/abs/2110.15211} {arXiv:2110.15211 [hep-ph]}\BibitemShut
 {NoStop}%
\bibitem [{\citenamefont {Nason}\ and\ \citenamefont
 {Salam}(2022)}]{Nason:2021xke}%
 \BibitemOpen
 \bibfield {author} {\bibinfo {author} {\bibfnamefont {P.}~\bibnamefont
 {Nason}}\ and\ \bibinfo {author} {\bibfnamefont {G.~P.}\ \bibnamefont
 {Salam}},\ }\href {\doibase 10.1007/JHEP01(2022)067} {\bibfield {journal}
 {\bibinfo {journal} {JHEP}\ }\textbf {\bibinfo {volume} {01}},\ \bibinfo
 {pages} {067} (\bibinfo {year} {2022})},\ \Eprint
 {http://arxiv.org/abs/2111.03553} {arXiv:2111.03553 [hep-ph]}\BibitemShut
 {NoStop}%
\bibitem [{\citenamefont {Andersen}\ \emph {et~al.}(2023)\citenamefont
 {Andersen}, \citenamefont {Maier},\ and\ \citenamefont
 {Ma{\^\i}tre}}]{Andersen:2023cku}%
 \BibitemOpen
 \bibfield {author} {\bibinfo {author} {\bibfnamefont {J.~R.}\ \bibnamefont
 {Andersen}}, \bibinfo {author} {\bibfnamefont {A.}~\bibnamefont {Maier}}, \
 and\ \bibinfo {author} {\bibfnamefont {D.}~\bibnamefont {Ma{\^\i}tre}},\
 }\href {\doibase 10.1140/epjc/s10052-023-11905-0} {\bibfield {journal}
 {\bibinfo {journal} {Eur. Phys. J. C}\ }\textbf {\bibinfo {volume} {83}},\
 \bibinfo {pages} {835} (\bibinfo {year} {2023})},\ \Eprint
 {http://arxiv.org/abs/2303.15246} {arXiv:2303.15246 [hep-ph]}\BibitemShut
 {NoStop}%
\bibitem [{\citenamefont {Frederix}\ and\ \citenamefont
 {Torrielli}(2023)}]{Frederix:2023hom}%
 \BibitemOpen
 \bibfield {author} {\bibinfo {author} {\bibfnamefont {R.}~\bibnamefont
 {Frederix}}\ and\ \bibinfo {author} {\bibfnamefont {P.}~\bibnamefont
 {Torrielli}},\ }\href {\doibase 10.1140/epjc/s10052-023-12243-x} {\bibfield
 {journal} {\bibinfo {journal} {Eur. Phys. J. C}\ }\textbf {\bibinfo {volume}
 {83}},\ \bibinfo {pages} {1051} (\bibinfo {year} {2023})},\ \Eprint
 {http://arxiv.org/abs/2310.04160} {arXiv:2310.04160 [hep-ph]}\BibitemShut
 {NoStop}%
\bibitem [{\citenamefont {Sarmah}\ \emph
 {et~al.}(2025{\natexlab{a}})\citenamefont {Sarmah}, \citenamefont
 {Si{\'o}dmok},\ and\ \citenamefont {Whitehead}}]{Sarmah:2024hdk}%
 \BibitemOpen
 \bibfield {author} {\bibinfo {author} {\bibfnamefont {P.}~\bibnamefont
 {Sarmah}}, \bibinfo {author} {\bibfnamefont {A.}~\bibnamefont {Si{\'o}dmok}},
 \ and\ \bibinfo {author} {\bibfnamefont {J.}~\bibnamefont {Whitehead}},\
 }\href {\doibase 10.1007/JHEP01(2025)062} {\bibfield {journal} {\bibinfo
 {journal} {JHEP}\ }\textbf {\bibinfo {volume} {01}},\ \bibinfo {pages} {062}
 (\bibinfo {year} {2025}{\natexlab{a}})},\ \Eprint
 {http://arxiv.org/abs/2409.16417} {arXiv:2409.16417 [hep-ph]}\BibitemShut
 {NoStop}%
\bibitem [{\citenamefont {Andersen}\ \emph {et~al.}(2024)\citenamefont
 {Andersen}, \citenamefont {Cueto}, \citenamefont {Jones},\ and\ \citenamefont
 {Maier}}]{Andersen:2024mqh}%
 \BibitemOpen
 \bibfield {author} {\bibinfo {author} {\bibfnamefont {J.~R.}\ \bibnamefont
 {Andersen}}, \bibinfo {author} {\bibfnamefont {A.}~\bibnamefont {Cueto}},
 \bibinfo {author} {\bibfnamefont {S.~P.}\ \bibnamefont {Jones}}, \ and\
 \bibinfo {author} {\bibfnamefont {A.}~\bibnamefont {Maier}},\ }\href@noop {}
 {\ (\bibinfo {year} {2024})},\ \Eprint {http://arxiv.org/abs/2411.11651}
 {arXiv:2411.11651 [hep-ph]}\BibitemShut {NoStop}%
\bibitem [{\citenamefont
 {Shyamsundar}(2025{\natexlab{a}})}]{Shyamsundar:2025nzn}%
 \BibitemOpen
 \bibfield {author} {\bibinfo {author} {\bibfnamefont {P.}~\bibnamefont
 {Shyamsundar}},\ }\href@noop {} {\ (\bibinfo {year} {2025}{\natexlab{a}})},\
 \Eprint {http://arxiv.org/abs/2502.08052} {arXiv:2502.08052 [hep-ph]}\BibitemShut {NoStop}%
\bibitem [{\citenamefont
 {Shyamsundar}(2025{\natexlab{b}})}]{Shyamsundar:2025mfw}%
 \BibitemOpen
 \bibfield {author} {\bibinfo {author} {\bibfnamefont {P.}~\bibnamefont
 {Shyamsundar}},\ }\href@noop {} {\ (\bibinfo {year} {2025}{\natexlab{b}})},\
 \Eprint {http://arxiv.org/abs/2502.08053} {arXiv:2502.08053 [hep-ph]}\BibitemShut {NoStop}%
\bibitem [{\citenamefont {Nachman}\ and\ \citenamefont
 {Noll}(2025)}]{Nachman:2025lid}%
 \BibitemOpen
 \bibfield {author} {\bibinfo {author} {\bibfnamefont {B.}~\bibnamefont
 {Nachman}}\ and\ \bibinfo {author} {\bibfnamefont {D.}~\bibnamefont {Noll}},\
 }\href {\doibase 10.1103/yx16-h7n9} {\bibfield {journal} {\bibinfo
 {journal} {Phys. Rev. D}\ }\textbf {\bibinfo {volume} {112}},\ \bibinfo
 {pages} {096009} (\bibinfo {year} {2025})},\ \Eprint
 {http://arxiv.org/abs/2505.03724} {arXiv:2505.03724 [hep-ph]}\BibitemShut
 {NoStop}%
\bibitem [{\citenamefont {Palmer}\ and\ \citenamefont
 {Kronheim}(2026)}]{Palmer:2025jmb}%
 \BibitemOpen
 \bibfield {author} {\bibinfo {author} {\bibfnamefont {C.}~\bibnamefont
 {Palmer}}\ and\ \bibinfo {author} {\bibfnamefont {B.}~\bibnamefont
 {Kronheim}},\ }\href {\doibase 10.1103/k8w6-wn37} {\bibfield {journal}
 {\bibinfo {journal} {Phys. Rev. D}\ }\textbf {\bibinfo {volume} {113}},\
 \bibinfo {pages} {012003} (\bibinfo {year} {2026})},\ \Eprint
 {http://arxiv.org/abs/2510.16217} {arXiv:2510.16217 [hep-ex]}\BibitemShut
 {NoStop}%
\bibitem [{\citenamefont {Sarmah}\ \emph
 {et~al.}(2025{\natexlab{b}})\citenamefont {Sarmah}, \citenamefont
 {Si{\'o}dmok},\ and\ \citenamefont {Whitehead}}]{Sarmah:2025vnb}%
 \BibitemOpen
 \bibfield {author} {\bibinfo {author} {\bibfnamefont {P.}~\bibnamefont
 {Sarmah}}, \bibinfo {author} {\bibfnamefont {A.}~\bibnamefont {Si{\'o}dmok}},
 \ and\ \bibinfo {author} {\bibfnamefont {J.}~\bibnamefont {Whitehead}},\
 }\href@noop {} {\ (\bibinfo {year} {2025}{\natexlab{b}})},\ \Eprint
 {http://arxiv.org/abs/2511.16605} {arXiv:2511.16605 [hep-ph]}\BibitemShut
 {NoStop}%
\bibitem [{\citenamefont {Gambhir}\ and\ \citenamefont
 {Mastandrea}(2025)}]{Gambhir:2025lka}%
 \BibitemOpen
 \bibfield {author} {\bibinfo {author} {\bibfnamefont {R.}~\bibnamefont
 {Gambhir}}\ and\ \bibinfo {author} {\bibfnamefont {R.}~\bibnamefont
 {Mastandrea}},\ }\href@noop {} {\ (\bibinfo {year} {2025})},\ \Eprint
 {http://arxiv.org/abs/2512.04160} {arXiv:2512.04160 [hep-ph]}\BibitemShut
 {NoStop}%
\bibitem [{\citenamefont {Mazzitelli}\ \emph {et~al.}(2021)\citenamefont
 {Mazzitelli}, \citenamefont {Monni}, \citenamefont {Nason}, \citenamefont
 {Re}, \citenamefont {Wiesemann},\ and\ \citenamefont
 {Zanderighi}}]{Mazzitelli:2020jio}%
 \BibitemOpen
 \bibfield {author} {\bibinfo {author} {\bibfnamefont {J.}~\bibnamefont
 {Mazzitelli}}, \bibinfo {author} {\bibfnamefont {P.~F.}\ \bibnamefont
 {Monni}}, \bibinfo {author} {\bibfnamefont {P.}~\bibnamefont {Nason}},
 \bibinfo {author} {\bibfnamefont {E.}~\bibnamefont {Re}}, \bibinfo {author}
 {\bibfnamefont {M.}~\bibnamefont {Wiesemann}}, \ and\ \bibinfo {author}
 {\bibfnamefont {G.}~\bibnamefont {Zanderighi}},\ }\href {\doibase
 10.1103/PhysRevLett.127.062001} {\bibfield {journal} {\bibinfo {journal}
 {Phys. Rev. Lett.}\ }\textbf {\bibinfo {volume} {127}},\ \bibinfo {pages}
 {062001} (\bibinfo {year} {2021})},\ \Eprint
 {http://arxiv.org/abs/2012.14267} {arXiv:2012.14267 [hep-ph]}\BibitemShut
 {NoStop}%
\bibitem [{\citenamefont {Mazzitelli}\ \emph {et~al.}(2022)\citenamefont
 {Mazzitelli}, \citenamefont {Monni}, \citenamefont {Nason}, \citenamefont
 {Re}, \citenamefont {Wiesemann},\ and\ \citenamefont
 {Zanderighi}}]{Mazzitelli:2021mmm}%
 \BibitemOpen
 \bibfield {author} {\bibinfo {author} {\bibfnamefont {J.}~\bibnamefont
 {Mazzitelli}}, \bibinfo {author} {\bibfnamefont {P.~F.}\ \bibnamefont
 {Monni}}, \bibinfo {author} {\bibfnamefont {P.}~\bibnamefont {Nason}},
 \bibinfo {author} {\bibfnamefont {E.}~\bibnamefont {Re}}, \bibinfo {author}
 {\bibfnamefont {M.}~\bibnamefont {Wiesemann}}, \ and\ \bibinfo {author}
 {\bibfnamefont {G.}~\bibnamefont {Zanderighi}},\ }\href {\doibase
 10.1007/JHEP04(2022)079} {\bibfield {journal} {\bibinfo {journal} {JHEP}\
 }\textbf {\bibinfo {volume} {04}},\ \bibinfo {pages} {079} (\bibinfo {year}
 {2022})},\ \Eprint {http://arxiv.org/abs/2112.12135} {arXiv:2112.12135
 [hep-ph]}\BibitemShut {NoStop}%
\bibitem [{\citenamefont {Alwall}\ \emph {et~al.}(2007)\citenamefont {Alwall}
 \emph {et~al.}}]{Alwall:2006yp}%
 \BibitemOpen
 \bibfield {author} {\bibinfo {author} {\bibfnamefont {J.}~\bibnamefont
 {Alwall}} \emph {et~al.},\ }\href {\doibase 10.1016/j.cpc.2006.11.010}
 {\bibfield {journal} {\bibinfo {journal} {Comput. Phys. Commun.}\ }\textbf
 {\bibinfo {volume} {176}},\ \bibinfo {pages} {300} (\bibinfo {year}
 {2007})},\ \Eprint {http://arxiv.org/abs/hep-ph/0609017}
 {arXiv:hep-ph/0609017}\BibitemShut {NoStop}%
\bibitem [{\citenamefont {Melia}\ \emph {et~al.}(2011)\citenamefont {Melia},
 \citenamefont {Nason}, \citenamefont {R{\"o}ntsch},\ and\ \citenamefont
 {Zanderighi}}]{Melia:2011tj}%
 \BibitemOpen
 \bibfield {author} {\bibinfo {author} {\bibfnamefont {T.}~\bibnamefont
 {Melia}}, \bibinfo {author} {\bibfnamefont {P.}~\bibnamefont {Nason}},
 \bibinfo {author} {\bibfnamefont {R.}~\bibnamefont {R{\"o}ntsch}}, \ and\
 \bibinfo {author} {\bibfnamefont {G.}~\bibnamefont {Zanderighi}},\ }\href
 {\doibase 10.1007/JHEP11(2011)078} {\bibfield {journal} {\bibinfo {journal}
 {JHEP}\ }\textbf {\bibinfo {volume} {11}},\ \bibinfo {pages} {078} (\bibinfo
 {year} {2011})},\ \Eprint {http://arxiv.org/abs/1107.5051} {arXiv:1107.5051
 [hep-ph]}\BibitemShut {NoStop}%
\bibitem [{\citenamefont {Degrande}\ \emph {et~al.}(2013)\citenamefont
 {Degrande}, \citenamefont {Greiner}, \citenamefont {Kilian}, \citenamefont
 {Mattelaer}, \citenamefont {Mebane}, \citenamefont {Stelzer}, \citenamefont
 {Willenbrock},\ and\ \citenamefont {Zhang}}]{Degrande:2012wf}%
 \BibitemOpen
 \bibfield {author} {\bibinfo {author} {\bibfnamefont {C.}~\bibnamefont
 {Degrande}}, \bibinfo {author} {\bibfnamefont {N.}~\bibnamefont {Greiner}},
 \bibinfo {author} {\bibfnamefont {W.}~\bibnamefont {Kilian}}, \bibinfo
 {author} {\bibfnamefont {O.}~\bibnamefont {Mattelaer}}, \bibinfo {author}
 {\bibfnamefont {H.}~\bibnamefont {Mebane}}, \bibinfo {author} {\bibfnamefont
 {T.}~\bibnamefont {Stelzer}}, \bibinfo {author} {\bibfnamefont
 {S.}~\bibnamefont {Willenbrock}}, \ and\ \bibinfo {author} {\bibfnamefont
 {C.}~\bibnamefont {Zhang}},\ }\href {\doibase 10.1016/j.aop.2013.04.016}
 {\bibfield {journal} {\bibinfo {journal} {Annals Phys.}\ }\textbf {\bibinfo
 {volume} {335}},\ \bibinfo {pages} {21} (\bibinfo {year} {2013})},\ \Eprint
 {http://arxiv.org/abs/1205.4231} {arXiv:1205.4231 [hep-ph]}\BibitemShut
 {NoStop}%
\bibitem [{\citenamefont {Falkowski}\ \emph {et~al.}(2016)\citenamefont
 {Falkowski}, \citenamefont {Gonzalez-Alonso}, \citenamefont {Greljo},\ and\
 \citenamefont {Marzocca}}]{Falkowski:2015jaa}%
 \BibitemOpen
 \bibfield {author} {\bibinfo {author} {\bibfnamefont {A.}~\bibnamefont
 {Falkowski}}, \bibinfo {author} {\bibfnamefont {M.}~\bibnamefont
 {Gonzalez-Alonso}}, \bibinfo {author} {\bibfnamefont {A.}~\bibnamefont
 {Greljo}}, \ and\ \bibinfo {author} {\bibfnamefont {D.}~\bibnamefont
 {Marzocca}},\ }\href {\doibase 10.1103/PhysRevLett.116.011801} {\bibfield
 {journal} {\bibinfo {journal} {Phys. Rev. Lett.}\ }\textbf {\bibinfo
 {volume} {116}},\ \bibinfo {pages} {011801} (\bibinfo {year} {2016})},\
 \Eprint {http://arxiv.org/abs/1508.00581} {arXiv:1508.00581 [hep-ph]}\BibitemShut {NoStop}%
\bibitem [{\citenamefont {Falkowski}\ \emph {et~al.}(2017)\citenamefont
 {Falkowski}, \citenamefont {Gonzalez-Alonso}, \citenamefont {Greljo},
 \citenamefont {Marzocca},\ and\ \citenamefont {Son}}]{Falkowski:2016cxu}%
 \BibitemOpen
 \bibfield {author} {\bibinfo {author} {\bibfnamefont {A.}~\bibnamefont
 {Falkowski}}, \bibinfo {author} {\bibfnamefont {M.}~\bibnamefont
 {Gonzalez-Alonso}}, \bibinfo {author} {\bibfnamefont {A.}~\bibnamefont
 {Greljo}}, \bibinfo {author} {\bibfnamefont {D.}~\bibnamefont {Marzocca}}, \
 and\ \bibinfo {author} {\bibfnamefont {M.}~\bibnamefont {Son}},\ }\href
 {\doibase 10.1007/JHEP02(2017)115} {\bibfield {journal} {\bibinfo {journal}
 {JHEP}\ }\textbf {\bibinfo {volume} {02}},\ \bibinfo {pages} {115} (\bibinfo
 {year} {2017})},\ \Eprint {http://arxiv.org/abs/1609.06312} {arXiv:1609.06312
 [hep-ph]}\BibitemShut {NoStop}%
\bibitem [{\citenamefont {Helset}\ and\ \citenamefont
 {Trott}(2018)}]{Helset:2017mlf}%
 \BibitemOpen
 \bibfield {author} {\bibinfo {author} {\bibfnamefont {A.}~\bibnamefont
 {Helset}}\ and\ \bibinfo {author} {\bibfnamefont {M.}~\bibnamefont {Trott}},\
 }\href {\doibase 10.1007/JHEP04(2018)038} {\bibfield {journal} {\bibinfo
 {journal} {JHEP}\ }\textbf {\bibinfo {volume} {04}},\ \bibinfo {pages} {038}
 (\bibinfo {year} {2018})},\ \Eprint {http://arxiv.org/abs/1711.07954}
 {arXiv:1711.07954 [hep-ph]}\BibitemShut {NoStop}%
\bibitem [{\citenamefont {Baglio}\ \emph {et~al.}(2017)\citenamefont {Baglio},
 \citenamefont {Dawson},\ and\ \citenamefont {Lewis}}]{Baglio:2017bfe}%
 \BibitemOpen
 \bibfield {author} {\bibinfo {author} {\bibfnamefont {J.}~\bibnamefont
 {Baglio}}, \bibinfo {author} {\bibfnamefont {S.}~\bibnamefont {Dawson}}, \
 and\ \bibinfo {author} {\bibfnamefont {I.~M.}\ \bibnamefont {Lewis}},\ }\href
 {\doibase 10.1103/PhysRevD.96.073003} {\bibfield {journal} {\bibinfo
 {journal} {Phys. Rev. D}\ }\textbf {\bibinfo {volume} {96}},\ \bibinfo
 {pages} {073003} (\bibinfo {year} {2017})},\ \Eprint
 {http://arxiv.org/abs/1708.03332} {arXiv:1708.03332 [hep-ph]}\BibitemShut
 {NoStop}%
\bibitem [{\citenamefont {Azatov}\ \emph
 {et~al.}(2017{\natexlab{a}})\citenamefont {Azatov}, \citenamefont
 {Elias-Miro}, \citenamefont {Reyimuaji},\ and\ \citenamefont
 {Venturini}}]{Azatov:2017kzw}%
 \BibitemOpen
 \bibfield {author} {\bibinfo {author} {\bibfnamefont {A.}~\bibnamefont
 {Azatov}}, \bibinfo {author} {\bibfnamefont {J.}~\bibnamefont {Elias-Miro}},
 \bibinfo {author} {\bibfnamefont {Y.}~\bibnamefont {Reyimuaji}}, \ and\
 \bibinfo {author} {\bibfnamefont {E.}~\bibnamefont {Venturini}},\ }\href
 {\doibase 10.1007/JHEP10(2017)027} {\bibfield {journal} {\bibinfo {journal}
 {JHEP}\ }\textbf {\bibinfo {volume} {10}},\ \bibinfo {pages} {027} (\bibinfo
 {year} {2017}{\natexlab{a}})},\ \Eprint {http://arxiv.org/abs/1707.08060}
 {arXiv:1707.08060 [hep-ph]}\BibitemShut {NoStop}%
\bibitem [{\citenamefont {Panico}\ \emph {et~al.}(2018)\citenamefont {Panico},
 \citenamefont {Riva},\ and\ \citenamefont {Wulzer}}]{Panico:2017frx}%
 \BibitemOpen
 \bibfield {author} {\bibinfo {author} {\bibfnamefont {G.}~\bibnamefont
 {Panico}}, \bibinfo {author} {\bibfnamefont {F.}~\bibnamefont {Riva}}, \ and\
 \bibinfo {author} {\bibfnamefont {A.}~\bibnamefont {Wulzer}},\ }\href
 {\doibase 10.1016/j.physletb.2017.11.068} {\bibfield {journal} {\bibinfo
 {journal} {Phys. Lett. B}\ }\textbf {\bibinfo {volume} {776}},\ \bibinfo
 {pages} {473} (\bibinfo {year} {2018})},\ \Eprint
 {http://arxiv.org/abs/1708.07823} {arXiv:1708.07823 [hep-ph]}\BibitemShut
 {NoStop}%
\bibitem [{\citenamefont {Franceschini}\ \emph {et~al.}(2018)\citenamefont
 {Franceschini}, \citenamefont {Panico}, \citenamefont {Pomarol},
 \citenamefont {Riva},\ and\ \citenamefont {Wulzer}}]{Franceschini:2017xkh}%
 \BibitemOpen
 \bibfield {author} {\bibinfo {author} {\bibfnamefont {R.}~\bibnamefont
 {Franceschini}}, \bibinfo {author} {\bibfnamefont {G.}~\bibnamefont
 {Panico}}, \bibinfo {author} {\bibfnamefont {A.}~\bibnamefont {Pomarol}},
 \bibinfo {author} {\bibfnamefont {F.}~\bibnamefont {Riva}}, \ and\ \bibinfo
 {author} {\bibfnamefont {A.}~\bibnamefont {Wulzer}},\ }\href {\doibase
 10.1007/JHEP02(2018)111} {\bibfield {journal} {\bibinfo {journal} {JHEP}\
 }\textbf {\bibinfo {volume} {02}},\ \bibinfo {pages} {111} (\bibinfo {year}
 {2018})},\ \Eprint {http://arxiv.org/abs/1712.01310} {arXiv:1712.01310
 [hep-ph]}\BibitemShut {NoStop}%
\bibitem [{\citenamefont {Chiesa}\ \emph {et~al.}(2018)\citenamefont {Chiesa},
 \citenamefont {Denner},\ and\ \citenamefont {Lang}}]{Chiesa:2018lcs}%
 \BibitemOpen
 \bibfield {author} {\bibinfo {author} {\bibfnamefont {M.}~\bibnamefont
 {Chiesa}}, \bibinfo {author} {\bibfnamefont {A.}~\bibnamefont {Denner}}, \
 and\ \bibinfo {author} {\bibfnamefont {J.-N.}\ \bibnamefont {Lang}},\ }\href
 {\doibase 10.1140/epjc/s10052-018-5949-z} {\bibfield {journal} {\bibinfo
 {journal} {Eur. Phys. J. C}\ }\textbf {\bibinfo {volume} {78}},\ \bibinfo
 {pages} {467} (\bibinfo {year} {2018})},\ \Eprint
 {http://arxiv.org/abs/1804.01477} {arXiv:1804.01477 [hep-ph]}\BibitemShut
 {NoStop}%
\bibitem [{\citenamefont {Liu}\ and\ \citenamefont {Wang}(2019)}]{Liu:2018pkg}%
 \BibitemOpen
 \bibfield {author} {\bibinfo {author} {\bibfnamefont {D.}~\bibnamefont
 {Liu}}\ and\ \bibinfo {author} {\bibfnamefont {L.-T.}\ \bibnamefont {Wang}},\
 }\href {\doibase 10.1103/PhysRevD.99.055001} {\bibfield {journal} {\bibinfo
 {journal} {Phys. Rev. D}\ }\textbf {\bibinfo {volume} {99}},\ \bibinfo
 {pages} {055001} (\bibinfo {year} {2019})},\ \Eprint
 {http://arxiv.org/abs/1804.08688} {arXiv:1804.08688 [hep-ph]}\BibitemShut
 {NoStop}%
\bibitem [{\citenamefont {Grojean}\ \emph {et~al.}(2019)\citenamefont
 {Grojean}, \citenamefont {Montull},\ and\ \citenamefont
 {Riembau}}]{Grojean:2018dqj}%
 \BibitemOpen
 \bibfield {author} {\bibinfo {author} {\bibfnamefont {C.}~\bibnamefont
 {Grojean}}, \bibinfo {author} {\bibfnamefont {M.}~\bibnamefont {Montull}}, \
 and\ \bibinfo {author} {\bibfnamefont {M.}~\bibnamefont {Riembau}},\ }\href
 {\doibase 10.1007/JHEP03(2019)020} {\bibfield {journal} {\bibinfo {journal}
 {JHEP}\ }\textbf {\bibinfo {volume} {03}},\ \bibinfo {pages} {020} (\bibinfo
 {year} {2019})},\ \Eprint {http://arxiv.org/abs/1810.05149} {arXiv:1810.05149
 [hep-ph]}\BibitemShut {NoStop}%
\bibitem [{\citenamefont {Baglio}\ \emph
 {et~al.}(2019{\natexlab{a}})\citenamefont {Baglio}, \citenamefont {Dawson},\
 and\ \citenamefont {Lewis}}]{Baglio:2018bkm}%
 \BibitemOpen
 \bibfield {author} {\bibinfo {author} {\bibfnamefont {J.}~\bibnamefont
 {Baglio}}, \bibinfo {author} {\bibfnamefont {S.}~\bibnamefont {Dawson}}, \
 and\ \bibinfo {author} {\bibfnamefont {I.~M.}\ \bibnamefont {Lewis}},\ }\href
 {\doibase 10.1103/PhysRevD.99.035029} {\bibfield {journal} {\bibinfo
 {journal} {Phys. Rev. D}\ }\textbf {\bibinfo {volume} {99}},\ \bibinfo
 {pages} {035029} (\bibinfo {year} {2019}{\natexlab{a}})},\ \Eprint
 {http://arxiv.org/abs/1812.00214} {arXiv:1812.00214 [hep-ph]}\BibitemShut
 {NoStop}%
\bibitem [{\citenamefont {Azatov}\ \emph {et~al.}(2019)\citenamefont {Azatov},
 \citenamefont {Barducci},\ and\ \citenamefont {Venturini}}]{Azatov:2019xxn}%
 \BibitemOpen
 \bibfield {author} {\bibinfo {author} {\bibfnamefont {A.}~\bibnamefont
 {Azatov}}, \bibinfo {author} {\bibfnamefont {D.}~\bibnamefont {Barducci}}, \
 and\ \bibinfo {author} {\bibfnamefont {E.}~\bibnamefont {Venturini}},\ }\href
 {\doibase 10.1007/JHEP04(2019)075} {\bibfield {journal} {\bibinfo {journal}
 {JHEP}\ }\textbf {\bibinfo {volume} {04}},\ \bibinfo {pages} {075} (\bibinfo
 {year} {2019})},\ \Eprint {http://arxiv.org/abs/1901.04821} {arXiv:1901.04821
 [hep-ph]}\BibitemShut {NoStop}%
\bibitem [{\citenamefont {Baglio}\ \emph
 {et~al.}(2019{\natexlab{b}})\citenamefont {Baglio}, \citenamefont {Dawson},\
 and\ \citenamefont {Homiller}}]{Baglio:2019uty}%
 \BibitemOpen
 \bibfield {author} {\bibinfo {author} {\bibfnamefont {J.}~\bibnamefont
 {Baglio}}, \bibinfo {author} {\bibfnamefont {S.}~\bibnamefont {Dawson}}, \
 and\ \bibinfo {author} {\bibfnamefont {S.}~\bibnamefont {Homiller}},\ }\href
 {\doibase 10.1103/PhysRevD.100.113010} {\bibfield {journal} {\bibinfo
 {journal} {Phys. Rev. D}\ }\textbf {\bibinfo {volume} {100}},\ \bibinfo
 {pages} {113010} (\bibinfo {year} {2019}{\natexlab{b}})},\ \Eprint
 {http://arxiv.org/abs/1909.11576} {arXiv:1909.11576 [hep-ph]}\BibitemShut
 {NoStop}%
\bibitem [{\citenamefont {Baglio}\ \emph {et~al.}(2020)\citenamefont {Baglio},
 \citenamefont {Dawson}, \citenamefont {Homiller}, \citenamefont {Lane},\ and\
 \citenamefont {Lewis}}]{Baglio:2020oqu}%
 \BibitemOpen
 \bibfield {author} {\bibinfo {author} {\bibfnamefont {J.}~\bibnamefont
 {Baglio}}, \bibinfo {author} {\bibfnamefont {S.}~\bibnamefont {Dawson}},
 \bibinfo {author} {\bibfnamefont {S.}~\bibnamefont {Homiller}}, \bibinfo
 {author} {\bibfnamefont {S.~D.}\ \bibnamefont {Lane}}, \ and\ \bibinfo
 {author} {\bibfnamefont {I.~M.}\ \bibnamefont {Lewis}},\ }\href {\doibase
 10.1103/PhysRevD.101.115004} {\bibfield {journal} {\bibinfo {journal}
 {Phys. Rev. D}\ }\textbf {\bibinfo {volume} {101}},\ \bibinfo {pages}
 {115004} (\bibinfo {year} {2020})},\ \Eprint
 {http://arxiv.org/abs/2003.07862} {arXiv:2003.07862 [hep-ph]}\BibitemShut
 {NoStop}%
\bibitem [{\citenamefont {Ellis}\ \emph {et~al.}(2021)\citenamefont {Ellis},
 \citenamefont {Madigan}, \citenamefont {Mimasu}, \citenamefont {Sanz},\ and\
 \citenamefont {You}}]{Ellis:2020unq}%
 \BibitemOpen
 \bibfield {author} {\bibinfo {author} {\bibfnamefont {J.}~\bibnamefont
 {Ellis}}, \bibinfo {author} {\bibfnamefont {M.}~\bibnamefont {Madigan}},
 \bibinfo {author} {\bibfnamefont {K.}~\bibnamefont {Mimasu}}, \bibinfo
 {author} {\bibfnamefont {V.}~\bibnamefont {Sanz}}, \ and\ \bibinfo {author}
 {\bibfnamefont {T.}~\bibnamefont {You}},\ }\href {\doibase
 10.1007/JHEP04(2021)279} {\bibfield {journal} {\bibinfo {journal} {JHEP}\
 }\textbf {\bibinfo {volume} {04}},\ \bibinfo {pages} {279} (\bibinfo {year}
 {2021})},\ \Eprint {http://arxiv.org/abs/2012.02779} {arXiv:2012.02779
 [hep-ph]}\BibitemShut {NoStop}%
\bibitem [{\citenamefont {Degrande}\ and\ \citenamefont
 {Touch\`eque}(2022)}]{Degrande:2021zpv}%
 \BibitemOpen
 \bibfield {author} {\bibinfo {author} {\bibfnamefont {C.}~\bibnamefont
 {Degrande}}\ and\ \bibinfo {author} {\bibfnamefont {J.}~\bibnamefont
 {Touch\`eque}},\ }\href {\doibase 10.1007/JHEP04(2022)032} {\bibfield
 {journal} {\bibinfo {journal} {JHEP}\ }\textbf {\bibinfo {volume} {04}},\
 \bibinfo {pages} {032} (\bibinfo {year} {2022})},\ \Eprint
 {http://arxiv.org/abs/2110.02993} {arXiv:2110.02993 [hep-ph]}\BibitemShut
 {NoStop}%
\bibitem [{\citenamefont {Degrande}\ and\ \citenamefont
 {Li}(2023)}]{Degrande:2023iob}%
 \BibitemOpen
 \bibfield {author} {\bibinfo {author} {\bibfnamefont {C.}~\bibnamefont
 {Degrande}}\ and\ \bibinfo {author} {\bibfnamefont {H.-L.}\ \bibnamefont
 {Li}},\ }\href {\doibase 10.1007/JHEP06(2023)149} {\bibfield {journal}
 {\bibinfo {journal} {JHEP}\ }\textbf {\bibinfo {volume} {06}},\ \bibinfo
 {pages} {149} (\bibinfo {year} {2023})},\ \Eprint
 {http://arxiv.org/abs/2303.10493} {arXiv:2303.10493 [hep-ph]}\BibitemShut
 {NoStop}%
\bibitem [{\citenamefont {Aoude}\ \emph {et~al.}(2023)\citenamefont {Aoude},
 \citenamefont {Madge}, \citenamefont {Maltoni},\ and\ \citenamefont
 {Mantani}}]{Aoude:2023hxv}%
 \BibitemOpen
 \bibfield {author} {\bibinfo {author} {\bibfnamefont {R.}~\bibnamefont
 {Aoude}}, \bibinfo {author} {\bibfnamefont {E.}~\bibnamefont {Madge}},
 \bibinfo {author} {\bibfnamefont {F.}~\bibnamefont {Maltoni}}, \ and\
 \bibinfo {author} {\bibfnamefont {L.}~\bibnamefont {Mantani}},\ }\href
 {\doibase 10.1007/JHEP12(2023)017} {\bibfield {journal} {\bibinfo {journal}
 {JHEP}\ }\textbf {\bibinfo {volume} {12}},\ \bibinfo {pages} {017} (\bibinfo
 {year} {2023})},\ \Eprint {http://arxiv.org/abs/2307.09675} {arXiv:2307.09675
 [hep-ph]}\BibitemShut {NoStop}%
\bibitem [{\citenamefont {Degrande}\ and\ \citenamefont
 {Maltoni}(2024)}]{Degrande:2024bmd}%
 \BibitemOpen
 \bibfield {author} {\bibinfo {author} {\bibfnamefont {C.}~\bibnamefont
 {Degrande}}\ and\ \bibinfo {author} {\bibfnamefont {M.}~\bibnamefont
 {Maltoni}},\ }\href {\doibase 10.1016/j.physletb.2024.138970} {\bibfield
 {journal} {\bibinfo {journal} {Phys. Lett. B}\ }\textbf {\bibinfo {volume}
 {856}},\ \bibinfo {pages} {138970} (\bibinfo {year} {2024})},\ \Eprint
 {http://arxiv.org/abs/2403.16894} {arXiv:2403.16894 [hep-ph]}\BibitemShut
 {NoStop}%
\bibitem [{\citenamefont {El~Faham}\ \emph {et~al.}(2024)\citenamefont
 {El~Faham}, \citenamefont {Pelliccioli},\ and\ \citenamefont
 {Vryonidou}}]{ElFaham:2024uop}%
 \BibitemOpen
 \bibfield {author} {\bibinfo {author} {\bibfnamefont {H.}~\bibnamefont
 {El~Faham}}, \bibinfo {author} {\bibfnamefont {G.}~\bibnamefont
 {Pelliccioli}}, \ and\ \bibinfo {author} {\bibfnamefont {E.}~\bibnamefont
 {Vryonidou}},\ }\href {\doibase 10.1007/JHEP08(2024)087} {\bibfield
 {journal} {\bibinfo {journal} {JHEP}\ }\textbf {\bibinfo {volume} {08}},\
 \bibinfo {pages} {087} (\bibinfo {year} {2024})},\ \Eprint
 {http://arxiv.org/abs/2405.19083} {arXiv:2405.19083 [hep-ph]}\BibitemShut
 {NoStop}%
\bibitem [{\citenamefont {Banerjee}\ \emph {et~al.}(2024)\citenamefont
 {Banerjee}, \citenamefont {Reichelt},\ and\ \citenamefont
 {Spannowsky}}]{Banerjee:2024eyo}%
 \BibitemOpen
 \bibfield {author} {\bibinfo {author} {\bibfnamefont {S.}~\bibnamefont
 {Banerjee}}, \bibinfo {author} {\bibfnamefont {D.}~\bibnamefont {Reichelt}},
 \ and\ \bibinfo {author} {\bibfnamefont {M.}~\bibnamefont {Spannowsky}},\
 }\href {\doibase 10.1103/PhysRevD.110.115012} {\bibfield {journal} {\bibinfo
 {journal} {Phys. Rev. D}\ }\textbf {\bibinfo {volume} {110}},\ \bibinfo
 {pages} {115012} (\bibinfo {year} {2024})},\ \Eprint
 {http://arxiv.org/abs/2406.15640} {arXiv:2406.15640 [hep-ph]}\BibitemShut
 {NoStop}%
\bibitem [{\citenamefont {Thomas}\ and\ \citenamefont
 {Vryonidou}(2025)}]{Thomas:2024dwd}%
 \BibitemOpen
 \bibfield {author} {\bibinfo {author} {\bibfnamefont {M.~O.~A.}\
 \bibnamefont {Thomas}}\ and\ \bibinfo {author} {\bibfnamefont
 {E.}~\bibnamefont {Vryonidou}},\ }\href {\doibase 10.1007/JHEP03(2025)038}
 {\bibfield {journal} {\bibinfo {journal} {JHEP}\ }\textbf {\bibinfo
 {volume} {03}},\ \bibinfo {pages} {038} (\bibinfo {year} {2025})},\ \Eprint
 {http://arxiv.org/abs/2411.00959} {arXiv:2411.00959 [hep-ph]}\BibitemShut
 {NoStop}%
\bibitem [{\citenamefont {Haisch}\ \emph {et~al.}(2025)\citenamefont {Haisch},
 \citenamefont {Linder}, \citenamefont {Pelliccioli}, \citenamefont {Re},\
 and\ \citenamefont {Zanderighi}}]{Haisch:2025jqr}%
 \BibitemOpen
 \bibfield {author} {\bibinfo {author} {\bibfnamefont {U.}~\bibnamefont
 {Haisch}}, \bibinfo {author} {\bibfnamefont {J.}~\bibnamefont {Linder}},
 \bibinfo {author} {\bibfnamefont {G.}~\bibnamefont {Pelliccioli}}, \bibinfo
 {author} {\bibfnamefont {E.}~\bibnamefont {Re}}, \ and\ \bibinfo {author}
 {\bibfnamefont {G.}~\bibnamefont {Zanderighi}},\ }\href {\doibase
 10.1007/JHEP11(2025)080} {\bibfield {journal} {\bibinfo {journal} {JHEP}\
 }\textbf {\bibinfo {volume} {11}},\ \bibinfo {pages} {080} (\bibinfo {year}
 {2025})},\ \Eprint {http://arxiv.org/abs/2507.21768} {arXiv:2507.21768
 [hep-ph]}\BibitemShut {NoStop}%
\bibitem [{\citenamefont {El~Faham}\ \emph {et~al.}(2026)\citenamefont
 {El~Faham}, \citenamefont {Ventura},\ and\ \citenamefont
 {Vryonidou}}]{ElFaham:2025fow}%
 \BibitemOpen
 \bibfield {author} {\bibinfo {author} {\bibfnamefont {H.}~\bibnamefont
 {El~Faham}}, \bibinfo {author} {\bibfnamefont {G.}~\bibnamefont {Ventura}}, \
 and\ \bibinfo {author} {\bibfnamefont {E.}~\bibnamefont {Vryonidou}},\ }\href
 {\doibase 10.1007/JHEP04(2026)050} {\bibfield {journal} {\bibinfo {journal}
 {JHEP}\ }\textbf {\bibinfo {volume} {04}},\ \bibinfo {pages} {050} (\bibinfo
 {year} {2026})},\ \Eprint {http://arxiv.org/abs/2511.04338} {arXiv:2511.04338
 [hep-ph]}\BibitemShut {NoStop}%
\bibitem [{\citenamefont {Pelliccioli}\ and\ \citenamefont
 {Re}(2026)}]{Pelliccioli:2026ltl}%
 \BibitemOpen
 \bibfield {author} {\bibinfo {author} {\bibfnamefont {G.}~\bibnamefont
 {Pelliccioli}}\ and\ \bibinfo {author} {\bibfnamefont {E.}~\bibnamefont
 {Re}},\ }\href@noop {} {\ (\bibinfo {year} {2026})},\ \Eprint
 {http://arxiv.org/abs/2601.09540} {arXiv:2601.09540 [hep-ph]}\BibitemShut
 {NoStop}%
\bibitem [{\citenamefont {Azatov}\ \emph
 {et~al.}(2017{\natexlab{b}})\citenamefont {Azatov}, \citenamefont {Contino},
 \citenamefont {Machado},\ and\ \citenamefont {Riva}}]{Azatov:2016sqh}%
 \BibitemOpen
 \bibfield {author} {\bibinfo {author} {\bibfnamefont {A.}~\bibnamefont
 {Azatov}}, \bibinfo {author} {\bibfnamefont {R.}~\bibnamefont {Contino}},
 \bibinfo {author} {\bibfnamefont {C.~S.}\ \bibnamefont {Machado}}, \ and\
 \bibinfo {author} {\bibfnamefont {F.}~\bibnamefont {Riva}},\ }\href {\doibase
 10.1103/PhysRevD.95.065014} {\bibfield {journal} {\bibinfo {journal} {Phys.
 Rev. D}\ }\textbf {\bibinfo {volume} {95}},\ \bibinfo {pages} {065014}
 (\bibinfo {year} {2017}{\natexlab{b}})},\ \Eprint
 {http://arxiv.org/abs/1607.05236} {arXiv:1607.05236 [hep-ph]}\BibitemShut
 {NoStop}%
\bibitem [{\citenamefont {Hagiwara}\ \emph {et~al.}(1987)\citenamefont
 {Hagiwara}, \citenamefont {Peccei}, \citenamefont {Zeppenfeld},\ and\
 \citenamefont {Hikasa}}]{Hagiwara:1986vm}%
 \BibitemOpen
 \bibfield {author} {\bibinfo {author} {\bibfnamefont {K.}~\bibnamefont
 {Hagiwara}}, \bibinfo {author} {\bibfnamefont {R.~D.}\ \bibnamefont
 {Peccei}}, \bibinfo {author} {\bibfnamefont {D.}~\bibnamefont {Zeppenfeld}},
 \ and\ \bibinfo {author} {\bibfnamefont {K.}~\bibnamefont {Hikasa}},\ }\href
 {\doibase 10.1016/0550-3213(87)90685-7} {\bibfield {journal} {\bibinfo
 {journal} {Nucl. Phys. B}\ }\textbf {\bibinfo {volume} {282}},\ \bibinfo
 {pages} {253} (\bibinfo {year} {1987})}\BibitemShut {NoStop}%
\bibitem [{\citenamefont {Navas}\ \emph {et~al.}(2024)\citenamefont {Navas}
 \emph {et~al.}}]{ParticleDataGroup:2024cfk}%
 \BibitemOpen
 \bibfield {author} {\bibinfo {author} {\bibfnamefont {S.}~\bibnamefont
 {Navas}} \emph {et~al.} (\bibinfo {collaboration} {Particle Data Group}),\
 }\href {\doibase 10.1103/PhysRevD.110.030001} {\bibfield {journal} {\bibinfo
 {journal} {Phys. Rev. D}\ }\textbf {\bibinfo {volume} {110}},\ \bibinfo
 {pages} {030001} (\bibinfo {year} {2024})}\BibitemShut {NoStop}%
\bibitem [{\citenamefont {Bardin}\ \emph {et~al.}(1988)\citenamefont {Bardin},
 \citenamefont {Leike}, \citenamefont {Riemann},\ and\ \citenamefont
 {Sachwitz}}]{Bardin:1988xt}%
 \BibitemOpen
 \bibfield {author} {\bibinfo {author} {\bibfnamefont {D.}~\bibnamefont
 {Bardin}}, \bibinfo {author} {\bibfnamefont {A.}~\bibnamefont {Leike}},
 \bibinfo {author} {\bibfnamefont {T.}~\bibnamefont {Riemann}}, \ and\
 \bibinfo {author} {\bibfnamefont {M.}~\bibnamefont {Sachwitz}},\ }\href
 {\doibase 10.1016/0370-2693(88)91627-9} {\bibfield {journal} {\bibinfo
 {journal} {Phys. Lett. B}\ }\textbf {\bibinfo {volume} {206}},\ \bibinfo
 {pages} {539} (\bibinfo {year} {1988})}\BibitemShut {NoStop}%
\bibitem [{\citenamefont {Denner}\ \emph {et~al.}(2000)\citenamefont {Denner},
 \citenamefont {Dittmaier}, \citenamefont {Roth},\ and\ \citenamefont
 {Wackeroth}}]{Denner:2000bj}%
 \BibitemOpen
 \bibfield {author} {\bibinfo {author} {\bibfnamefont {A.}~\bibnamefont
 {Denner}}, \bibinfo {author} {\bibfnamefont {S.}~\bibnamefont {Dittmaier}},
 \bibinfo {author} {\bibfnamefont {M.}~\bibnamefont {Roth}}, \ and\ \bibinfo
 {author} {\bibfnamefont {D.}~\bibnamefont {Wackeroth}},\ }\href {\doibase
 10.1016/S0550-3213(00)00511-3} {\bibfield {journal} {\bibinfo {journal}
 {Nucl. Phys. B}\ }\textbf {\bibinfo {volume} {587}},\ \bibinfo {pages} {67}
 (\bibinfo {year} {2000})},\ \Eprint {http://arxiv.org/abs/hep-ph/0006307}
 {arXiv:hep-ph/0006307 [hep-ph]}\BibitemShut {NoStop}%
\bibitem [{\citenamefont {Buckley}\ \emph {et~al.}(2015)\citenamefont
 {Buckley}, \citenamefont {Ferrando}, \citenamefont {Lloyd}, \citenamefont
 {Nordstr{\"o}m}, \citenamefont {Page}, \citenamefont {R{\"u}fenacht},
 \citenamefont {Sch{\"o}nherr},\ and\ \citenamefont {Watt}}]{Buckley:2014ana}%
 \BibitemOpen
 \bibfield {author} {\bibinfo {author} {\bibfnamefont {A.}~\bibnamefont
 {Buckley}}, \bibinfo {author} {\bibfnamefont {J.}~\bibnamefont {Ferrando}},
 \bibinfo {author} {\bibfnamefont {S.}~\bibnamefont {Lloyd}}, \bibinfo
 {author} {\bibfnamefont {K.}~\bibnamefont {Nordstr{\"o}m}}, \bibinfo {author}
 {\bibfnamefont {B.}~\bibnamefont {Page}}, \bibinfo {author} {\bibfnamefont
 {M.}~\bibnamefont {R{\"u}fenacht}}, \bibinfo {author} {\bibfnamefont
 {M.}~\bibnamefont {Sch{\"o}nherr}}, \ and\ \bibinfo {author} {\bibfnamefont
 {G.}~\bibnamefont {Watt}},\ }\href {\doibase 10.1140/epjc/s10052-015-3318-8}
 {\bibfield {journal} {\bibinfo {journal} {Eur. Phys. J. C}\ }\textbf
 {\bibinfo {volume} {75}},\ \bibinfo {pages} {132} (\bibinfo {year} {2015})},\
 \Eprint {http://arxiv.org/abs/1412.7420} {arXiv:1412.7420 [hep-ph]}\BibitemShut {NoStop}%
\bibitem [{\citenamefont {Ball}\ \emph {et~al.}(2017)\citenamefont {Ball} \emph
 {et~al.}}]{NNPDF:2017mvq}%
 \BibitemOpen
 \bibfield {author} {\bibinfo {author} {\bibfnamefont {R.~D.}\ \bibnamefont
 {Ball}} \emph {et~al.} (\bibinfo {collaboration} {NNPDF}),\ }\href {\doibase
 10.1140/epjc/s10052-017-5199-5} {\bibfield {journal} {\bibinfo {journal}
 {Eur. Phys. J. C}\ }\textbf {\bibinfo {volume} {77}},\ \bibinfo {pages} {663}
 (\bibinfo {year} {2017})},\ \Eprint {http://arxiv.org/abs/1706.00428}
 {arXiv:1706.00428 [hep-ph]}\BibitemShut {NoStop}%
\bibitem [{\citenamefont {Aaboud}\ \emph {et~al.}(2019)\citenamefont {Aaboud}
 \emph {et~al.}}]{ATLAS:2019rob}%
 \BibitemOpen
 \bibfield {author} {\bibinfo {author} {\bibfnamefont {M.}~\bibnamefont
 {Aaboud}} \emph {et~al.} (\bibinfo {collaboration} {ATLAS}),\ }\href
 {\doibase 10.1140/epjc/s10052-019-7371-6} {\bibfield {journal} {\bibinfo
 {journal} {Eur. Phys. J. C}\ }\textbf {\bibinfo {volume} {79}},\ \bibinfo
 {pages} {884} (\bibinfo {year} {2019})},\ \Eprint
 {http://arxiv.org/abs/1905.04242} {arXiv:1905.04242 [hep-ex]}\BibitemShut
 {NoStop}%
\bibitem [{\citenamefont {Lombardi}\ \emph {et~al.}(2021)\citenamefont
 {Lombardi}, \citenamefont {Wiesemann},\ and\ \citenamefont
 {Zanderighi}}]{Lombardi:2021rvg}%
 \BibitemOpen
 \bibfield {author} {\bibinfo {author} {\bibfnamefont {D.}~\bibnamefont
 {Lombardi}}, \bibinfo {author} {\bibfnamefont {M.}~\bibnamefont {Wiesemann}},
 \ and\ \bibinfo {author} {\bibfnamefont {G.}~\bibnamefont {Zanderighi}},\
 }\href {\doibase 10.1007/JHEP11(2021)230} {\bibfield {journal} {\bibinfo
 {journal} {JHEP}\ }\textbf {\bibinfo {volume} {11}},\ \bibinfo {pages} {230}
 (\bibinfo {year} {2021})},\ \Eprint {http://arxiv.org/abs/2103.12077}
 {arXiv:2103.12077 [hep-ph]}\BibitemShut {NoStop}%
\bibitem [{\citenamefont {Cacciari}\ \emph {et~al.}(2012)\citenamefont
 {Cacciari}, \citenamefont {Salam},\ and\ \citenamefont
 {Soyez}}]{Cacciari:2011ma}%
 \BibitemOpen
 \bibfield {author} {\bibinfo {author} {\bibfnamefont {M.}~\bibnamefont
 {Cacciari}}, \bibinfo {author} {\bibfnamefont {G.~P.}\ \bibnamefont {Salam}},
 \ and\ \bibinfo {author} {\bibfnamefont {G.}~\bibnamefont {Soyez}},\ }\href
 {\doibase 10.1140/epjc/s10052-012-1896-2} {\bibfield {journal} {\bibinfo
 {journal} {Eur. Phys. J. C}\ }\textbf {\bibinfo {volume} {72}},\ \bibinfo
 {pages} {1896} (\bibinfo {year} {2012})},\ \Eprint
 {http://arxiv.org/abs/1111.6097} {arXiv:1111.6097 [hep-ph]}\BibitemShut
 {NoStop}%
\bibitem [{\citenamefont {Cacciari}\ \emph {et~al.}(2008)\citenamefont
 {Cacciari}, \citenamefont {Salam},\ and\ \citenamefont
 {Soyez}}]{Cacciari:2008gp}%
 \BibitemOpen
 \bibfield {author} {\bibinfo {author} {\bibfnamefont {M.}~\bibnamefont
 {Cacciari}}, \bibinfo {author} {\bibfnamefont {G.~P.}\ \bibnamefont {Salam}},
 \ and\ \bibinfo {author} {\bibfnamefont {G.}~\bibnamefont {Soyez}},\ }\href
 {\doibase 10.1088/1126-6708/2008/04/063} {\bibfield {journal} {\bibinfo
 {journal} {JHEP}\ }\textbf {\bibinfo {volume} {04}},\ \bibinfo {pages} {063}
 (\bibinfo {year} {2008})},\ \Eprint {http://arxiv.org/abs/0802.1189}
 {arXiv:0802.1189 [hep-ph]}\BibitemShut {NoStop}%
 \bibitem [{\citenamefont {Sj\"ostrand}\ \emph {et~al.}(2015)\citenamefont
 {Sj\"ostrand}, \citenamefont {Ask}, \citenamefont {Christiansen},
 \citenamefont {Corke}, \citenamefont {Desai}, \citenamefont {Ilten},
 \citenamefont {Mrenna}, \citenamefont {Prestel}, \citenamefont {Rasmussen},\
 and\ \citenamefont {Skands}}]{Sjostrand:2014zea}%
 \BibitemOpen
 \bibfield {author} {\bibinfo {author} {\bibfnamefont {T.}~\bibnamefont
 {Sj\"ostrand}}, \bibinfo {author} {\bibfnamefont {S.}~\bibnamefont {Ask}},
 \bibinfo {author} {\bibfnamefont {J.~R.}\ \bibnamefont {Christiansen}},
 \bibinfo {author} {\bibfnamefont {R.}~\bibnamefont {Corke}}, \bibinfo
 {author} {\bibfnamefont {N.}~\bibnamefont {Desai}}, \bibinfo {author}
 {\bibfnamefont {P.}~\bibnamefont {Ilten}}, \bibinfo {author} {\bibfnamefont
 {S.}~\bibnamefont {Mrenna}}, \bibinfo {author} {\bibfnamefont
 {S.}~\bibnamefont {Prestel}}, \bibinfo {author} {\bibfnamefont {C.~O.}\
 \bibnamefont {Rasmussen}}, \ and\ \bibinfo {author} {\bibfnamefont {P.~Z.}\
 \bibnamefont {Skands}},\ }\href {\doibase 10.1016/j.cpc.2015.01.024}
 {\bibfield {journal} {\bibinfo {journal} {Comput. Phys. Commun.}\ }\textbf
 {\bibinfo {volume} {191}},\ \bibinfo {pages} {159} (\bibinfo {year}
 {2015})},\ \Eprint {http://arxiv.org/abs/1410.3012} {arXiv:1410.3012
 [hep-ph]}\BibitemShut {NoStop}%
\bibitem [{\citenamefont {Monni}\ \emph {et~al.}(2020)\citenamefont {Monni},
  \citenamefont {Nason}, \citenamefont {Re}, \citenamefont {Wiesemann},\ and\
  \citenamefont {Zanderighi}}]{Monni:2019whf}%
  \BibitemOpen
  \bibfield  {author} {\bibinfo {author} {\bibfnamefont {P.~F.}\ \bibnamefont
  {Monni}}, \bibinfo {author} {\bibfnamefont {P.}~\bibnamefont {Nason}},
  \bibinfo {author} {\bibfnamefont {E.}~\bibnamefont {Re}}, \bibinfo {author}
  {\bibfnamefont {M.}~\bibnamefont {Wiesemann}}, \ and\ \bibinfo {author}
  {\bibfnamefont {G.}~\bibnamefont {Zanderighi}},\ }\href {\doibase
  10.1007/JHEP05(2020)143} {\bibfield  {journal} {\bibinfo  {journal} {JHEP}\
  }\textbf {\bibinfo {volume} {05}},\ \bibinfo {pages} {143} (\bibinfo {year}
  {2020})},\ \bibinfo {note} {[Erratum: JHEP {\bf 02}, 031 (2022)]},\ \Eprint
  {http://arxiv.org/abs/1908.06987} {arXiv:1908.06987 [hep-ph]}\BibitemShut
  {NoStop}%
\bibitem [{\citenamefont {Ellis}(2017)}]{Ellis:2016jkw}%
 \BibitemOpen
 \bibfield {author} {\bibinfo {author} {\bibfnamefont {J.}~\bibnamefont
 {Ellis}},\ }\href {\doibase 10.1016/j.cpc.2016.08.019} {\bibfield {journal}
 {\bibinfo {journal} {Comput. Phys. Commun.}\ }\textbf {\bibinfo {volume}
 {210}},\ \bibinfo {pages} {103} (\bibinfo {year} {2017})},\ \Eprint
 {http://arxiv.org/abs/1601.05437} {arXiv:1601.05437 [hep-ph]}\BibitemShut
 {NoStop}%
\end{thebibliography}


%

\end{document}